\documentclass[12pt,a4paper]{article}
\pdfoutput=1

\usepackage[T1]{fontenc}
\usepackage[utf8]{inputenc}
\usepackage[nomath]{lmodern}
\usepackage{microtype}
\usepackage{jheppub}
\hypersetup{pdfencoding=unicode, bookmarksopen=true, bookmarksnumbered}
\usepackage{amsmath}
\usepackage{amsfonts}
\usepackage{amssymb}
\usepackage{mathrsfs}
\usepackage{mathtools}
\usepackage{physics}
\usepackage{graphicx}
\usepackage{caption}
\usepackage{subcaption}
\usepackage{enumitem}
\usepackage{framed,float}
\usepackage{booktabs}
\usepackage{array}
\usepackage{soul} 
\usepackage{tikz}
\usepackage{mathdots}
\usetikzlibrary{calc,arrows,decorations.markings,decorations.pathreplacing,decorations.pathmorphing}
\usepackage{multirow}
\usepackage{anyfontsize}
\newenvironment{Tabular}[1]
{\tabular{#1}}
{\endtabular}

\def\CN{{\cal N}}

\newcommand{\PE}{\operatorname{PE}}

\newcolumntype{L}[1]{>{\raggedright\let\newline\\\arraybackslash\hspace{0pt}}m{#1}}
\newcolumntype{C}[1]{>{\centering\let\newline\\\arraybackslash\hspace{0pt}}m{#1}}
\newcolumntype{R}[1]{>{\raggedleft\let\newline\\\arraybackslash\hspace{0pt}}m{#1}}

\tikzset{%
	show curve controls/.style={
		postaction={
			decoration={
				show path construction,
				curveto code={
					\draw [blue] 
					(\tikzinputsegmentfirst) -- (\tikzinputsegmentsupporta)
					(\tikzinputsegmentlast) -- (\tikzinputsegmentsupportb);
					\fill [red, opacity=0.5] 
					(\tikzinputsegmentsupporta) circle [radius=.5ex]
					(\tikzinputsegmentsupportb) circle [radius=.5ex];
				}
			},
			decorate
}}}

\title{Blowup Equations for Little Strings}

\author[a,b]{Hee-Cheol Kim,}
\author[a]{Minsung Kim,}
\author[a]{Yuji Sugimoto}

\affiliation[a]{Department of Physics, POSTECH, Pohang 790-784, Korea}
\affiliation[b]{Asia Pacific Center for Theoretical Physics, Postech, Pohang 37673, Korea}

\abstract{
We propose blowup equations for 6d little string theories which generalize Nakajima-Yoshioka's blowup equations for the 4d/5d instanton partition functions on Omega background.
We find that unlike the blowup equations for standard SQFTs, we need to sum over auxiliary magnetic fluxes on the blown-up $ \mathbb{P}^1$ for a non-dynamical 2-form gauge field which plays a role in canceling the mixed anomalies of the gauge symmetries. We demonstrate with explicit examples that the blowup equations, when combined with the modular properties, can be solved in order to determine the elliptic genera of little strings.
}

\begin{document}

\maketitle

\section{Introduction}

Since the introduction of supersymmetric quantum field theories (SQFTs), numerous advancements have been made.
From a classification perspective, five-dimensional supersymmetric field theories are classified by examining the consistency of physics on the Coulomb branch of moduli space \cite{Seiberg:1996bd,Morrison:1996xf,Intriligator:1997pq,Jefferson:2017ahm}, utilizing geometric descriptions \cite{Intriligator:1997pq,Xie:2017pfl,Jefferson:2018irk,Apruzzi:2019vpe,Apruzzi:2019opn,Apruzzi:2019enx,Bhardwaj:2019jtr}, and analyzing the RG-flows of 6d superconformal field theories (SCFTs) on a circle \cite{DelZotto:2017pti,Bhardwaj:2018yhy,Bhardwaj:2018vuu,Bhardwaj:2019fzv,Apruzzi:2019kgb,Bhardwaj:2020gyu,Bhardwaj:2020kim}.
Six-dimensional supersymmetric field theories are classified based on the types of non-compact bases and the methods of gluing them together in F-theory compactified on non-compact elliptically fibered Calabi-Yau threefolds \cite{Heckman:2013pva,Heckman:2015bfa, Bhardwaj:2015xxa,  Bhardwaj:2019hhd}.
The classification of 6d little string theories (LSTs) is also discussed in \cite{Bhardwaj:2015oru,Bhardwaj:2019hhd} which will be explained right after.

There have been significant quantitative studies conducted on higher dimensional SQFTs. For instance, it is possible to calculate the supersymmetric partition functions of these theories on $\Omega$-deformed $\mathbb{R}^4$ using various methods. This partition function is a type of Witten index that counts BPS states on the Coulomb branch. For theories with classical gauge groups, it can be computed using supersymmetric localization based on ADHM constructions of the instanton moduli space \cite{Nekrasov:2002qd,Nekrasov:2003rj} or using the topological vertex formalism introduced in \cite{Aganagic:2003db, Iqbal:2007ii, Awata:2008ed}.
We can also calculate the partition function in the presence of the codimension two or four defects in these theories \cite{Dimofte:2010tz, Awata:2010bz, Nekrasov:2013xda,Gaiotto:2014ina,Bullimore:2014upa,Bullimore:2014awa,Gaiotto:2015una, Nekrasov:2015wsu,Kim:2016qqs,Nekrasov:2016qym,Chang:2016iji,Mori:2016qof,Kimura:2017auj,Agarwal:2018tso,Assel:2018rcw, Kim:2020npz,Uhlemann:2020bek}.
While the SUSY localization and topological vertex formalism are effective methods for qualitatively studying higher dimensional SQFTs, one disadvantage of these methods is that we need to know ADHM constructions for instanton moduli space of gauge theories or brane web descriptions of these SQFTs in Type IIB string theory.

An alternative approach to calculating the instanton partition functions is by utilizing the blowup equation, which was initially developed for the study of Donaldson invariants in mathematics.
In \cite{Nakajima:2003pg},  a systematic method for formulating and solving the blowup equations to obtain Nekrasov's instanton partition functions of 4d $\mathcal{N}=2$ $SU(N)$ gauge theories was proposed, and this method was subsequently generalized to 5d $\mathcal{N}=1$ $SU(N)$ gauge theories in \cite{Nakajima:2005fg, Gottsche:2006bm}.
More recently, several extensions have been made to compute various observables in higher dimensional field theories:
the instanton partition functions of 5d SUSY gauge theories with generic gauge groups and matter representations were computed in \cite{Keller:2012da,Kim:2019uqw,Kim:2020hhh}, the blowup equations for refined topological strings on certain local Calabi-Yau 3-folds were formulated in \cite{Huang:2017mis,Kim:2020hhh}, and the elliptic genera of self-dual strings in 6d SCFTs on tensor branch were calculated using elliptic blowup equations in \cite{Gu:2019dan,Gu:2019pqj,Gu:2020fem,Kim:2020hhh}.
Also, as another extension, the blowup equations for 5d and 6d supersymmetric field theories with codimension four defects were proposed in \cite{Kim:2021gyj}.
One of the key benefits of the blowup approach is its capability to systematically calculate the partition functions, including non-perturbative contributions, through the use of the effective prepotential and consistent magnetic fluxes on the blowup background which can be systematically obtained for any 5d or 6d supersymmetric field theory \cite{Huang:2017mis,Kim:2020hhh}.
While we now have a large amount of examples for the blowup formalism, it remains to be verified whether it can be applied to little string theories.

Little string theories were originally introduced as worldvolume theories of NS5-branes in the gravity decoupling limit \cite{Berkooz:1997cq,Seiberg:1997zk, Losev:1997hx, Intriligator:1997dh, Brunner:1997gf}.
Depending on the space in which the NS5-branes reside, there are $\CN=(2,0),(1,1),$ and $(1,0)$ LSTs in 6 dimensions.
The decoupling limit is achieved by taking the string coupling constant to zero while maintaining an intrinsic string scaling finite.
The resulting theory becomes a non-local theory without gravity and it has some stringy properties such as T-duality.
In this sense, this theory is an intermediate theory between local quantum field theories and the usual string theories, so a deep understanding of the LSTs may help us understand both subjects.
Additionally, since NS5-branes are known to be one of the most challenging and interesting non-perturbative objects to study, a better understanding of these objects is desired.
Beside this, there are also various motivations for studying LSTs, including the investigation of discrete light-cone quantization and holography in a linear dilation background, etc.
For a brief overview of LSTs in these contexts, we recommend readers to see \cite{Aharony:1999ks, Kutasov:2001uf}.

A systematic construction of the LSTs is proposed based on the geometric phases of F-theory in \cite{Bhardwaj:2015oru}.
This construction allows us to classify LSTs according to the types of base curves and the manner in which they are connected, generalizing the geometric classification of 6d SCFTs. In quantitative studies, the partition functions of A-type LSTs engineered by NS5-branes on A-type singularities have been obtained using the localization method applied to the worldvolume theories of 2d instantonic strings \cite{Kim:2015gha}.
Similarly, the elliptic genera of LSTs on some of D-type singularities have been computed based on the localization method \cite{Kim:2017xan}.
Also, in \cite{Kim:2018gak}, the elliptic genera of some LSTs were calculated using T-dualities and the modular ansatz, which is based on the modular properties of the elliptic genera. However, these computations have only been carried out in a few simple cases, as the localization method requires ADHM-like constructions of little string worldvolume theories that are currently unavailable in most cases. 
The modular ansatz method also has some limitations, including a rapid increase in the number of unknown coefficients as the string number increases and the need for precise knowledge of T-duality for LSTs.
As such, it is important to find alternative methods for calculating the partition functions of LSTs.

In this paper, we propose a systematic method for constructing blowup equations for LSTs and provide examples of its application in the explicit calculation of partition functions. The blowup equations can be formulated by using two key ingredients: the effective prepotential evaluated on the $\Omega$-background, which can be obtained from the effective cubic and mixed Chern-Simons terms on the tensor branch, and a set of magnetic fluxes on the blown-up $\mathbb{P}^1$. We will explain how to obtain these ingredients for arbitrary LSTs.

It turns out that the blowup equations for LSTs are rather different from those for 5d/6d SQFTs. Interestingly, the blowup equations for LSTs involve summation over magnetic fluxes for an auxiliary gauge field as well as those for the dynamical gauge fields. This auxiliary gauge field is a non-dynamical 2-form field used to cancel mixed anomalies of gauge symmetries. We will explain the precise role of this auxiliary gauge field in the next section. However, we note that the summation over auxiliary magnetic fluxes is not convergent in terms of K\"ahler parameters. This is essentially due to the absence of a quadratic kinetic term for the auxiliary gauge field.  Despite this, we show that the partition functions of the LSTs calculated from other methods satisfy the blowup equations that we have proposed when certain upper and lower bounds are placed on the power of K\"ahler parameters coupled to the auxiliary magnetic fluxes.
Furthermore, we will show that elliptic genera of the LSTs can be calculated by solving the blowup equations when combined with the modular ansatz.
We illustrate our approach using $\hat{A}_1$ type IIA and IIB LSTs, $E_8\times E_8$ and $SO(32)$ Heterotic LSTs as rank-1 LSTs, and the $SU(3)$ gauge theory with a symmetric and an antisymmetric hypermultiplets as a rank-2 LST\footnote{Here we use a word ``rank'' as a number of dynamical parameters which is different from \cite{Bhardwaj:2015oru}.}.

The rest of this paper is organized as follows.
In Section~\ref{sec:blowupLST}, we provide a review of the blowup equations and modular bootstrap approach for 6d SCFTs, and present the proposed blowup equations for LSTs.
In Section~\ref{sec:example}, we demonstrate how our proposal works with several examples.
In Section~\ref{sec:conclusion}, we summarize our results and discuss some future directions.
In Appendix~\ref{app:elliptic}, we collect some facts about the elliptic functions used in the main context.
In Appendix~\ref{app:integral}, we present the computations of the elliptic genera of some LSTs using ADHM constructions of instantonic strings.

\section{Blowup equations and Modular bootstrap}\label{sec:blowupLST}

In this section, we propose the blowup equations for the partition functions of 6d little string theories on $T^2\times \mathbb{R}^4$. We begin by reviewing the formulation of blowup equations for 6d SCFTs and then extend this approach to construct the blowup equations for 6d LSTs. We also describe how to calculate the elliptic genera of strings in LSTs using their modular properties and by solving the blowup equations. 

The elliptic genera of strings in 6d LSTs on a torus $T^2$ times $\Omega$-deformed $\mathbb{R}^4$, which is a Witten index, is defined as
\begin{align}\label{ellipticgenus}
Z_k(\tau,\phi,m;\epsilon_{1,2}) = \Tr_{RR} \Big[ (-1)^F e^{2\pi i (\tau H_L - \bar{\tau} H_R)} e^{2\pi i \epsilon_1(J_1+J_R)} e^{2\pi i \epsilon_2(J_2+J_R)} e^{2\pi i \phi\cdot\Pi} e^{2\pi i m\cdot F} \Big] \, ,
\end{align}
where $ \tau $ is the complex structure of the torus, $ H_L $ and $ H_R $ are left-moving and right-moving Hamiltonians in the 2d worldsheet, $ J_1 $ and $ J_2 $ are Cartan generators of the $ SO(4) $ Lorentz rotation on $\mathbb{R}^4$, $ J_R $ is the Cartan for $ SU(2)_R $ R-symmetry, $ \epsilon_1 $ and $ \epsilon_2 $ are the $ \Omega $-deformation parameters, $ \Pi $ and $ F $ are gauge and flavor charges, and $ \phi $ and $ m $ collectively denote chemical potentials for gauge and flavor symmetries, repectively. The supercharge $ Q $ and its conjugate $ Q^\dagger $ commute with $ J_1+J_R $ and $ J_2+J_R $, and the right-moving Hamiltonian is given by $ 2H_R = \{ Q, Q^\dagger \} $. 
This elliptic genus counts BPS spectrum in the Ramond sector annihilated by $ Q $ and $ Q^\dagger $, so it is independent of $ \bar{\tau} $, in the 2d worldsheet SCFT living on strings with tensor charge denoted by $k$.

\subsection{Blowup equations for 6d SCFTs}\label{sec:blowup-CFT}

Let us review the blowup equations for the 6d SCFT which are functional equations for the full partition function defined by
\begin{align}\label{BPS partition function}
    Z(\tau,\varphi,\phi,m;\epsilon_{1,2}) = e^{-2\pi i \mathcal{E}} Z_{\rm pert} \times \sum_k e^{-k\varphi}Z_k \ ,
\end{align}
with $Z_{k=0}=1$ where $Z_{\rm pert}$ is the perturbative contribution of the partition function on $T^2\times \mathbb{R}^4$, $ \mathcal{E} $ is called the \emph{effective prepotential} and $\varphi$ here denotes the tension of the self-dual strings with charge $k$ parameterized by the scalar vacuum expectation values in the tensor multiplets. 
We note that the partition function \eqref{BPS partition function} can be written as a factorized expression of
\begin{align}\label{GV}
Z = e^{-2\pi i \mathcal{E}} Z_{\mathrm{GV}} = e^{-2\pi i \mathcal{E}} \PE\qty[\sum_{j_l,j_r,\mathbf{d}} (-1)^{2(j_l+j_r)} N_{j_l,j_r}^{\mathbf{d}} \frac{\sqrt{p_1 p_2} \chi_{j_l}(\epsilon_-) \chi_{j_r}(\epsilon_+)}{(1-p_1)(1-p_2)}  e^{2\pi i \mathbf{d}\cdot\mathbf{m}} ],
\end{align}
where $ Z_{\mathrm{GV}} $ is the refined Gopakumar-Vafa (GV) invariants \cite{Gopakumar:1998ii, Gopakumar:1998jq} counting the BPS degeneracies  $ N_{j_l,j_r}^{\mathbf{d}} $ on $\Omega$-background. Here, $ \PE[f(\mu)] = \exp[\sum_{n=1}^\infty \frac{1}{n}f(\mu^n)] $ is the Plethystic exponential, $ \mathbf{m} $ collectively denotes chemical potentials $(\varphi,\phi,m)$ for tensor, gauge and flavor symmetries, $ \mathbf{d} $ is the electric charge of the BPS state with Lorentz spin $ (j_l, j_r) = (\frac{J_1-J_2}{2}, \frac{J_1+J_2}{2}) $, $ \chi_j $ is the $ SU(2) $ character of spin $ j $ representation, $\epsilon_\pm = \frac{\epsilon_1 \pm \epsilon_2}{2}$ and $ p_{1,2} = e^{2\pi i \epsilon_{1,2}} $.

The effective prepotential $ \mathcal{E} $ in the prefactor arises from the classical
action and the regularization factors in the path integral. We can compute it by evaluating the low energy effective action on the $\Omega$-background in the presence of non-trivial background gauge fields. The low energy effective action of 6d SCFTs on a circle which was computed in \cite{Bonetti:2011mw,Bonetti:2013ela,Grimm:2015zea,Bhardwaj:2019fzv,Kim:2020hhh} consists of the classical (or tree-level) and the 1-loop contributions. The tree-level action for a 6d SCFT is given by
\begin{align}
S_{\mathrm{tree}} = \int \qty( -\frac{1}{2} \Omega^{\alpha\beta} G_\alpha \wedge *G_\beta - \Omega^{\alpha\beta} B_\alpha \wedge X_{4\beta} +\cdots ) \ 
\end{align}
where $\cdots$ stands for the SUSY completions, $ \Omega^{\alpha\beta} = \Sigma^\alpha \cdot \Sigma^\beta $ is the intersection form of the tensor multiplets corresponding to the intersection pairing of compact cycles $ \Sigma^\alpha $ in the base surface of the associated elliptic Calabi-Yau threefold,  and $ G_\alpha $ is the 3-form field strength for a self-dual 2-form tensor field $ B_\alpha $. The second term is the Green-Schwarz term which is required to cancel the 1-loop gauge anomalies via Green-Schwarz-Sagnotti mechanism \cite{Sagnotti:1992qw}. The Green-Schwarz term contributes to the 6d anomaly 8-form as
\begin{align}\label{I_GS}
I_{\mathrm{GS}} = -\frac{1}{2} \Omega^{\alpha\beta} X_{4\alpha} X_{4\beta} \, ,
\end{align}
where the 4-form $ X_{4\alpha} $, which appears in the Bianchi identity $dG_\alpha = X_{4\alpha}$, is given by
\begin{align}\label{X4}
X_{4\alpha} = -\frac{1}{4} a_\alpha p_1(T_6) + \frac{1}{4} \sum_a b_{a,\alpha} \Tr F_a^2 + c_\alpha c_2(R) \, .
\end{align}
Here, $ p_1(T_6) $ and $ c_2(R) $ are the first Pontryagin class of the 6d spacetime tangent bundle and the second Chern class of $ SU(2)_R $ R-symmetry bundle, respectively, and $ F_a $ is the field strength of the $ a $-th symmetry group including all gauge and flavor symmetries. The coefficients $ a_\alpha $, $ b_{a,\alpha} $ and $ c_\alpha $ are determined from anomaly cancellation conditions.

The classical action provides non-trivial contributions to the effective prepotential.
The characteristic classes in the Green-Schwarz terms can be replaced by the $\Omega$-deformation parameters as
\begin{align}
p_1(T_6) \mapsto -(\epsilon_1^2 + \epsilon_2^2) \, , \quad
c_2(R) \mapsto \epsilon_+^2 \, ,
\end{align}
with $ \epsilon_\pm = \frac{\epsilon_1 \pm \epsilon_2}{2} $. Using this, the tree-level effective prepotential on the $\Omega$-deformed background is evaluated as
\begin{align}\label{eq:tree}
\mathcal{E}_{\mathrm{tree}} = \frac{1}{\epsilon_1 \epsilon_2} \qty[ -\frac{\tau}{2} \Omega^{\alpha\beta} \phi_{\alpha,0} \phi_{\beta,0} - \Omega^{\alpha\beta} \phi_{\alpha,0} \qty( \frac{a_\beta}{4} (\epsilon_1^2 + \epsilon_2^2) + \frac{b_{a,\beta}}{2} K_{a,ij} \phi_{a,i} \phi_{a,j} + c_\beta \epsilon_+^2 ) ] \, ,
\end{align}
where $ \phi_{\alpha,0}= i\varphi_\alpha/2\pi $ denotes the scalar VEV in the tensor multiplet, $ K_{a,ij} $ is the Killing form of the $ a $-th symmetry group, and $ \phi_{a,i} $ are the holonomies for gauge and flavor symmetries.

There are also 1-loop contributions to the effective prepotential $ \mathcal{E} $ which can be calculated as follows. The 6d SCFT compactified on a circle leads to a 5d Kaluza-Klein (KK) theory. The low energy theory of this 5d KK theory on a Coulomb branch is characterized by topological Chern-Simons couplings which involve contributions from the Kaluza-Klein momentum states along the circle as well as zero momentum states. The 1-loop Chern-Simons terms at low energy can be written as
\begin{align}
S_{\mathrm{1-loop}} = \int \qty( \frac{C_{ijk}}{24\pi^2} A_i \wedge F_j \wedge F_k - \frac{1}{48} C_i^G A_i \wedge p_1(T_6) + \frac{1}{2} C_i^R A_i \wedge c_2(R) ) \, ,
\end{align}
where the first term is the cubic Chern-Simons term for the gauge and the flavor symmetries, the second and third terms are the mixed gauge-gravity and gauge-$ SU(2)_R $ R-symmetry Chern-Simons terms, respectively. 

The cubic Chern-Simons terms are determined by the cubic prepotential given by \cite{Intriligator:1997pq,Bonetti:2011mw, Bonetti:2013ela, Grimm:2015zea}
\begin{align}
\mathcal{F}_{\mathrm{1-loop}} = \frac{1}{12} \sum_{n\in\mathbb{Z}} \qty( \sum_{e \in \mathbf{R}} \abs{n\tau + e\cdot\phi}^3  - \sum_f \sum_{w\in\mathbf{w}_f} \abs{n\tau + w\cdot\phi + m_f}^3 ),
\end{align}
where $ \mathbf{R} $ is the set of root vectors of the 6d gauge group, $ \mathbf{w}_f $ and $ m_f $ are a set of weights and mass parameters, respectively, of the $ f $-th charged hypermultiplet. The summation over all integer KK charges $ n $ can be performed by using the zeta function regularization\footnote{For a 6d SCFT with twist, we can have fractional KK-momentum states. See \cite{Kim:2020hhh}.}. 
The mixed Chern-Simons coefficients $ C_i^G $ and $ C_i^R $ can be computed in a similar manner. At this time, the contributions of the positive KK charge states and negative KK charge states cancel each other, and so we find
\begin{align}
C_i^G = -\partial_i \qty( \sum_{e\in\mathbf{R}} \abs{e\cdot\phi} - \sum_f \sum_{w\in\mathbf{w}_f} \abs{w\cdot \phi + m_f} ) \, , \quad
C_i^R = \frac{1}{2}\partial_i \sum_{e \in \mathbf{R}} \abs{e\cdot\phi} \, .
\end{align}
The 1-loop contribution to the effective prepotential is comprised of the collection of the Chern-Simons contributions.
\begin{align}\label{eq:1-loop-E}
\mathcal{E}_{\mathrm{1-loop}} = \frac{1}{\epsilon_1 \epsilon_2} \qty( \mathcal{F}_{\mathrm{1-loop}} + \frac{\epsilon_1^2 + \epsilon_2^2}{48} C_i^G \phi_i + \frac{\epsilon_+^2}{2} C_i^R \phi_i ) \, .
\end{align}

The full effective prepotential of a 6d SCFT on a torus times $\Omega$-deformed $\mathbb{R}^4$ is then given by the sum of the classical and the 1-loop contributions:
\begin{align}
\mathcal{E} = \mathcal{E}_{\mathrm{tree}} + \mathcal{E}_{\mathrm{1-loop}} \, .
\end{align}
This explains how to compute the effective prepotentials $ \mathcal{E} $ for arbitrary 6d SCFTs. We remark that when the 6d SCFT is compactified on a circle with automorphism twists, the intersection form $\Omega^{\alpha\beta}$, the Killing form $K_{a,ij}$, and the gauge and flavor algebra appearing in the effective prepotential should be replaced by those of the twisted theory. See \cite{Kim:2020hhh} for explicit calculations of $ \mathcal{E} $ for many interesting 6d SCFTs on $T^2\times \mathbb{R}^4$ with/without twists.

Let us now explain how to formulate the blowup equations for 6d SCFTs. Consider a 6d SCFT on a blowup geometry $ \hat{\mathbb{C}}^2 $ obtained by replacing the origin of the $ \mathbb{C}^2 $ by a 2-sphere $ \mathbb{P}^1 $. The partition function on this $ \hat{\mathbb{C}}^2 $ background, which we will call $\hat{Z}$, is factorized under the supersymmetric localization as a product of two contributions coming from the north and south pole of the $\mathbb{P}^1$ at the origin \cite{Nakajima:2003pg,Nakajima:2005fg}. It turns out that the partition function is independent of the volume of the $\mathbb{P}^1$, and thus blowing down the $\mathbb{P}^1$ results in a smooth transition from $\hat{Z}$ to the ordinary partition function $Z$ on $\mathbb{C}^2$ without the $\mathbb{P}^1$ at the origin. More precisely,
as indicated in \cite{Kim:2020hhh}, the partition function $ \hat{Z} $ defined on $ \hat{\mathbb{C}}^2 $ is related after the blowdown transition to the ordinary partition function $ Z $ on $ \mathbb{C}^2 $ by replacing $ (-1)^F $ in \eqref{ellipticgenus} and \eqref{BPS partition function} by $ (-1)^{2J_R} $. This replacement of the fermion number operator can be implemented by shifting the $\Omega$-deformation parameter $ \epsilon_1 $ for the angular momentum to $ \epsilon_1+1 $. Thus one finds
\begin{align}\label{hat Z}
\begin{aligned}
\hat{Z}(\phi, m, \epsilon_1, \epsilon_2) &= e^{-2\pi i \mathcal{E}(\phi, m, \epsilon_1, \epsilon_2)} \hat{Z}_{\mathrm{GV}}(\phi, m, \epsilon_1, \epsilon_2) \, , \\
\hat{Z}_{\mathrm{GV}}(\phi, m, \epsilon_1, \epsilon_2) &= Z_{\mathrm{GV}}(\phi, m, \epsilon_1+1, \epsilon_2) \, .
\end{aligned}
\end{align}
Note that $ \epsilon_1 $ in the prefactor $ \mathcal{E} $ remains the same because shifting $\epsilon_1$ in the GV-invariant does not affect the regularization factor.

Now, by identifying the blowup partition function $\hat{Z}$ on $\hat{\mathbb{C}}^2$, which takes a factorized expression under localization, with the partition function $Z$ on the ordinary $\mathbb{C}^2$ background, we can find a functional equation so-called a blowup equation as follows \cite{Nakajima:2003pg, Nakajima:2005fg, Gottsche:2006bm} (See also \cite{Huang:2017mis, Gu:2018gmy, Gu:2019dan, Gu:2019pqj, Gu:2020fem, Kim:2019uqw, Kim:2020hhh, Kim:2021gyj, Sun:2021lsq} for various generalizations) :
\begin{align}\label{blowup}
\Lambda(m, \epsilon_1, \epsilon_2) \hat{Z}(\phi, m, \epsilon_1, \epsilon_2)
= \sum_{\vec{n}} (-1)^{\abs{\vec{n}}} \hat{Z}^{(N)}(\vec{n}, \vec{B}) \hat{Z}^{(S)}(\vec{n}, \vec{B}) \, ,
\end{align}
where $ \abs{\vec{n}}=\sum_i n_i $ denotes the sum of magnetic fluxes $n_i$ for the dynamical tensors and gauge symmetry groups on $ \mathbb{P}^1 $, $ \vec{B} $ denotes the background magnetic fluxes  for the global symmetries, and $ \Lambda $ is a constant prefactor independent of the dynamical K\"ahler parameters $ \phi $. Here, $ \hat{Z}^{(N)} $ and $ \hat{Z}^{(S)} $ are localized partition functions near the north and south poles of the $ \mathbb{P}^1 $. They can be obtained, since the local geometries can be approximated as $\mathbb{C}^2$, from the ordinary partition function $Z$ by shifting the chemical potentials as
\begin{align}\label{Z-shift}
    \begin{aligned}
        \hat{Z}^{(N)}(\vec{n}, \vec{B}) &= \hat{Z}(\phi_i + \epsilon_1 n_i, m_j + \epsilon_1 B_j, \epsilon_1, \epsilon_2 - \epsilon_1)  \ ,\\
        \hat{Z}^{(S)}(\vec{n}, \vec{B}) &= \hat{Z}(\phi_i + \epsilon_2 n_i, m_j + \epsilon_2 B_j, \epsilon_1 - \epsilon_2, \epsilon_2) \, .
    \end{aligned}
\end{align}
Here $\phi_i$ collectively denotes the scalar VEVs in  the tensor and gauge multiplets.

The prefactor $\Lambda$ can be zero. In this case, the blowup equation is called vanishing blowup equation. For example, when the 6d SCFT contains a half hypermultiplet that does not form a full hypermultiplet, the theory admits only vanishing blowup equations. We will not discuss vanishing blowup equations in this paper.

We cannot turn on arbitrary magenetic fluxes $ (\vec{n}, \vec{B}) $ on the $ \mathbb{P}^1$, but they must be correctly quantized. The proper quantization conditions for the maginetic fluxes are \cite{Huang:2017mis}
\begin{align}\label{eq:flux-quantization}
(\vec{n}, \vec{B}) \cdot e \text{ is integral/half-integral}\ \Leftrightarrow\ 2(j_l+j_r) \text{ is odd/even},
\end{align}
for all BPS particles of the gauge and flavor charge $ e $ and spin $ (j_l, j_r) $. Among $ (\vec{n}, \vec{B}) $ satisfying the quantization conditions, a set of special magnetic fluxes called \emph{consistent magnetic fluxes} can give a blowup equation \eqref{blowup} which the partition function $Z$ obeys.
We refer the reader to \cite{Kim:2020hhh} for a detailed discussion on the process of identifying the consistent magnetic fluxes and solving the blowup equations to calculate the BPS spectra of 5d and 6d SQFTs.

It is more convenient to express the blowup equation \eqref{blowup} in terms of the GV-invariant as follows:
\begin{align}\label{eq:blowup-GV}
\Lambda(m, \epsilon_1, \epsilon_2) \hat{Z}_{\mathrm{GV}}(\phi, m, \epsilon_1, \epsilon_2)
= \sum_{\vec{n}} (-1)^{|\vec{n}|} e^{-2\pi i V} \hat{Z}_{\mathrm{GV}}^{(N)}(\vec{n}, \vec{B}) \hat{Z}_{\mathrm{GV}}^{(S)}(\vec{n}, \vec{B}),
\end{align}
where
\begin{align}
\begin{aligned}
V &= \mathcal{E}(\phi_i, m_j, \epsilon_1, \epsilon_2) - \mathcal{E}(\phi_i + \epsilon_1 n_i, m_j+\epsilon_1 B_j, \epsilon_1, \epsilon_2-\epsilon_1) \\
&\quad - \mathcal{E}(\phi_i + \epsilon_2 n_i, m_j+\epsilon_2 B_j, \epsilon_1-\epsilon_2, \epsilon_2) \, .
\end{aligned}
\end{align}
For a 6d SCFT, we can split the GV-invariant part as
\begin{align}\label{elliptic-genus-CFT}
Z_{\mathrm{GV}} = Z_{\mathrm{pert}} \times Z_{\mathrm{str}} = Z_{\mathrm{pert}} \times \sum_{\vec{k}} e^{-2\pi i \Omega^{\alpha\beta} k_\alpha \phi_{\beta,0}} Z_{\vec{k}} \, .
\end{align}
Here, $ Z_{\mathrm{pert}} $ is the 1-loop perturbative contributions from the tensor, vector and hypermultiplets. $ Z_{\mathrm{str}} $ is the self-dual string contributions which are given by a summation over the elliptic genera $ Z_{\vec{k}} $ of the worldsheet SCFTs on self-dual strings with tensor charge $\vec{k}\equiv (k_1,k_2,\cdots ,k_N)$ where $k_\alpha\in \mathbb{Z}_{\ge 0}$ for all $\alpha$. 
The explicit form of the 1-loop contributions is given by
\begin{align}\label{eq:pert-Z}
    Z_{\mathrm{pert}} = \PE \qty[ I_{\mathrm{tensor}} + I_{\mathrm{vector}} + I_{\mathrm{hyper}} ] \, ,
\end{align}
where the single-letter contributions $I_{\mathrm{tensor}}$, $I_{\mathrm{vector}}$, $I_{\mathrm{hyper}}$ of tensor, vector and hypermultiplet are
\begin{align}
    I_{\mathrm{tensor}} &= - \frac{p_1+p_2}{(1-p_1)(1-p_2)} \frac{1}{1-q} \, , \\
    I_{\mathrm{vector}} &= - \frac{1+p_1p_2}{(1-p_1)(1-p_2)} \frac{1}{2} \sum_{n \in \mathbb{Z}} \sum_{\rho \in \mathbf{R}} e^{2\pi i |n \tau + \rho \cdot \phi|} \, , \label{Ivec} \\
    I_{\mathrm{hyper}} &= \frac{\sqrt{p_1p_2}}{(1-p_1)(1-p_2)} \sum_{n \in \mathbb{Z}} \sum_f \sum_{w \in \mathbf{w}_f} e^{2\pi i |n\tau + w\cdot \phi + m_f|} \, , \label{Ihyp}
\end{align}
where $ q=e^{2\pi i \tau}$.

For a given 6d SCFT, we can systematically compute the effective prepotential $ \mathcal{E} $ and  find the consistent magnetic fluxes $ (\vec{n}, \vec{B}) $, and thus formulate the blowup equations as in (\ref{eq:blowup-GV}).
We then expand the blowup equations in terms of the K\"ahler parameters $ e^{2\pi i \mathbf{d}\cdot\mathbf{m}} $ and solve them iteratively to calculate the BPS degeneracies of $ N_{j_l,j_r}^{\mathbf{d}} $ of the 6d theory.

\subsection{Blowup equations for LSTs} \label{subsec:blowup-LST}

We will now extend the blowup formalism for 6d SCFTs to 6d LSTs. Little string theories are characterized by a collection of 2-form tensor fields whose intersection pairing, represented by $\Omega^{\alpha\beta}$, is negative semi-definite and has a single null direction. This means that there exists a unit vector $\ell_\alpha$ in the string charge lattice such that $\Omega^{\alpha\beta}\ell_\beta=0$.  As a result, the tensor field corresponding to the null direction $\ell_\alpha$ in the string charge lattice is non-dynamical. We will call the strings with tensor charges propotional to $\ell_\alpha$ as the full winding strings. The tension of these full winding strings, represented by $T\sim M_{\rm string}^2$, is always finite, and it defines the intrinsic scale of the LST. The full winding strings in LSTs are therefore distinguised from the self-dual strings in 6d SCFTs which have tensionless limit.

We are interested in the elliptic genera of 2d worldsheet SCFTs of LSTs on tensor branch. We can write the contributions from the dynamical strings to the partition function as a  collection of the elliptic genera of the strings as follows:
\begin{align}\label{eq:LST-partition}
Z_{\mathrm{str}} = \sum_{\vec{k}} v_1^{k_1} \cdots v_N^{k_N} Z_{\vec{k}} \ ,
\end{align}
where
\begin{align}
v_\alpha \equiv e^{-2\pi i \Omega^{\alpha\beta} \phi_{\beta,0}} \  , \quad
v_N \equiv e^{2\pi i (w - \Omega^{N\beta} \phi_{\beta,0})} \ ,
\end{align}
with $\alpha = 1, \cdots, N-1$ and $\beta=1,\cdots,N$.
Here, the scalar VEVs of the $N-1$ tensor multiplets $ \phi_{\beta,0} $ and the little string tension $ w \sim T $ play the role of chemical potentials for the string charges. The full winding string states are represented by the fugacity $e^{2\pi i w}$, but are independent of other tensor scalar VEVs $\phi_{\beta,0}$.

There is a natural limit $w \rightarrow i\infty$  while keeping $\phi_{\alpha,0}$ finite. In this limit the full winding string states are truncated and the LST is reduced to a 6d SCFT with $N-1$ tensor multiplets. From this point of view, the LST can be considered as an affine extension of the 6d SCFT by attaching an affine tensor node to the tensor quiver diagram. This leads to the intersection form $\Omega^{\alpha\beta}$ with $N$ tensor nodes of the LST \cite{Bhardwaj:2015oru}. The partition function \eqref{eq:LST-partition} under this 6d SCFT limit becomes that of the self-dual strings in the 6d SCFT and it satisfies the blowup equation discussed in the previous subsection.

We now proceed to construct the blowup equations for the partition function of the LSTs and use them to compute their elliptic genera.
One of the distinguished features of the LSTs from 6d SCFTs is that the mixed gauge-global anomalies in LSTs are not completely canceled by the standard Green-Schwarz mechanism. The anomaly 8-form for the mixed anomalies should take a factorized form as
\begin{align}
	I_8^{\rm mixed} = Y_4\wedge X_{4,0} \ ,
\end{align}
where the first factor $Y_4$ is a 4-form given in terms of the second Chern classes for the dynamical gauge fields
\begin{align}
	Y_4 = \frac{1}{4}\sum_{\alpha=1}^N \ell_\alpha{\rm Tr} F_{G_\alpha}^2 \ ,
\end{align}
and the second factor $X_{4,0}$ is a 4-form independent of the dynamical gauge field which can be written as
\begin{align}\label{X40}
    X_{4,0} = -\frac{1}{4} a_0 p_1(T_6) + \frac{1}{4} \sum_a b_{a,0} \Tr F_{a}^2 + c_0 c_2(R) \, .
\end{align}
Here, $F_{G_\alpha}$ and $F_a$ are the field strength for the gauge group $G_\alpha$ and that for the $a$-th flavor group respectively. We normalize the instanton number as $k_\alpha=\frac{1}{4}{\rm Tr}F_{G_\alpha}^2 \in \mathbb{Z}$ when integrated over a 4-manifold and it parametrizes the $\alpha$-th direction in the string charge lattice.  The coefficents $a_0$, $b_{a,0}$, and $c_0$ are fixed by the 1-loop and the Green-Schwarz anomaly calculations. When the theory has a F-theory construction on an elliptic Calabi-Yau threefold, we can identify them as 
\begin{align}
	a_0 = K\cdot \Sigma_{\rm LST}\ , \quad b_{a,0}=\Sigma_{F_a}\cdot\Sigma_{\rm LST}\ , \quad c_0=\ell_\alpha h^\vee_{\alpha} \, ,
\end{align}
where $K$ is the canonical class of the base $B$ in the CY 3-fold, $\Sigma_{\rm LST}$ is the curve class associated to the little string scale satisfying $\Omega \cdot \Sigma_{\rm LST}=0$, $\Sigma_{F_a}$ is the curve class supporting the 7-brane with $a$-th flavor symmetry, $h^\vee_{\alpha}$ is the dual Coxeter number of the group $G_\alpha$,  and the dot $\cdot$ between two curve classes stands for the intersection number of the curves.

The non-vanishing mixed anomalies are inconsistent with the dynamical gauge symmetries in the presence of background gauge fields for the global symmetries. Therefore, there must be a regularization scheme that cancels these mixed gauge anomalies while preserving the dynamical gauge symmetries, even in the presence of non-trivial background fields for the global symmetries. 

There are some choices of regularization scheme.
For instance, in \cite{Cordova:2020tij}, the mixed gauge anomalies were canceled by adding Green-Schwarz counterterms involving a 2-form background gauge field coupled to the 2-form instanton currents $J\sim \star {\rm Tr}F_G\wedge F_G$. Here, the 2-form background gauge field transforms under the background global symmetry transformation and also under the local Lorentz transformation. This results in a continuous 2-group global symmetry. 

In this paper we will introduce another counterterm which leads to the consistent blowup equations for LSTs as we will explain below.
We shall introduce the counterterm defined as
\begin{align}\label{B0-S}
    \Delta S = -\int B_0 \wedge X_{4,0}\ ,
\end{align}
with a 2-form gauge field $B_0$ which transforms under the dynamical gauge transformation parametrized by $\Lambda_{G}$ as
\begin{align}
	 B_0 \ \rightarrow \ B_0+\frac{1}{4} \ell_\alpha\, {\rm Tr} \Lambda_{G_\alpha} F_{G_\alpha} \ .
\end{align}
This modifies the Bianchi identity for the 3-form field strength $H_0=dB_0$ as
\begin{align}
	dH_0 = Y_4 \ .
\end{align}
Then the gauge variation of the counterterm cancels the gauge anomalies arising from $I_8^{\rm mixed}$ in the presence of the background fields for the global symmetries. Let us emphasize that the 2-form field $B_0$ here is not a fixed background field since it transforms non-trivially under the dynamical gauge transformation. We need to integrate this field in the path integral although it has no kinetic term in the action. This 2-form field can be considered as a kind of Lagrange multiplier introduced to cancel the mixed gauge-global anomalies. 

We are now ready to formulate the blowup equations for the LSTs. We first need to prepare the effective prepotential on the $\Omega$-background. The tree-level action for a LST is almost the same as that of 6d SCFTs, but now there are additional contributions from the gauge kinetic terms coupled to the little string tension $w$ and the counterterm \eqref{B0-S}. We propose that the tree-level effective prepotential for a LST is
\begin{align}
    \mathcal{E}_{\rm tree}^{\rm LST} &= \mathcal{E}_{\rm tree}^{\rm SCFT} + \mathcal{E}_{\rm tree}^{(0)} \ , \\
    \mathcal{E}_{\mathrm{tree}}^{(0)}
&= \frac{1}{\epsilon_1 \epsilon_2} \qty[ \frac{w}{2} \ell_\alpha K_{\alpha,ij} \phi_{\alpha,i} \phi_{\alpha,j} -  \phi_{0,0}\qty( \frac{a_0}{4} (\epsilon_1^2 + \epsilon_2^2) + \frac{b_{a,0}}{2} K_{a,ij} m_{a,i} m_{a,j} + c_0 \epsilon_+^2) ] \, . \nonumber
\end{align}
Here, we introduced an auxiliary scalar VEV $\phi_{0,0}$ to take into account the magnetic flux of the 2-form $B_0$ in \eqref{B0-S} on the blowup background, which will be explained in more detail below.
The first term with $w$ in $\mathcal{E}_{\rm tree}^{(0)}$ is the gauge kinetic terms evaluated on the $\Omega$-background and the second term with $\phi_{0,0}$ is the contribution from the counterterm \eqref{B0-S}.
The 1-loop contributions to the effective prepotential $\mathcal{E}_{\rm 1-loop}$ and to the GV-invariant $Z_{\rm pert}$ can be calculated in the same way as those for 6d SCFTs presented in \eqref{eq:1-loop-E} and in \eqref{eq:pert-Z} respectively in the preivous subsection.

Now we claim that the partition function $ Z = e^{-2\pi i \mathcal{E}} \times Z_{\mathrm{GV}} $ of a little string theory satisfies the blowup equation
\begin{align}
    \begin{aligned}
        \Lambda(m; \epsilon_1, \epsilon_2) \hat{Z}(\phi, m; \epsilon_1, \epsilon_2) &= \sum_{\vec{n}} (-1)^{\abs{\vec{n}}} \hat{Z}^{(N)}(\vec{n}, \vec{B}) \hat{Z}^{(S)}(\vec{n}, \vec{B}) \\
        \Leftrightarrow \Lambda(m, \epsilon_1, \epsilon_2) \hat{Z}_{\mathrm{GV}}(\phi, m, \epsilon_1, \epsilon_2) &= \sum_{\vec{n}} (-1)^{|\vec{n}|} e^{-2\pi i V} \hat{Z}_{\mathrm{GV}}^{(N)}(\vec{n}, \vec{B}) \hat{Z}_{\mathrm{GV}}^{(S)}(\vec{n}, \vec{B}) \, ,
    \end{aligned}
\end{align}
with a set of consistent magnetic fluxes $\vec{n},\vec{B}$ satisfying the quantization in \eqref{eq:flux-quantization}. $ \hat{Z} $ is again the partition function with the $\epsilon_1$ shift given in \eqref{hat Z}, and $ \hat{Z}^{(N)} $ and $ \hat{Z}^{(S)} $ are the local partition functions near the north pole and south pole of the $ \mathbb{P}^1 $ defined by \eqref{Z-shift}.

There are a few remarks for the blowup equations for LSTs. Firstly, the magnetic fluxes $\vec{n}$ on the blownup $\mathbb{P}^1$ in the blowup equation involve not only the magnetic fluxes for the dynamical tensor and gauge symmetry groups, but also the magnetic flux for the 2-form gauge field $B_0$ that is added to cancel the mixed gauge-global anomalies. As explained above, the 2-form field $B_0$ behaves like a Lagrange multiplier and we should sum over its magnetic fluxes on the blowup background. Otherwise, it will not be possible to activate background fields for the symmetries that have mixed anomalies with gauge symmetries. In the blowup equation, turning on a flux $n_{0,0}$ for $B_0$ is implemented by a shift of the auxiliary scalar field in the form $\phi_{0,0}\rightarrow \phi_{0,0} +n_{0,0}\epsilon_{1,2}$ with $n_{0,0}\in \mathbb{Z}$. The summation of these auxiliary magnetic fluxes is crucial to construct a consistent blowup equation for LSTs that have mixed gauge anomalies.
We note that the auxiliary field $\phi_{0,0}$ only  serves the purpose of activating the fluxes $n_{0,0}\epsilon_{1,2}$ and ultimately disappears in the blowup equation through the use of the combination $ V=\mathcal{E}^{(N)} + \mathcal{E}^{(S)} - \mathcal{E} $.

Secondly, the sum over the auxiliary magnetic flux $n_{0,0}$ on $\mathbb{P}^1$ in the blowup equation is not convergent. It turns out that the $n_{0,0}$ dependent terms appear only in the exponent $V$ in the blowup equation and they are all linear in $n_{0,0}$. Namely, the right side of the blowup equation contains a sum over $n_{0,0}$ of the form
\begin{align}
	\sum_{n_{0,0}\in \mathbb{Z}} e^{- n_{0,0} f(m;\epsilon_{1,2})+ \cdots} \times \cdots \ ,
\end{align}
with a function $f(m;\epsilon_{1,2})$ independent of the dynamical K\"ahler parameters $\phi$. This sum is obviously divergent, so it seems that the blowup equation is not well-defined. 

Nevertheless, we assert that the blowup equation of a LST is still valid  in the following sense.  As we will demonstrate explicitly with examples in the next section, the LST partition functions satisfy the blowup equations if we first expand them in terms of K\"ahler parameters and then sum over the auxiliary magnetic fluxes $n_{0,0}$. Surprisingly, if one sums up the fluxes $|n_{0,0}|\le n_{\rm max}$, one finds that every order in the K\"ahler parameter expansion is exactly canceled, leaving a few terms coming from the maximum flux $|n_{0,0}|=n_{\rm max}$. These remaining terms are also canceled iteratively by new terms appearing when the maximum flux is increased such as $n_{\rm max}\rightarrow n_{\rm max}+1\rightarrow n_{\rm max}+2\rightarrow n_{\rm max}+3$, and so on. Hence, if sufficiently large enough fluxes are summed up, all terms arising from smaller $n_{0,0}$ fluxes are canceled out. This is how the blowup equation works for LSTs and is rather different from the structure of the typical blowup equations for 4d/5d/6d SCFTs. 

In particular, without the sum over the auxiliary flux $n_{0,0}$, the above blowup equation does not hold at all. This is related to the fact that the LSTs possess mixed gauge anomalies in the presence of background fields such as $\vec{B}$ and $\epsilon_{1,2}$ for the global and Lorentz symmetries which we need to activate to formulate a consistent blowup equation and that we need to introduce the 2-form $B_0$ and the counterterm \eqref{B0-S} associated to the auxiliary flux to cancel such mixed gauge anomalies. We have checked this for a number of examples that we will discuss in detail in the next section. Therefore, we propose that the auxiliary magnetic flux $n_{0,0}$ must be taken into account in the construction of the blowup equations for LSTs. The counterterm \eqref{B0-S} with the auxiliary 2-form field $B_0$ is required in this sense.

Importantly, we can use the blowup equations, combining them with the modular ansatz, to determine the elliptic genera of 2d worldsheet SCFTs on strings in LSTs. To show this, let us now illustrate how to bootstrap the BPS spectra of LSTs using the blowup equations and the modular properties of the elliptic genera.

\subsection{Bootstrapping LSTs}

We first review how to formulate a general ansatz for the elliptic genus of BPS strings in 6d theories by exploiting its properties under the modular transformation. 
The modular property of the elliptic genus defined in \eqref{ellipticgenus} is governed by the 't Hooft anomalies of the worldsheet SCFT.
Under the modular transformation, the elliptic genus transforms as \cite{Benini:2013xpa},
\begin{align}\label{Z-modular-transf}
Z_{\vec{k}}\qty(\frac{a \tau + b}{c\tau + d}, \frac{z}{c\tau+d}) = \epsilon(a,b,c,d)^{c_R-c_L} \exp( \frac{2\pi i c}{c\tau+d}f(z)) Z_{\vec{k}}(\tau, z) \, ,
\end{align}
where $ \smqty(a & b \\ c & d) \in \mathrm{SL}(2, \mathbb{Z}) $, $ \epsilon(a,b,c,d) $ is a phase factor, $ c_{L,R} $ are chiral central charges of the worldsheet SCFT, and $z$ collectively denotes chemical potentials for the symmetries. The \emph{modular anomaly} $ f(z) $ is closely related with the anomaly polynomial $ I_4 $ of the 2d SCFT \cite{Gu:2017ccq, DelZotto:2017mee}. In fact, it agrees with the supersymmetric Casimir energy of the 2d SCFT defined in \cite{Bobev:2015kza} which is given by an equivariant integral of the anomaly polynomial $I_4$,
\begin{align}
	f(z) = \int_{\rm eq}I_4 \ .
\end{align}
The equivariant integration here can be implemented by the replacement rules for the characteristic classes as
\begin{align}
&p_1(T_2) \mapsto 0 \, , \quad c_2(l) \mapsto \epsilon_-^2 \, , \quad c_2(r) ,
c_2(R) \mapsto \epsilon_+^2 \, , \quad \frac{1}{2}\Tr F_a^2 \mapsto K_{a,ij} \phi_{a,i} \phi_{a,j} \, .
\label{eqint}
\end{align}
Knowing the anomaly polynomial of the worldsheet SCFT and the modular transformation in \eqref{Z-modular-transf}, we can formulate an ansatz for the elliptic genus in terms of elliptic functions.

The anomaly polynomial of the 2d SCFTs living on self-dual strings in 6d SCFTs has been calculated in \cite{Kim:2016foj,Shimizu:2016lbw} by using anomaly inflow mechanism. For a 6d SCFT with an intersection form $\Omega^{\alpha\beta}_{\rm cft}$, the anomaly polynomial of the worldsheet CFT on a self-dual string with charge $\vec{k}=\{k_\alpha\}$ is
\begin{align}\label{eq:I4-CFT}
    I_4 = \Omega^{\alpha\beta}_{\rm cft} k_\alpha \qty(X_{4\beta} + \frac{1}{2} k_\beta \chi_4(T_4))  \, ,
\end{align}
where $X_{4\beta}$ is a 4-form defined in \eqref{X4},
$\chi_4(T_4)$ is the Euler class of the transverse $SO(4)=SU(2)_l\times SU(2)_r$ Lorentz rotation which can be written as $\chi_4(T_4)=c_2(l)-c_2(r)$ in terms of the second Chern classes for the $SU(2)_l\times SU(2)_r$ bundle. The first Pontryagin class $p_1(T_6)$ of the 6d tangent bundle in $X_{4\alpha}$ is decomposed as $p_1(T_6) = p_1(T_2)-2c_2(l)-2c_2(r)$.

Similarly, the anomaly polynomials of 2d SCFTs on BPS strings in a number of LSTs were calculated in \cite{Kim:2018gak}. We will generalize this computation and provide a universal expression for the anomaly polynomials of the 2d SCFTs on strings in LSTs. 

The 't~Hooft anomalies on the 2d worldsheet of the self-dual strings in the 6d SCFTs embedded in a LST should be the same as \eqref{eq:I4-CFT}. However, there is another contribution to the 't Hooft anomalies coming from the full winding strings in the LST. This extra contribution can be captured by integrating the mixed gauge anomaly 8-form $I_8^{\rm mixed}$ on the full winding string background \cite{Cordova:2020tij}. 
Let us define the number of full winding strings $\kappa\in \mathbb{Z}$  as a maximal integer satisfying $k_\alpha-\kappa \ell_\alpha\ge0$ for all $\alpha$.
We propose that the anomaly polynomial of the 2d SCFT on strings in a little string theory is
\begin{align}\label{eq:I4-LST}
    I_4 = \Omega^{\alpha\beta} k_\alpha \qty(X_{4\beta} + \frac{1}{2} k_\beta \chi_4(T_4)) + \kappa X_{4,0} \, .
\end{align}
Here, $\Omega^{\alpha\beta}$ is the Dirac pairing of the $N$-dimensional string charge lattice and $X_{4,0}$ is defined in \eqref{X40}. We can use this anomaly polynomial to compute the modular anomaly $f(z)$ in \eqref{Z-modular-transf} for the worldsheet SCFTs for strings with a charge $\vec{k}$.

A function which transforms as \eqref{Z-modular-transf} under the modular transformation is known as \emph{Jacobi form}\footnote{We summarize the definition, terminologies, and properties of Jacobi forms in Appendix~\ref{app:elliptic}}. In the language of Jacobi forms, \eqref{Z-modular-transf} implies that elliptic genus has weight 0 and its indices are fixed by 't Hooft anomaly coefficients of the global symmetries on the worldsheet theory.
To write down an ansatz for the elliptic genus of $\vec{k}$ strings using the Jacobi forms, we need to use the  modular property in \eqref{Z-modular-transf} and the pole structure of $ Z_{\vec{k}} $.

We propose a \emph{modular ansatz} for the elliptic genus for the strings in LSTs, which will be of the form 
\begin{align}\label{modular-ansatz}
    Z_{\vec{k}} = \frac{1}{\eta(\tau)^{2|c_L-c_R|}} \frac{\Phi_{\vec{k}}(\tau, \epsilon_\pm, \phi, m)}{\prod_\alpha \mathcal{D}_{\alpha}^{\mathrm{cm}}(\tau, \epsilon_\pm) \mathcal{D}_{\alpha}^{\mathfrak{g}_\alpha}(\tau, \epsilon_\pm, \phi)} \qty(\frac{1}{\mathcal{D}_{\kappa}^{\mathrm{bulk}}(\tau, \epsilon_\pm, m_0)}) \, ,
\end{align}
where $ \phi $ and $ m $ collectively denote chemical potentials for gauge and flavor symmetries, respectively. This is an extension of the ansatzes introduced in \cite{Huang:2015sta, DelZotto:2016pvm, Gu:2017ccq, DelZotto:2017mee, Kim:2018gak}.
Here the denominator factors in the ansatz will be fixed by pole structure expected for the moduli space of strings, which we will explain now.

Firstly, the factor $ \eta(\tau)^{2|c_L-c_R|} $, where $ \eta(\tau) $ is the Dedekind eta function defined in \eqref{dedekind-eta}, fixes the leading behavior of the elliptic genus in $ q $-expansion which is determined by the vacuum Casimir energy of the 2d SCFT. 

The second factor of the form 
\begin{align}\label{modular-cm}
    \mathcal{D}_{\alpha}^{\mathrm{cm}}(\tau, \epsilon_\pm) = \prod_{s=1}^{k_\alpha} \varphi_{-1,1/2}(\tau, s\epsilon_1) \varphi_{-1,1/2}(\tau, s \epsilon_2) \, ,
\end{align}
is the contribution coming from the transverse motions of the strings \cite{Haghighat:2015ega, DelZotto:2018tcj}, where $ \varphi_{-1,1/2} $ is the weight $ -1 $ and index $ 1/2 $ Jacobi form
\begin{align}
    \varphi_{-1,1/2}(\tau, z) = i \frac{\theta_1(\tau, z)}{\eta(\tau)^3} \ ,
\end{align}
with the Jacobi theta function $ \theta_1(\tau, z) $ given in \eqref{Jacobi-theta}. Notice that the leading order of $\varphi_{-1,1/2}(\tau, z)$ in $q$-expansion is $\mathcal{O}(1)$, so that this does not change the leading behavior of the elliptic genus in $q$-expansion.

The third factor, $ \mathcal{D}_{\alpha}^{\mathfrak{g}_\alpha} $, arises from the bosonic zero modes of instantons for the 6d gauge algebra $ \mathfrak{g}_\alpha $.
For instance, when the gauge algebra is $ \mathfrak{g}_\alpha=\mathfrak{su}(2) $, we have
\begin{align}\label{su2-denom}
    \mathcal{D}_{\alpha}^{A_1} = \prod_{s=1}^{k_\alpha} \prod_{l=0}^{s-1} \varphi_{-1,1/2}((s+1)\epsilon_+ + (s-1-2l)\epsilon_- \pm e \cdot \phi) \, ,
\end{align}
where $ e $ is the positive root of $ \mathfrak{su}(2) $ and $ \varphi(x \pm y) \equiv \varphi(\tau, x+y) \varphi(\tau, x-y) $. For a general gauge algebra $ \mathfrak{g}_\alpha $, we use an embedding $ \mathfrak{su}(2)\subset \mathfrak{g}_\alpha $ which maps three $ SU(2) $ generators to generators $ T_e^{a=1,2,3} $ associated with a positive root $ e $ of $ \mathfrak{g}_\alpha $ \cite{Belavin:1975fg, Bernard:1977nr}. This embedding gives the denominator factor $ \mathcal{D}_{\alpha}^{\mathfrak{g}_\alpha} $ as \cite{Kim:2018gak}
\begin{align}
    \mathcal{D}_{\alpha}^{\mathfrak{g}_\alpha} = \prod_{e \in \mathbf{R}^+_{\mathfrak{g}_\alpha}} \mathcal{D}_{\lfloor k_\alpha/\xi_e \rfloor, e}^{A_1} \, ,
\end{align}
where $ \mathbf{R}^+_{\mathfrak{g}_\alpha} $ is the set of positive roots of $ \mathfrak{g}_\alpha $, $ \mathcal{D}_{k,e}^{A_1} $ is \eqref{su2-denom} by replacing an $ \mathfrak{su}(2) $ positive root with a given root of $ \mathfrak{g}_\alpha $, $ \lfloor \cdot \rfloor $ is the floor function and
\begin{align}
    \tr(T_e^a T_e^b) = \xi_e \delta^{ab} \, , \quad
    \xi_e = \left\{ \begin{array}{ll}
            1 & \quad \text{if $e$ is a long root or $ \mathfrak{g} = A_n, D_n, E_n $} \\
            2 & \quad \text{if $e$ is a short root and $\mathfrak{g} = B_n, C_n, F_4$} \\
            3 & \quad \text{if $e$ is a short root and $\mathfrak{g} = G_2$}
    \end{array} \right. \ .
\end{align}

Lastly, some LSTs have a denominator factor $\mathcal{D}_{\kappa}^{\mathrm{bulk}}$, which depends on the winding number $\kappa$, as mentioned in \cite{Kim:2018gak}. This factor is associated to certain full winding string states that are decoupled from the 6d LST, which means that these states do not possess dynamical gauge charges, and presumably escape to the bulk spacetime in which the LST is embedded. In the examples we present in section~\ref{subsec:heterotic-LST}, for example, the LSTs can be embedded in 10d heterotic string theories and the modular ansatz for strings in these theories includes a denominator factor of the form
\begin{align}\label{eq:bulk-denominator}
    \mathcal{D}_{\kappa}^{\mathrm{bulk}} = \prod_{s=1}^{\kappa} \varphi_{-1,1/2}(\pm s\lambda(m_0 - \epsilon_+)) \ ,
\end{align}
where $m_0$ is the chemical potential for the $SU(2)_m \subset SU(2)_R\times SU(2)_m$ rotational symmetry transverse to the 6d spacetime of the LSTs. These are chosen to match the ADHM constructions for the strings.
Specifically, the value of $\lambda$ is set to 1 for $ SO(32) $ heterotic LST in section~\ref{subsubsec:SO32LST} and  to 2 for $ E_8\times E_8 $ heterotic LST in section~\ref{subsec:E8xE8}. However, we find that  it is possible to select alternative denominator factors that do not alter the dynamical string spectrum, while leading to different full winding string states that decouple from the LST. For example, as we will show in section~\ref{subsec:E8xE8}, the same string spectrum for the $ E_8\times E_8 $ heterotic LST can be obtained by using a different denominator factor with $\lambda=1$. We postulate that, for a generic LST, it is always possible to choose the factor $\mathcal{D}_{\kappa}^{\mathrm{bulk}}$  to be either trivial or in the form specified in \eqref{eq:bulk-denominator} with a certain $\lambda\in\mathbb{Z}$, provided that a modular ansatz of the form \eqref{modular-ansatz} can be established. This modular ansatz will be consistent with the dynamical string spectrum of the LST, though the decoupled states that are independent of dynamical gauge symmetries may vary.

After factoring out the denominator factors, the numerator $ \Phi_{\vec{k}} $ in the modular ansatz \eqref{modular-ansatz} starts at $ q^0 $ order in $ q $-expansion and becomes a Weyl-invariant Jacobi form whose weight and indices are fixed by the 't Hooft anomalies of a given theory and the structure of the denominator in the modular ansatz. For every simple Lie algebra, the Weyl-invariant Jacobi forms with given weight and index can be written as a linear combination of finite generators \cite{Wirthmuller, Wang:2018fil}. Thus, the numerator is given by the finite linear combination of the generators $ \varphi_{k_{jl},m_{jl}}(z_l) $ of the Weyl invariant Jacobi forms with weight $ k_{jl} $ and index $ m_{jl} $ for $ l $'th symmetry algebra $ \mathfrak{g}_l $:
\begin{align}\label{numerator-ansatz}
    \Phi_{\vec{k}} = \sum_i C_i E_4^{a_4^{(i)}} E_6^{a_6^{(i)}} \prod_{j,l} \varphi_{k_{jl} m_{jl}}(z_l)^{b_{jl}^{(i)}} \, ,
\end{align}
where $ z_l $ denotes a chemical potential for $ l $-th symmetry, $ C_i \in \mathbb{C} $, and $ E_4 $ and $ E_6 $ denote the Eisenstein series of weight 4 and 6, respectively. The exponents $ a_{4,6}^{(i)} $ and $ b_{jl}^{(i)} $ are constrained by the condition requiring that the elliptic genus $Z_{\vec{k}}$ be transformed as \eqref{Z-modular-transf} under the modular transformation. They are thus non-negative integers satisfying two conditions:
\begin{equation}
    4a_4^{(i)} + 6a_6^{(i)} + \sum_{j,l} k_{jl} b_{jl}^{(i)} - |c_L-c_R| + \sum_\alpha 2k_\alpha + \sum_\alpha \sum_{e \in \mathbf{R}_{\mathfrak{g}_\alpha}^+} \sum_{s=1}^{\lfloor k_\alpha/\xi_e \rfloor} 2s - (\operatorname{weight}(\mathcal{D}_\kappa^{\mathrm{bulk}}) ) = 0 \, , \label{ansatz-weight} \\
\end{equation}
and
\begin{align}
    f(z) &= \sum_{j,l} m_{jl} b_{jl}^{(i)} \langle z_l, z_l \rangle_{\mathfrak{g}_l} - \frac{1}{2} \sum_\alpha \sum_{s=1}^{k_\alpha} s^2 (\epsilon_1^2 + \epsilon_2^2)  \label{ansatz-index}\\
    & \qquad - \frac{1}{2} \sum_\alpha \sum_{e \in \mathbf{R}^+_{\mathfrak{g}_\alpha}} \sum_{s=1}^{\lfloor k_\alpha/\xi_e \rfloor} \qty( (s+1)\epsilon_+ + (s-1-2l)\epsilon_- \pm e \cdot \phi)^2 - (\operatorname{index}(\mathcal{D}_\kappa^{\mathrm{bulk}})) \nonumber
\end{align}
for each $ i $, where $ \langle \cdot, \cdot \rangle_{\mathfrak{g}_l} $ is a symmetric bilinear form for $ \mathfrak{g}_l$ defined by its Killing form.
The index $l$ runs for all symmetries of the 2d worldsheet SCFT, while $\mathfrak{g}_\alpha $ denotes gauge symmetry for $ \alpha $-th node.
Hence the modularity fixes the elliptic genus up to finitely many constants $ C_i $ in \eqref{numerator-ansatz}.

The unknown constants $ C_i $'s can be fixed by imposing the GV-invariant ansatz \eqref{GV} of the elliptic genus and by solving the blowup equation. Note that the modular ansatz \eqref{modular-ansatz} for $ \sum_\alpha k_\alpha > 1 $ can have higher order poles at $ \epsilon_1=0 $ and $ \epsilon_2=0 $ arising from the center of mass contribution \eqref{modular-cm} for generic $ C_i $'s. However, the single letter index in the Plethystic exponential of the GV-invariant ansatz can have only simple poles at $ \epsilon_1=0 $ and $ \epsilon_2=0 $. 
This imposes strong constraints on $ C_i $'s. 

Furthermore, we demand that the partition function satisfies the blowup equation. In contrast to the 5d/6d SCFT cases, the blowup equations for LSTs involve a divergent summation over an auxiliary magnetic flux $ n_{0,0} $, as explained in the previous subsection. Due to this structure, it seems that the partition function of an LST cannot be determined solely by solving the blowup equations and the GV-invariant ansatz \eqref{GV}. However, the modular ansatz in \eqref{modular-ansatz} places further constraints on the partition function and, by combining it with the blowup equations, it should be feasible to completely determine the partition functions of LSTs in K\"ahler parameter expansion. We will demonstrate this with several interesting examples in the next section.

\section{Examples}\label{sec:example} 

In this section, the partition functions of several low rank LSTs are calculated.  We first compute elliptic genera of strings using the ADHM constructions.
We then construct the blowup equations for the partition functions of the LSTs and verify that the results from the ADHM constructions satisfy the blowup equations. Lastly, we formulate the modular ansatz for the elliptic genera of strings in the LSTs and fix the unknown coefficients in the ansatz by solving the blowup equations. We show that the partition functions obtained through this method are consistent with the results obtained using ADHM constructions.

\subsection{\texorpdfstring{$ \hat{A}_1 $}{\^A1} LSTs}

Our first example is the little string theories on $ N $ parallel NS5-branes in type II string theories in gravity decoupling limit introduced in \cite{Berkooz:1997cq, Seiberg:1997zk}. In the IIA theory, the little string theory is the $ \mathcal{N}=(2,0) $ LST with $N$ tensor multiplets. This LST is realized in F-theory by an elliptically fibered Calabi-Yau threefold whose base surface contains a loop of $ N $ rational curves of self-intersection number $ -2 $ \cite{Bhardwaj:2015oru}. The intersection matrix $ \Omega^{\alpha\beta} $ of the $-2$ curves is given by the minus of the Cartan matrix of the affine Lie algebra $ \hat{A}_{N-1} = A_{N-1}^{(1)} $. 
On the other hand, the LST in the IIB theory is the $ \mathcal{N}=(1,1) $ Yang-Mills theory with $U(N)$ gauge group which is realized in F-theory by an elliptic CY 3-fold with a base containing a genus one curve of self-intersection number $ 0 $.
These two LSTs in IIA and in IIB, which we call $ \hat{A}_{N-1} $ LSTs, are related via T-duality under a circle compactification. In this subsection, we consider the partition functions and the blowup equations of these LSTs for $ N=2 $.

\subsubsection{IIA picture}\label{subsubsec:IIA}

Let us first consider the $ \mathcal{N}=(2,0) $ $\hat{A}_1$ little string theory for 2 NS5-branes in type IIA string theory. 
This theory has two tensor multiplets (for one dynamical tensor field and one free tensor field) with the intersection form
\begin{align}\label{(2,0)-omega}
\Omega^{\alpha\beta} = \mqty(-2 & 2 \\ 2 & -2) \, .
\end{align}
The index part of the partition function is factorized as
\begin{align}
Z^{\mathrm{IIA}}_{\mathrm{GV}} = Z_{\mathrm{pert}}^{\mathrm{IIA}} \cdot Z_{\mathrm{str}}^{\mathrm{IIA}} \, ,
\end{align}
where the perturbative partition function is given by the contributions coming from two $ \mathcal{N}=(2,0) $ tensor multiplets
\begin{align}
Z_{\mathrm{pert}}^{\mathrm{IIA}} = \PE\qty[ \qty( \frac{\sqrt{p_1 p_2}}{(1-p_1)(1-p_2)} \qty(M+M^{-1}) - \frac{p_1 + p_2}{(1-p_1)(1-p_2)} ) \frac{2 q}{1-q} ] \, .
\end{align}
Here  $ M=e^{2\pi i m} $ is the fugacity for the $SU(2)_m\subset SU(2)_R\times SU(2)_m$ rotational symmetry of the $ \mathbb{R}^4 $ plane transverse to the NS5-branes.

The partition function $ Z_{\mathrm{str}}^{\mathrm{IIA}} $ is from the strings carrying tensor charges defined as
\begin{align}\label{(2,0)-Zstr}
    Z_{\mathrm{str}}^{\mathrm{IIA}} = \sum_{k_1,k_2\ge0} Q^{k_1} \qty(\frac{e^{2\pi i w}}{Q})^{k_2} Z_{(k_1,k_2)}^{\mathrm{IIA}},
\end{align}
where $ Q \equiv e^{2\pi i (2\phi_{1,0} - 2\phi_{2,0})} $ is the fugacity for the dynamical tensor charge and $ w $ is the chemical potential for the winding number. We will now study two distinct methods for calculating the elliptic genus $ Z_{(k_1,k_2)}^{\mathrm{IIA}} $: the ADHM construction based on 2d gauged linear sigma model (GLSM) on the strings, and the blowup approach with the modular ansatz.

\paragraph{GLSM}

\begin{figure}
    \centering
    \begin{subfigure}[b]{0.6\textwidth}
        \centering
        \begin{tabular}{c|cccccccccl}
            & 0 & 1 & 2 & 3 & 4 & 5 & 6 & 7 & 8 & 9($ S^1 $) \\ \hline
            NS5 & \textbullet & \textbullet & \textbullet & \textbullet & \textbullet & \textbullet \\ 
            D2 & \textbullet & \textbullet & & & & & & & & \textbullet \\
            D6 & \textbullet & \textbullet & \textbullet & \textbullet & \textbullet & \textbullet & & & & \textbullet
        \end{tabular}
        \subcaption{}
    \end{subfigure}
    \begin{subfigure}[b]{0.39\textwidth}
        \centering
        \begin{tikzpicture}
            \draw [thick, orange] (0, 1.1) -- (2.5, 1.1);
            \draw [thick, orange] (0, 1) -- (2.5, 1);
            \draw [thick, orange] (0, 0.9) -- (2.5, 0.9);
            \draw [thick, orange] (2.5, 1.05) -- (3.5, 1.05);
            \draw [thick, orange] (2.5, 0.95) -- (3.5, 0.95);
            \draw [thick, orange] (0, 1.05) -- (-1, 1.05);
            \draw [thick, orange] (0, 0.95) -- (-1, 0.95);
            \draw [thick] (0, 0) -- (0, 2);
            \draw [thick] (2.5, 0) -- (2.5, 2);
            \draw (0, -0.3) node {NS5};
            \draw (2.5, -0.3) node {NS5};
            \draw (1.25, 1.4) node {$ k_1 $ D2's};
            \draw (3.3, 0.5) node {$ k_2 $ D2's};
            \draw [thick, ->] (-0.5, 2.3) -- (0.5, 2.3);
            \draw (0.8, 2.37) node {\small $ x^9 $};
            \draw (-0.8, 1) node {\small $ \Arrowvert $};
            \draw (3.3, 1) node {\small $ \Arrowvert $};
        \end{tikzpicture}
        \subcaption{}
    \end{subfigure}
    \phantom{a}
    \begin{subfigure}{1\textwidth}
        \centering
        \begin{tabular}{c|c|cc}
            Field & Type & $ U(k_1) \times U(k_2) $ & $ U(1)_{m} $ \\ \hline
            $ (A_\mu^{(i)}, \lambda_{+(i)}^{\dot{\alpha}A}) $ & vector & $ (\mathbf{adj}, \mathbf{1}), (\mathbf{1}, \mathbf{adj}) $ \\
            $ (a_{\alpha\dot{\beta}}^{(i)}, \lambda_{-(i)}^{\alpha A}) $ & hyper & $ (\mathbf{adj}, \mathbf{1}), (\mathbf{1}, \mathbf{adj}) $ \\
            $ (\varphi_{A}^{(i)}, \chi_{-(i)}^{\dot{\alpha}}) $ & twisted hyper & $ (\mathbf{k}_1, \overline{\mathbf{k}}_2) $, $ (\overline{\mathbf{k}}_1, \mathbf{k}_2) $ & $ +1 $ \\
            $ (\chi_{+(i)}^{\alpha}) $ & Fermi & $ (\mathbf{k}_1, \overline{\mathbf{k}}_2) $, $ (\overline{\mathbf{k}}_1, \mathbf{k}_2) $ & $ +1 $ \\
            $ (q_{\dot{\alpha}}^{(i)}, \psi_-^{A(i)}) $ & hyper & $ (\mathbf{k}_1, \mathbf{1}) $, $ (\mathbf{1}, \mathbf{k}_2) $ \\
            $ (\Psi_+^{(i)}) $ & Fermi & $ (\mathbf{k}_1, \mathbf{1}), (\mathbf{1}, \mathbf{k}_2) $ & $ +1 $ \\
            $ (\tilde{\Psi}_+^{(i)}) $ & Fermi & $ (\overline{\mathbf{k}}_1, \mathbf{1}), (\mathbf{1}, \overline{\mathbf{k}}_2) $ & $ -1 $
        \end{tabular}
        \subcaption{}
    \end{subfigure}
    \caption{(a) and (b) are brane configurations for $ \hat{A}_1 $ LST in the type IIA string theory where the $x^9$-direction is compactified on a circle. (c) is the $ \mathcal{N}=(0,4) $ matter contents in the 2d GLSM on the worldsheet of strings.}\label{fig:IIA-brane}
\end{figure}

We start with the brane construction studied in \cite{Kim:2015gha}. The brane construction for the $\hat{A}_1$ LST is depicted in Figure~\ref{fig:IIA-brane}(a) and (b). 
Here, we compactify the 9-th direction on a circle, and put two NS5-branes extended along $ 012345$ directions at $ x^9 = \phi_{1,0} $ and $ x^9 = \phi_{2,0} $, respectively.
The strings in the LST arise from the $ k_1 $ D2-branes and $ k_2 $ D2-branes stretched between two NS5-branes. We also put a single D6-brane, which becomes trivial in the M-theory uplift, to explicitly provide $U(1)_m$ symmetry in the 2d GLSM.
See \cite{Kim:2015gha} for a detailed study of this brane configuration. 

The 2d GLSM on D2-branes has $ U(k_1) \times U(k_2) $ gauge symmetry, $ SU(2)_l \times SU(2)_r $ symmetry which rotates $2345$ directions, $ SO(3) $ symmetry for $678$ directions, and $ U(1)_m $ symmetry. At low energy, we expect the $ SO(3) $ and $ U(1)_m $ symmetries to be enhanced to $ SU(2)_R \times SU(2)_m $. 
There are an $ \mathcal{N}=(0,4) $ vector multiplet $ (A_\mu^{(i)}, \lambda_{+(i)}^{\dot{\alpha}A}) $ and adjoint hypermultiplets $ (a_{\alpha\dot{\beta}}^{(i)}, \lambda_{-(i)}^{\alpha A}) $ for each gauge node, bifundamental twisted hypermultiplets $ (\varphi_{A}^{(i)}, \chi_{-(i)}^{\dot{\alpha}}) $ and Fermi multiplets $ (\chi_{+(i)}^{\alpha}) $ from D2-D2 string modes, and hypermultiplets $ (q_{\dot{\alpha}}^{(i)}, \psi_-^{A(i)}) $ and Fermi multiplets $ (\Psi_+^{(i)}) $, $ (\tilde{\Psi}_+^{(i)}) $ from the D2-D6 string modes. Here, $ i=1,2 $ denotes each gauge node, $ \pm $ represent 2d chirality of fermions, $ \{\alpha, \beta, \cdots\} $, $ \{\dot{\alpha}, \dot{\beta}, \cdots \} $ and $ \{A, B, \cdots\} $ are doublet indices for the $ SU(2)_l $, $ SU(2)_r $, and $ SU(2)_R $, respectively. We summarize the matter content of the 2d GLSM in Figure~\ref{fig:IIA-brane}(c). 

The gauge theory description for the 2d worldsheet theory allows us to express the elliptic genus by a contour integral of 1-loop determinants from the supermultiplets, and the contour integral can be evaluated by using the JK-residue prescription as discussed in \cite{Benini:2013nda, Benini:2013xpa}. The result is \cite{Kim:2015gha}
\begin{align}\label{(2,0)-GLSM}
Z_{(k_1,k_2)}^{\mathrm{IIA}} = \sum_{\{Y_1,Y_2\}, \abs{Y_i}=k_i} \prod_{i=1}^2 \prod_{(a,b) \in Y_i} \frac{\theta_1(\tau, E_{i,i+1}^{(a,b)} - m + \epsilon_-) \theta_1(\tau, E_{i,i-1}^{(a,b)} + m + \epsilon_-)}{\theta_1(\tau, E_{i,i}^{(a,b)} + \epsilon_1) \theta_1(\tau, E_{i,i}^{(a,b)} - \epsilon_2)} \, ,
\end{align}
with
\begin{align}
E_{i,j}^{(a,b)} = (Y_{i,a}-b) \epsilon_1 - (Y_{j,b}^T - a)\epsilon_2 \, ,
\end{align}
where $ Y_1 $ and $ Y_2 $ are Young diagrams, and $ Y_{i,a} $ and $ Y_{i,b}^T $ are the length of $ a $-th row and $ b $-th column of $ Y_i $, respectively.

\paragraph{Modularity}

The modular properties of the elliptic genus can be obtained from the anomalies of the 2d worldsheet CFT. The chiral fermions in the GLSM contribute to the anomaly polynomial as
\begin{align}
    \lambda_{+(i)}^{\dot{\alpha}A} + \lambda_{-(i)}^{\alpha A} \ &\to \ \sum_{i=1}^2 2k_i^2 \qty(\frac{c_2(r)+c_2(R)}{2} - \frac{c_2(l)+c_2(R)}{2} ) \, , \\
    \chi_{-(i)}^{\dot{\alpha}} + \chi_{+(i)}^{\alpha} \ &\to \ 4k_1 k_2 \frac{c_2(l)-c_2(r)}{2} \, , \\
    \psi_-^{A(i)} + \Psi_+^{(i)} + \tilde{\Psi}_+^{(i)} \ &\to \ (k_1+k_2) \qty(c_2(R) + \frac{1}{4} \Tr F_m^2) \, ,
\end{align}
where $ F_m $ is the field strength for $ U(1)_m $ symmetry.
The same anomaly polynomial can also be obtained from the anomaly inflow given in \eqref{eq:I4-LST}:
\begin{align}
    I_4 = -(k_1-k_2)^2 (c_2(l) - c_2(r)) + (k_1+k_2) \qty( -c_2(R) + \frac{1}{4} \Tr F_m^2) \, .
\end{align}
Hence, the modular anomaly of the $ (k_1, k_2) $ elliptic genus is
\begin{align}
    \int_{\mathrm{eq}} I_4 = (k_1-k_2)^2 (-\epsilon_-^2 + \epsilon_+^2) + (k_1+k_2) (-\epsilon_+^2 + m^2) \, ,
\end{align}
where we use the replacement rule in \eqref{eqint}.

We can then establish a modular ansatz for $ (k_1, k_2) $-string elliptic genus as
\begin{align}\label{(2,0)-modular-ansatz}
    Z_{(k_1,k_2)}^{\mathrm{IIA}} = \frac{\Phi_{(k_1,k_2)}(\tau, \epsilon_{\pm}, m)}{\prod_{s_1=1}^{k_1} \varphi_{-1,1/2}(s_1 \epsilon_{1,2}) \cdot \prod_{s_2=1}^{k_2} \varphi_{-1,1/2}(s_2 \epsilon_{1,2})} \, .
\end{align}
The numerator $ \Phi_{(k_1,k_2)} $ can be written in terms of the Eisenstein series $ E_4 $, $ E_6 $ and the $ SU(2) $ Weyl invariant Jacobi forms $ \varphi_{-2,1} $, $ \varphi_{0,1} $ for $ \epsilon_\pm $ and $ m $ as we explained in \eqref{numerator-ansatz}:
\begin{align}\label{A1-numerator-ansatz}
    \begin{aligned}
        \Phi_{(k_1,k_2)} = \sum_i C_i^{(k_1, k_2)} E_4^{a^{(i)}_4} E_6^{a^{(i)}_6} \varphi_{-2,1}(\epsilon_+)^{b^{(i)}_1} \varphi_{0,1}(\epsilon_+)^{b^{(i)}_2} \varphi_{-2,1}(\epsilon_-)^{b^{(i)}_3} \\
        \cdot \varphi_{0,1}(\epsilon_-)^{b^{(i)}_4} \varphi_{-2,1}(m)^{b^{(i)}_5} \varphi_{0,1}(m)^{b^{(i)}_6} \, .
    \end{aligned}
\end{align}

We need to determine the unknown coefficients in the modular ansatz for the numerator $ \Phi_{(k_1,k_2)} $. For this, we first impose the consistency conditions \eqref{ansatz-weight} and \eqref{ansatz-index} and then use the GV-invariant ansatz \eqref{GV}. 
For instance, let us consider $ (k_1, k_2)=(1, 0) $ case.
By using \eqref{ansatz-weight} and \eqref{ansatz-index}, the numerator has weight $ -2 $ and the modular anomaly $ f(z)=\epsilon_+^2 + m^2 $.
Then the ansatz reduces to
\begin{align}\label{A1-(1,0)-modular}
    \Phi_{(1,0)} = C_1^{(1,0)} \varphi_{0,1}(\epsilon_+) \varphi_{-2,1}(m) + C_2^{(1,0)} \varphi_{-2,1}(\epsilon_+) \varphi_{0,1}(m) \, ,
\end{align}
where $ C_1^{(1,0)} $ and $ C_2^{(1,0)} $ are unknown constants. Now by expanding $ Z_{(1,0)} $ in terms of $ q=e^{2\pi i \tau} $ up to $ q^1 $ order and comparing it with the GV-invariant form \eqref{GV}, one can find that BPS state degeneracies $ N_{j_l,j_r}^{\mathbf{d}} $ appearing in the $ (1,0) $-string elliptic genus can be non-negative integers only if
\begin{align}
    C_1^{(1,0)} = -C_2^{(1,0)} \in \mathbb{Z}/12 \, , \quad C_1^{(1,0)} \geq 0 \, .
\end{align}
Similarly, $ \Phi_{(1,1)} $ has weight $ -4 $ and the modular anomaly $ f(z)=2\epsilon_-^2 + 2m^2 $, and thus it can be written with 4 unknown constants as
\begin{align}\label{A1-(1,1)-modular}
    \begin{aligned}
        \Phi_{(1,1)} &= C_1^{(1,1)} \varphi_{0,1}(\epsilon_-)^2 \varphi_{-2,1}(m)^2 + C_2^{(1,1)} \varphi_{-2,1}(\epsilon_-) \varphi_{0,1}(\epsilon_-) \varphi_{-2,1}(m) \varphi_{0,1}(m) \\
        & \quad + C_3^{(1,1)} \varphi_{-2,1}(\epsilon_-)^2 \varphi_{0,1}(m) + C_4^{(1,1)} E_4 \varphi_{-2,1}(\epsilon_-)^2 \varphi_{-2,1}(m)^2 \, .
    \end{aligned}
\end{align}
In order to have only simple poles at $ \epsilon_1=0 $ and $ \epsilon_2=0 $ in $(1,1)$-order after taking Plethystic logarithm as \eqref{GV}, the coefficient $ C_i^{(1,1)} $'s should satisfy
\begin{align}
    C_1^{(1,1)} = \qty(C_1^{(1,0)})^2 \, , \
    C_2^{(1,1)} = 2C_1^{(1,0)} C_2^{(1,0)} \, , \
    C_3^{(1,1)} = \qty(C_2^{(1,0)})^2 \, , \
    C_4^{(1,1)} = 0 \, .
\end{align}
Therefore, all the coefficients are fixed by one coefficient $C_1^{(1,0)}$.

We can perform a similar computation for $ (k_0, k_2)=(2,0) $, which has 44 unknown constants in the modular ansatz. Requiring the partition function has correct GV-invariant form \eqref{GV} at this order, we can express all $ C_i^{(2,0)} $ in terms of $ C_1^{(1,0)} $. Moreover, we find only two solutions at this order: 
one is $ C_1^{(1,0)}=0 $ which leads to the trivial solution $ Z_{(1,0)}^{\mathrm{IIA}} = Z_{(1,1)}^{\mathrm{IIA}} = Z_{(2,0)}^{\mathrm{IIA}} = 0 $, and another one is $ C_1^{(1,0)} = 1/12 $. The latter non-trivial solution reproduces the result \eqref{(2,0)-GLSM} from the ADHM computation at $ (k_1, k_2)=(1,0), (1,1), (2,0) $. 

Furthermore, we also check that 110 unknown constants in the $ (k_1, k_2)=(2,1) $ modular ansatz can be completely fixed by requiring the GV-invariant ansatz \eqref{GV}.
We report the results in Table~\ref{table:A1-LST-modular}. In the table, we write an ordered list of $ C_i^{(k_1, k_2)} $, where $ C_i^{(k_1, k_2)} $ appears earlier than $ C_j^{(k_1, k_2)} $ in the list if $ (a_4^{(i)}, a_6^{(i)}, b_1^{(i)}, \cdots, b_6^{(i)}) $ in the ansatz \eqref{A1-numerator-ansatz} appears before $ (a_4^{(j)}, a_6^{(j)}, b_1^{(j)}, \cdots, b_6^{(j)}) $ in an ascending order\footnote{
We define the ascending order as follows. 
Suppose the modular ansatz is given as \eqref{numerator-ansatz}, where we label weights and indices of the Jacobi forms as $ j_1 < j_2 $ if $ k_{j_1l} < k_{j_2,l} $ or $ k_{j_1l}=k_{j_2l} $, $ m_{j_1l} < m_{j_2l} $.
We also define a set of the exponents in the ansatz as
\begin{align*}
L^{(i)} := \big\{ a_1^{(i)}, a_2^{(i)}, b_{j_1,l_1}^{(i)}, \cdots, b_{j_{N_1}, l_1}^{(i)}, \cdots, b_{j_1, l_n}^{(i)}, \cdots, b_{j_{N_n}, l_n}^{(i)} \big\}.
\end{align*}
Then, if we have $(L^{(i)})_{1}=(L^{(j)})_{1},...,(L^{(i)})_{s-1}=(L^{(j)})_{s-1}$, and $(L^{(i)})_{s}<(L^{(j)})_{s}$, we order $L^{(i)}$ and $L^{(j)}$ as $\{L^{(i)}, L^{(j)} \}$.
In this way, we fix the ordering of $L^{(i)}$, and we define their set $\{L^{(1)},...,L^{(N)} \}$ that we call ascending order. 
The ordering of $ C_i$ follows the ordering of $L^{(i)}$.
\label{ascending}}.
For instance, in the case of $\Phi_{(1,1)}$, we have
\begin{align}
    (a_4^{(i)}, a_6^{(i)}, b_1^{(i)}, \cdots, b_6^{(i)}) = \left\{
    \begin{array}{ll}
        (0, 0, 0, 0, 0, 2, 2, 0) & \quad (i=1) \\
        (0, 0, 0, 0, 1, 1, 1, 1) & \quad (i=2) \\
        (0, 0, 0, 0, 2, 0, 0, 1) & \quad (i=3) \\
        (1, 0, 0, 0, 2, 0, 2, 0) & \quad (i=4)
    \end{array} \right.
\end{align}
When we look at $a_4^{(i)}$, $a_4^{(4)}$ is the biggest value, so $C_{4}^{(1,1)}$ is the last element.
Similarly, we find $b_3^{(i)}$, $b_3^{(1)}<b_3^{(2)}<b_3^{(3)}=b_3^{(4)}$, so $C_{1}^{(1,1)}$ is the first element, and $C_{2}^{(1,1)}$ is the second element.
Therefore, the ascending order of $\{ C_i \}$ in this case is
\begin{align}
\{C_{1}^{(1,1)}, C_{2}^{(1,1)}, C_{3}^{(1,1)}, C_{4}^{(1,1)} \}.
\end{align}

\begin{table}
	\centering
	\begin{Tabular}{|c|L{70ex}|} \hline
        $ (k_1, k_2) $ & \multicolumn{1}{c|}{$ \big\{C_i^{(k_1, k_2)} \big\} $} \\ \hline \hline
		$ (1, 0) $ & $ \frac{1}{2^2 \cdot 3} \{ 1, -1 \} $ \\ \hline
		$ (1, 1) $ & $ \frac{1}{2^4 \cdot 3^2} \{ 1, -2, 1, 0 \} $ \\ \hline
		$ (2, 0) $ & \scriptsize $ \frac{1}{2^{15} \cdot 3^8} \{1, \allowbreak -1, \allowbreak -1, \allowbreak 1, \allowbreak 0, \allowbreak 0, \allowbreak 40, \allowbreak -40, \allowbreak -40, \allowbreak 40, \allowbreak 0, \allowbreak 0, \allowbreak -32, \allowbreak 32, \allowbreak 32, \allowbreak -32, \allowbreak 0, \allowbreak -15, \allowbreak 15, \allowbreak 15, \allowbreak -15, \allowbreak 0, \allowbreak 0, \allowbreak 24, \allowbreak -24, \allowbreak -24, \allowbreak 24, \allowbreak 0, \allowbreak 0, \allowbreak 0, \allowbreak -45, \allowbreak 45, \allowbreak 45, \allowbreak -45, \allowbreak 0, \allowbreak 0, \allowbreak 0, \allowbreak 0, \allowbreak 0, \allowbreak 27, \allowbreak -27, \allowbreak -27, \allowbreak 27, \allowbreak 0\} $ \\ \hline
		$ (2, 1) $ & \scriptsize $ \frac{1}{2^{17} \cdot 3^9} \{ 1, \allowbreak -5, \allowbreak 4, \allowbreak 0, \allowbreak 2, \allowbreak 2, \allowbreak -4, \allowbreak -3, \allowbreak 3, \allowbreak 0, \allowbreak -16, \allowbreak 8, \allowbreak 8, \allowbreak 0, \allowbreak -32, \allowbreak 40, \allowbreak -8, \allowbreak 48, \allowbreak -48, \allowbreak 48, \allowbreak -48, \allowbreak 8, \allowbreak -40, \allowbreak 32, \allowbreak 0, \allowbreak -8, \allowbreak -8, \allowbreak 16, \allowbreak 0, \allowbreak 96, \allowbreak -96, \allowbreak -128, \allowbreak 64, \allowbreak 64, \allowbreak 0, \allowbreak 128, \allowbreak -160, \allowbreak 32, \allowbreak 0, \allowbreak 6, \allowbreak 6, \allowbreak -12, \allowbreak 0, \allowbreak -12, \allowbreak 24, \allowbreak -24, \allowbreak 12, \allowbreak -9, \allowbreak 9, \allowbreak -18, \allowbreak 18, \allowbreak 6, \allowbreak 6, \allowbreak -12, \allowbreak 3, \allowbreak -3, \allowbreak 0, \allowbreak -24, \allowbreak 24, \allowbreak 0, \allowbreak 0, \allowbreak 0, \allowbreak 48, \allowbreak -48, \allowbreak 24, \allowbreak -24, \allowbreak 0, \allowbreak -48, \allowbreak 48, \allowbreak 0, \allowbreak 0, \allowbreak 0, \allowbreak 0, \allowbreak 9, \allowbreak -9, \allowbreak 36, \allowbreak -18, \allowbreak -18, \allowbreak -54, \allowbreak 54, \allowbreak -27, \allowbreak 27, \allowbreak -36, \allowbreak 72, \allowbreak -72, \allowbreak 36, \allowbreak 0, \allowbreak 36, \allowbreak -18, \allowbreak -18, \allowbreak 0, \allowbreak 0, \allowbreak 0, \allowbreak 0, \allowbreak 0, \allowbreak 0, \allowbreak 0, \allowbreak -81, \allowbreak 81, \allowbreak 108, \allowbreak -54, \allowbreak -54, \allowbreak 0, \allowbreak -108, \allowbreak 135, \allowbreak -27, \allowbreak 0, \allowbreak 0, \allowbreak 0, \allowbreak 0 \} $ \\ \hline
	\end{Tabular}
	\caption{Coefficients $C_i^{(k_1,k_2)}$ in the modular ansatz of $ \mathcal{N}=(2,0) $ $ \hat{A}_1 $ LST.} \label{table:A1-LST-modular}
\end{table}

\paragraph{Blowup equation}

Finally, we consider the blowup equation for the (2,0) $\hat{A}_1$ LST. As explained in the section~\ref{subsec:blowup-LST}, the tree level contribution to the effective prepotential consists of two parts. The first one is from the Green-Schwarz term for the dynamical tensor multiplet and the second one is the contribution from the auxiliary 2-form field $ B_0 $ to cancel the mixed gauge-global anomalies. We can write the effective prepotential as
\begin{align}\label{IIA-prepotential}
    \mathcal{E} = \frac{1}{\epsilon_1 \epsilon_2} \qty( \tau(\phi_{1,0}-\phi_{2,0})^2 + (\phi_{1,0} - \phi_{2,0}) (-m^2 + \epsilon_+^2) ) + \mathcal{E}_{\mathrm{tree}}^{(0)} \, ,
\end{align}
where $ \phi_{1,0} - \phi_{2,0} $ is the scalar vacuum expectation value (VEV) of the dynamical tensor multiplet.
The second contribution $\mathcal{E}_{\mathrm{tree}}^{(0)}$ from the auxiliary 2-form field $ B_0 $ is given by
\begin{align}
    \mathcal{E}_{\mathrm{tree}}^{(0)} = \frac{1}{\epsilon_1 \epsilon_2} \qty( -2m^2 + 2\epsilon_+^2 )\phi_{0,0} \, ,
\end{align}
where $ \phi_{0,0} $ is the auxiliary scalar associated with the $ B_0 $ field. 

To formulate the blowup equation, we have to sum over magnetic fluxes for both the dynamical tensor field and the auxiliary 2-form field which can be realized by shifting the parameters as
\begin{align}
    \phi_{1,0} - \phi_{2,0} \to \phi_{1,0} - \phi_{2,0} + n_{1,0} \epsilon_{1,2} \, , \quad 
    \phi_{0,0} \to \phi_{0,0} + n_{0,0} \epsilon_{1,2} \, , 
    \quad n_{1,0}, n_{0,0} \in \mathbb{Z} \, .
\end{align}
We do not turn on the background magnetic fluxes for $ \tau $ and $ w $: $ B_\tau = B_w = 0 $.
We propose the blowup equation for this LST as
\begin{align}\label{A1-(2,0)-blowup}
    \Lambda \hat{Z}_{\mathrm{str}}^{\mathrm{IIA}} = \sum_{n_1,n_2 \in \mathbb{Z}} (-1)^{n_1+n_2} q^{(n_1-n_2)^2} \qty(M \sqrt{p_1 p_2})^{n_1+n_2} \hat{Z}_{\mathrm{str}}^{\mathrm{IIA}(N)} \hat{Z}_{\mathrm{str}}^{\mathrm{IIA}(S)} \, ,
\end{align}
where $  n_1\equiv n_{0,0}+n_{1,0}  $ and $ n_2\equiv n_{0,0} $. 
We absorbed the perturbative part of the partition function into $ \Lambda $ as it is independent of the parameters $ \phi_{0,0}, \phi_{1,0}-\phi_{2,0} $.

To begin with, we will demonstrate how the known elliptic genera, as given in \eqref{(2,0)-GLSM}, can be a solution to the blowup equation, although this equation becomes singular along the summation direction $ n_1=n_2 $, as mentioned in section~\ref{subsec:blowup-LST}. 
At $ (k_1, k_2) = (1, 0) $ order, the blowup equation \eqref{A1-(2,0)-blowup} is given by
\begin{align}\label{A1-(2,0)-blowup-(1,0)}
\sum_{n_1,n_2 \in \mathbb{Z}} F(n_1, n_2) \coloneqq \sum_{n_1,n_2 \in \mathbb{Z}} & (-1)^{n_1+n_2} q^{(n_1-n_2)^2} \qty(M \sqrt{p_1 p_2})^{n_1+n_2} \\
& \cdot \qty( p_1^{2(n_1-n_2)} \hat{Z}_{(1,0)}^{\mathrm{IIA}(N)} + p_2^{2(n_1-n_2)} \hat{Z}_{(1,0)}^{\mathrm{IIA}(S)} - \hat{Z}_{(1,0)}^{\mathrm{IIA}}) = 0 \, , \nonumber
\end{align}
where we choose $ \Lambda = \sum_k \Lambda_k e^{2\pi i k w} $ with
\begin{align}
\Lambda_0 = \sum_{n_1,n_2 \in \mathbb{Z}} (-1)^{n_1+n_2} q^{(n_1-n_2)^2} \qty(M \sqrt{p_1 p_2})^{n_1+n_2} \, .
\end{align}
Suppose we consider only magnetic fluxes $ (n_1, n_2) = (0,0) $. Then \eqref{A1-(2,0)-blowup-(1,0)} becomes
\begin{align}
\sum_{(n_1,n_2)=(0,0)} F(n_1, n_2) 
&= \bigg( \frac{1}{M^2} +  \frac{(1+p_1)(1+p_2)}{M \sqrt{p_1p_2}} + \qty(2+p_1+p_1^{-1}+p_2+p_2^{-1}) \nonumber \\
&\qquad + \frac{M (1+p_1)(1+p_2)}{\sqrt{p_1p_2}} + M^2 \bigg) q + \mathcal{O}(q^2),
\end{align}
in the double expansion of $ q $ and $ M $. 
Now we add the contributions coming from the magnetic fluxes $ |n_{1,2}| \le  1 $. We then find that $ M^0 $ and $ M^{\pm 1} $ terms at $ q^1 $ order are all canceled and the remaining terms are
\begin{align}
\sum_{\abs{n_{1,2}} \leq 1} F(n_1, n_2)
&= \bigg( \frac{1}{M^4 p_1p_2} + \frac{(1+p_1)(1+p_2)}{M^3 (p_1p_2)^{3/2}} + \frac{1+p_1+p_2}{M^2 p_1p_2} + M^2(p_1+p_2+p_1p_2) \nonumber \\
&\quad + M^3(1+p_1)(1+p_2)\sqrt{p_1p_2} + M^4 p_1 p_2 \bigg) q + \mathcal{O}(q^2) \, .
\end{align}
Again, if we consider the summation of the magnetic fluxes up to $ \abs{n_1}, \abs{n_2} \leq 2 $, $ M^{\pm 2} $ and $ M^{\pm 3} $ terms are canceled and only higher order terms with $ M^{\pm 4, \pm 5, \pm 6} $ remain at $ q^1 $ order. In this way, if we sum over sufficiently large magnetic fluxes, every order of the K\"ahler parameter expansion is satisfied. Using the elliptic genera \eqref{(2,0)-GLSM} and
\begin{align}
\Lambda = \frac{e^{\pi i w/12}}{\eta(w)} \Lambda_0 \, ,
\end{align}
we checked that such cancellation occurs up to $ k_1, k_2 \leq 2 $ string numbers and $ q^3 $ order.

Now, we will solve the blowup equation and determine the unknown coefficients in the modular ansatz. For $ Z_{(k,0)} $ and $ Z_{(0,k)} $, we can use the elliptic genera for $k$ M-strings in \cite{Haghighat:2013gba} and they satisfy the blowup equations for the M-string theory as discussed in \cite{Gu:2019pqj, Kim:2020hhh}.
The $ (k_1, k_2)=(1,1) $ order in the blowup equation is independent of dynamical K\"ahler parameter and thus we cannot fix four unknown coefficients in the modular ansatz  at this order. What we can determine is $ \Lambda$ at this order  expressed as $ \tau $, $ m $, $ \epsilon_{1,2} $, and the unknown constants in the ansatz. Next, we solve the $(k_1, k_2)=(2,1) $ order in the blowup equation which now contains dynamical K\"ahler parameter $ \phi_{1,0}-\phi_{2,0} $.
We need to fix $ 4+110 $ undetermined coefficients arising from $ (k_1, k_2)=(1,1), (2,1) $ elliptic genera.
For this we substitute the modular ansatz and $ \Lambda_1 $ into the blowup equation at $ (k_1, k_2)=(2,1) $ order, and solve it order by order in the $q$ and $M$ double expansion as previously described.
This allows us to determine all $ 4+110 $ unknown coefficients as well as the $ \Lambda_1 $ factor. The result is in perfect agreement with Table~\ref{table:A1-LST-modular}. We expect that higher order elliptic genera can be similarly calculated.

\subsubsection{IIB picture}

The LST theory on two NS5-branes in type IIB string theory is the $ \mathcal{N}=(1,1) $ $ U(2) $ Yang-Mills theory.
The partition function of this LST is factorized as
\begin{align}
Z^{\mathrm{IIB}}_{\mathrm{GV}} = Z_{\mathrm{pert}}^{\mathrm{IIB}} \cdot Z_{\mathrm{str}}^{\mathrm{IIB}} \, ,
\end{align}
where the perturbative contribution coming from the $ U(2) $ vector multiplet and an adjoint hypermultiplet is given by
\begin{align}
\begin{aligned}
Z_{\mathrm{pert}}^{\mathrm{IIB}}
&= \PE\qty[ -\frac{1+p_1 p_2}{(1-p_1)(1-p_2)} \qty(Q^2 + 2q + q Q^{-2}) \frac{1}{1-q}] \\
&\quad \cdot \PE\qty[\frac{\sqrt{p_1p_2}}{(1-p_1)(1-p_2)} \qty(Q^2 + 2q + qQ^{-2}) \qty(M + M^{-1}) \frac{1}{1-q} ] \, ,
\end{aligned}
\end{align}
where $ Q = e^{2\pi i \phi_1} $ is the $ SU(2) $ gauge fugacity and $ M=e^{2\pi i m} $ is the fugacity for the $SU(2)_m$ symmetry of the adjoint hypermultiplet. 
The partition function of the instanton strings is given by
\begin{align}
Z_{\mathrm{str}}^{\mathrm{IIB}} = \sum_{k=0}^\infty e^{2\pi i k w} Z_k^{\mathrm{IIB}} \, ,
\end{align}
where the little string tension $ w $ is identified with the square of the inverse gauge coupling $ 1/g_{\mathrm{YM}}^2 $ in the low energy Yang-Mills theory.

\paragraph{GLSM}

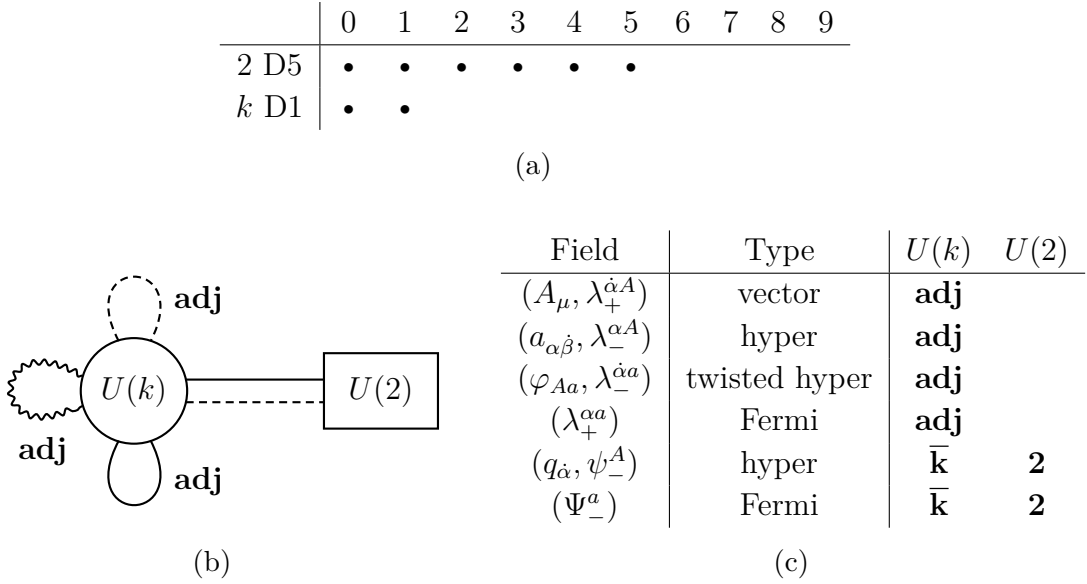
\begin{figure}
    \centering
    \begin{subfigure}[b]{0.6\textwidth}
        \centering
        \begin{tabular}{c|ccccccccccc}
            & 0 & 1 & 2 & 3 & 4 & 5 & 6 & 7 & 8 & 9 \\ \hline
            2 D5 & \textbullet & \textbullet & \textbullet & \textbullet & \textbullet & \textbullet \\ 
            $k$ D1 & \textbullet & \textbullet & & & & 
        \end{tabular}
        \subcaption{}
    \end{subfigure}
    
    \phantom{a}

\begin{subfigure}[b]{0.44\textwidth}
        \centering
        \begin{tikzpicture}
            \draw (0, 0) node {$ U(k) $};
            \draw [thick] (0, 0) circle (0.7);
		    \draw [thick] (-0.2, -0.66) .. controls (-0.8, -1.8) and (0.8, -1.8) .. (0.22, -0.66);
		    \draw [thick, decorate, decoration={snake, segment length=4pt, amplitude=1pt}] (-0.68, 0.2) .. controls (-1.9, 0.9) and (-1.9, -0.9) .. (-0.68, -0.2);
		    \draw [thick, densely dashed] (-0.2, 0.66) .. controls (-0.8, 1.8) and (0.8, 1.8) .. (0.22, 0.66);
            \draw (0.85, -1.2) node {\textbf{adj}};
            \draw (-1.2, -0.8) node {\textbf{adj}};
            \draw (0.85, 1.2) node {\textbf{adj}};
            \draw [thick] (0.68, 0.15) -- (2.5, 0.15);
            \draw [thick, densely dashed] (0.68, -0.15) -- (2.5, -0.15);
		    \draw [thick] (2.5, 0.5) rectangle (4, -0.5);
		    \draw (3.25, 0) node {$ U(2) $};
        \end{tikzpicture}
        \subcaption{}
    \end{subfigure}
    \begin{subfigure}[b]{0.55\textwidth}
        \centering
        \begin{tabular}{c|c|cc}
            Field & Type & $ U(k) $ & $ U(2) $ \\ \hline
            $ (A_\mu, \lambda_+^{\dot{\alpha}A}) $ & vector & $ \mathbf{adj} $ & \\
            $ (a_{\alpha\dot{\beta}}, \lambda_-^{\alpha A}) $ & hyper & $ \mathbf{adj} $ \\
            $ (\varphi_{Aa}, \lambda_-^{\dot{\alpha}a}) $ & twisted hyper & $ \mathbf{adj} $ & \\
            $ (\lambda_+^{\alpha a}) $ & Fermi & $ \mathbf{adj} $ & \\
            $ (q_{\dot\alpha}, \psi_-^A) $ & hyper & $ \overline{\mathbf{k}} $ & $ \mathbf{2} $ \\
            $ (\Psi_-^a) $ & Fermi & $ \overline{\mathbf{k}} $ & $ \mathbf{2} $
        \end{tabular}
        \subcaption{}
    \end{subfigure}
    \caption{(a) A brane configuration of the $ \hat{A}_1 $ LST in the type IIB string theory which consists of $ 2 $ D5-branes and $ k $ D1-branes. (b) The 2d $ \mathcal{N}=(0,4) $ gauge theory description for the $k$ $ D1 $-branes, where a circle represents gauge symmetry, a square means flavor symmetry, and solid, dashed and zigzag lines denote hypermultiplets, Fermi multiplets and twisted hypermultiplets, repectively. (c) The $ \mathcal{N}=(0,4) $ matter content of the 2d gauge theory.} \label{fig:IIB-brane}
\end{figure}

Upon applying S-duality, the system of NS5-branes and F1-strings in the type IIB string theory is transformed into a system of the D1- and D5-branes.
For $k$-instanton strings, we consider a configuration where $ k $ D1-branes are bound to $ 2 $ D5-branes as illustrated in Figure~\ref{fig:IIB-brane}(a). The 2d theory on the D1-branes is described by a $ \mathcal{N}=(4,4) $ $U(k)$ gauge theory with $ U(2) $ flavor symmetry.
The theory also has an $ SO(4) = SU(2)_l \times SU(2)_r $ symmetry which rotates the $2345$ directions, and an $ SO(4) = SU(2)_R \times SU(2)_{m} $ which rotates the $6789$ directions. 
This brane configuration is studied in \cite{Aharony:1999dw, Kim:2015gha}, and we summarize the 2d gauge theory description and its matter content in $ \mathcal{N}=(0,4) $ language in Figure~\ref{fig:IIB-brane}(b) and (c). 
In Figure~\ref{fig:IIB-brane}(c), we denote by $ \{a, b\cdots\} $ doublet indices for $ SU(2)_m $, and other indices have been already introduced in Section \ref{subsubsec:IIA}.

Based on the 2d gauge theory description, the elliptic genus of the LST can be computed by evaluating the JK-residue of the contour integral of the 1-loop determinants from all the supermultiplets. The result for $k$-strings is \cite{Kim:2015gha}
\begin{align}\label{(1,1)-GLSM}
Z_k^{\mathrm{IIB}} = \sum_{\abs{Y_1}+\abs{Y_2} = k} \prod_{i,j=1}^2 \prod_{s\in Y_i} \frac{\theta_1(\tau, E_{ij}(s) + m - \epsilon_-) \theta_1(\tau, E_{ij}(s) - m - \epsilon_-)}{\theta_1(\tau, E_{ij}(s) - \epsilon_1) \theta_1(\tau, E_{ij}(s) + \epsilon_2)} \, ,
\end{align}
where $ Y_1 $ and $ Y_2 $ are Young diagrams, $ s $ is a box in the Young diagram, and
\begin{align}
E_{ij}(s) = a_i - a_j - \epsilon_1 h_i(s) + \epsilon_2 v_j(s),
\end{align}
with $ a_1 = -a_2 = \phi_1 $.
Here, $ h_i(s) $ and $ v_j(s) $ are the arm length and leg length of a box $ s $ in $ Y_i $ and $ Y_j $, respectively. 

Since the IIA LST and the IIB LST are related by the T-duality, they should have the same BPS spectra when placed on a spatial circle.
This means that the partition function $ Z_{\mathrm{GV}}^{\mathrm{IIA}} $ of the type IIA LST is same as  $ Z_{\mathrm{GV}}^{\mathrm{IIB}} $  for the type IIB LST under the exchange $ \tau $ and $ w $ up to extra factors which are independent of the dynamical parameter. More precisely, the following relation has been checked explicitly by expanding both sides in terms of $ w $, $ q $, $ Q $ and $ M $ in \cite{Kim:2015gha}:
\begin{align}
    \left. Z^{\mathrm{IIA}}_{\mathrm{GV}} \right\vert_{\substack{\tau \leftrightarrow w \qquad \quad \! \\ \phi_{1,0}-\phi_{2,0} \to \phi_1}} = Z^{\mathrm{IIB}}_{\mathrm{GV}} \cdot \frac{q^{1/24}}{\eta(\tau)} \, .
\end{align}

\paragraph{Modularity}

The modular properties of the elliptic genus can be read off from the anomalies of the 2d chiral matters given in Figure~\ref{fig:IIB-brane}(c). The chiral fermions contribute to the 2d anomaly polynomial as
\begin{align}
    \lambda_+^{\dot{\alpha}A} + \lambda_-^{\alpha A} \ &\to \ 2k^2 \qty(\frac{c_2(r) + c_2(R)}{2} - \frac{c_2(l) + c_2(R)}{2}) \, , \\
    \lambda_+^{\alpha a} + \lambda_-^{\dot{\alpha}a} \ &\to \ 2k^2 \qty(\frac{c_2(l) + c_2(m)}{2} - \frac{c_2(r) + c_2(m)}{2}) \, , \\
    \Psi_+^a + \psi_-^A \ &\to \ 2k \qty(c_2(m) - c_2(R)) \, ,
\end{align}
where $ c_2(m) $ is the second Chern class of the $SU(2)_m$ symmetry.
Summing up these contributions gives the anomaly polynomial of the 2d gauge theory
\begin{align}
I_4 = 2k \qty(-c_2(R) + \frac{1}{4} \Tr F_m^2) \, .
\end{align}
This indeed agrees with the anomaly inflow result in \eqref{eq:I4-LST}, which takes into account the contribution from the counterterm in \eqref{B0-S}. This serves as indirect evidence to support the use of the counterterm in \eqref{B0-S} for cancelling the mixed gauge-global anomaly of the 6d LST.

From the modular anomaly $f(z)= \int I_4 = 2k(m^2 - \epsilon_+^2) $, we can set a modular ansatz for the $k$ instanton string as
\begin{align}\label{IIB-modular}
Z_k^{\mathrm{IIB}} = \frac{\Phi_k(\tau, \epsilon_\pm, 2\phi_1, m)}{\prod_{s=1}^k \varphi_{-1,1/2}(s \epsilon_{1,2}) \prod_{l=0}^{s-1} \varphi_{-1,1/2}((s+1)\epsilon_+ + (s-1-2l)\epsilon_- \pm 2\phi_1)} \, .
\end{align}
The numerator $ \Phi_k $ can be written in terms of the Eisenstein series $ E_4(\tau) $, $ E_6(\tau) $ and the $ SU(2) $ Weyl invariant Jacobi forms for $ \epsilon_\pm $, $ 2\phi_1 $ and $ m $. One can explicitly check that the elliptic genus \eqref{(1,1)-GLSM} has the same denominator structure as that in \eqref{IIB-modular}. We summarize the coefficients in the modular ansatz in Table~\ref{table:A1-B-LST-modular} obtained by comparing two expressions \eqref{(1,1)-GLSM} and \eqref{IIB-modular}. The ordering of the coefficients is ascending order with respect to $ \{ \epsilon_+, \epsilon_-, 2\phi_1, m \} $ for $ k=1 $ as defined in footnote~\ref{ascending}. Here, for $k=2$, we set $ \epsilon_+=0 $ for simplicity and the order of coefficients in the modular ansatz is ascending order with respect to $ \{\epsilon_-, 2\phi_1, m \} $.

\begin{table}
	\centering
	\begin{Tabular}{|C{8ex}|L{70ex}|} \hline
        $ k $ & \multicolumn{1}{c|}{$ \big\{ C_i^{(k)} \big\} $} \\ \hline \hline
		$ 1 $ & \small  $ \frac{1}{2^9 \cdot 3^5} \{ 1, \allowbreak -1, \allowbreak -1, \allowbreak 1, \allowbreak 3, \allowbreak 0, \allowbreak -3, \allowbreak 0, \allowbreak 0, \allowbreak 12, \allowbreak 0, \allowbreak -12, \allowbreak 8, \allowbreak 4, \allowbreak -8, \allowbreak -4, \allowbreak 0, \allowbreak -3, \allowbreak 0, \allowbreak 3, \allowbreak -9, \allowbreak -3, \allowbreak 9, \allowbreak 3, \allowbreak -3, \allowbreak 0, \allowbreak 3, \allowbreak 0, \allowbreak 0, \allowbreak -9, \allowbreak 0, \allowbreak 9 \} $ \\ \hline
		$ 2 $ ($ \epsilon_+ = 0 $) & \small $\frac{1}{2^{16} \cdot 3^9}$ $ \{-2, \allowbreak 4, \allowbreak -2, \allowbreak -1, \allowbreak 8, \allowbreak -10, \allowbreak 7, \allowbreak 6, \allowbreak -6, \allowbreak -4, \allowbreak -5, \allowbreak -9, \allowbreak 5, \allowbreak 10, \allowbreak 4, \allowbreak -5, \allowbreak 0, \allowbreak 8, \allowbreak 4, \allowbreak -4, \allowbreak 14, \allowbreak -24, \allowbreak -14, \allowbreak 16, \allowbreak 16, \allowbreak -28, \allowbreak 20, \allowbreak -24, \allowbreak -12, \allowbreak 28, \allowbreak -10, \allowbreak 0, \allowbreak -8, \allowbreak 20, \allowbreak 8, \allowbreak -10, \allowbreak 16, \allowbreak 80, \allowbreak -80, \allowbreak 0, \allowbreak -64, \allowbreak 16, \allowbreak 0, \allowbreak 0, \allowbreak -16, \allowbreak 48, \allowbreak 0, \allowbreak 0, \allowbreak -3, \allowbreak -9, \allowbreak 0, \allowbreak -6, \allowbreak 24, \allowbreak -6, \allowbreak -39, \allowbreak 63, \allowbreak -30, \allowbreak 0, \allowbreak 30, \allowbreak -24, \allowbreak 15, \allowbreak 18, \allowbreak -6, \allowbreak -30, \allowbreak -12, \allowbreak 15, \allowbreak 0, \allowbreak -24, \allowbreak -108, \allowbreak 84, \allowbreak -48, \allowbreak 108, \allowbreak 48, \allowbreak 0, \allowbreak 24, \allowbreak 36, \allowbreak -108, \allowbreak 0, \allowbreak 24, \allowbreak -36, \allowbreak 0, \allowbreak 0, \allowbreak 0, \allowbreak 0, \allowbreak -9, \allowbreak 36, \allowbreak 18, \allowbreak 9, \allowbreak 0, \allowbreak -72, \allowbreak -9, \allowbreak 18, \allowbreak -72, \allowbreak 63, \allowbreak 18, \allowbreak -36, \allowbreak 36, \allowbreak 0, \allowbreak -9, \allowbreak 9, \allowbreak 0, \allowbreak -18, \allowbreak -36, \allowbreak 72, \allowbreak 0, \allowbreak 36, \allowbreak -54, \allowbreak 27, \allowbreak 27, \allowbreak -81, \allowbreak 27, \allowbreak -54, \allowbreak 54, \allowbreak 0, \allowbreak -27, \allowbreak 27, \allowbreak 0, \allowbreak 0, \allowbreak 0, \allowbreak 0, \allowbreak 0\} $ \\ \hline
	\end{Tabular}
	\caption{Coefficients $C_i^{(k)}$ in the modular ansatz of $ \mathcal{N}=(1,1) $ $ \hat{A}_1 $ LST.} \label{table:A1-B-LST-modular}
\end{table}

\paragraph{Blowup equation}

We can fix the unknown constants in the modular ansatz using the blowup equation.
To begin with, let us evaluate the effective prepotential.
The 1-loop prepotential from the $ SU(2) $ vector multiplet, an adjoint hypermultiplet, and their KK towers is given by
\begin{equation}
\epsilon_1 \epsilon_2\mathcal{E}_{\rm 1-loop} = \frac{1}{12} \sum_{n \in \mathbb{Z}} \qty( \abs{n\tau \pm 2\phi_1}^3 - \abs{n\tau \pm 2\phi_1 + m}^3 ) +\epsilon_+^2\phi_1 = (-m^2+\epsilon_+^2)\phi_1 \: .
\end{equation}
Here, we use the zeta function regularization for the infinite sum. 
Then the effective prepotential is given by
\begin{align}\label{IIB-E}
    \mathcal{E} = \frac{1}{\epsilon_1 \epsilon_2} (-m^2 + \epsilon_+^2) \phi_1  + \mathcal{E}_{\mathrm{tree}}^{(0)} \, , \quad
    \mathcal{E}_{\mathrm{tree}}^{(0)} = \frac{1}{\epsilon_1 \epsilon_2} \qty( w \phi_1^2 + (-2m^2 + 2\epsilon_+^2) \phi_{0,0} ) \, ,
\end{align}
where the first term in $ \mathcal{E}_{\mathrm{tree}}^{(0)} $ is from the $ SU(2) $ gauge kinetic term and $ \phi_{0,0} $ is an auxiliary scalar for the non-dynamical 2-form field. One notices that under the reparametrization $ w \to \tau $, $ \phi_1 \to \phi_{1,0} - \phi_{2,0} $, the effective prepotential is the same as the type IIA prepotential in \eqref{IIA-prepotential}. 

To formulate the blowup equation, we choose magnetic fluxes for $ \phi_1 $, $ \phi_{0,0} $, $ m $, $ \tau $ and $ w $ as
\begin{align}
n_1 \in \mathbb{Z} \, , \quad
n_{0,0} \in \mathbb{Z} \, , \quad
B_m = 1/2 \, , \quad
B_\tau = B_w = 0 \, .
\end{align}
Since the effective prepotential in \eqref{IIB-E} and the elliptic genus in \eqref{(1,1)-GLSM} are the same as those for the type IIA picture up to reparametrization and overall factor, the same blowup equation should hold for the partition function in the type IIB picture:
\begin{align}
    \Lambda \hat{Z}_{\mathrm{str}}^{\mathrm{IIB}}
    = \sum_{n_{0,0},n_1 \in \mathbb{Z}} (-1)^{n_1} e^{-2\pi i V} \frac{\hat{Z}_{\mathrm{pert}}^{\mathrm{IIB}(N)} \hat{Z}_{\mathrm{pert}}^{\mathrm{IIB}(S)}}{\hat{Z}_{\mathrm{pert}}^{\mathrm{IIB}}} \hat{Z}_{\mathrm{str}}^{\mathrm{IIB}(N)} \hat{Z}_{\mathrm{str}}^{\mathrm{IIB}(S)} \, .
\end{align}
We checked that this blowup equation holds for up to 2-strings and the third order in $ q $-expansion. We also checked that inserting the 1-string modular ansatz \eqref{IIB-modular} into the blowup equation and solving it allows us to determine all 32 unknown constants given in Table~\ref{table:A1-B-LST-modular} within the ansatz.

\subsection{Heterotic LSTs} \label{subsec:heterotic-LST}

The second example is the $\mathcal{N}=(1,0)$ LSTs on $ N $ parallel NS5-branes in the $E_8\times E_8$ and $SO(32)$ heterotic string theories which we call rank $ N $ heterotic LSTs \cite{Berkooz:1997cq, Seiberg:1997zk}.
Again, these two LSTs are T-dual to each other under a circle compactification. 
In this subsection, we study the elliptic genera and the blowup equations of the rank 1 heterotic LSTs.

\subsubsection{\texorpdfstring{$ E_8 \times E_8 $}{E8×E8} picture}\label{subsec:E8xE8}

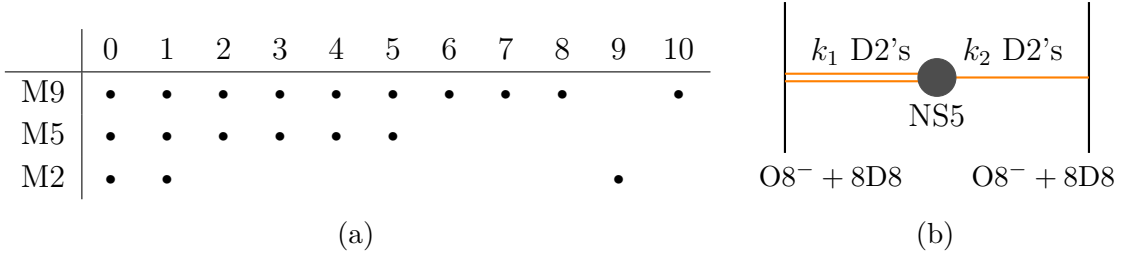
\begin{figure}
	\centering
	\begin{subfigure}[b]{0.64\textwidth}
		\centering
		\begin{tabular}{c|ccccccccccc}
			& 0 & 1 & 2 & 3 & 4 & 5 & 6 & 7 & 8 & 9 & 10 \\ \hline
			M9 & \textbullet & \textbullet & \textbullet & \textbullet & \textbullet & \textbullet & \textbullet & \textbullet & \textbullet & & \textbullet  \\
			M5 & \textbullet & \textbullet & \textbullet & \textbullet & \textbullet & \textbullet \\
			 M2 & \textbullet & \textbullet & & & & & & & & \textbullet 
		\end{tabular}
		\subcaption{}
	\end{subfigure}
	\begin{subfigure}[b]{0.35\textwidth}
		\centering
		\begin{tikzpicture}
		\draw [thick] (0, 1) -- (0, -1);
		\draw (0.6, -1.3) node {\small $ \mathrm{O8}^- + 8\mathrm{D8} $};
		\draw [thick] (4, 1) -- (4, -1);
		\draw (3.4, -1.3) node {\small $ \mathrm{O8}^- + 8\mathrm{D8} $};
		\draw (2, -0.5) node {NS5};
		\draw [thick, orange] (0, -0.05) -- (2, -0.05);
		\draw [thick, orange] (0, 0.05) -- (2, 0.05);
		\draw [thick, orange] (2, 0) -- (4, 0);
		\draw (1, 0.35) node {$ k_1 $ D2's};
		\draw (3, 0.35) node {$ k_2 $ D2's};
		\filldraw [black!70] (2, 0) circle (0.25);
		\end{tikzpicture}
		\subcaption{}
	\end{subfigure}
	\caption{(a) Branes for the $ E_8 \times E_8 $ LST in the M-theory setup. (b) The $ E_8 \times E_8 $ LST realized in type IIA string theory.}\label{fig:E8E8-Mtheory}
\end{figure}

The $E_8\times E_8$ heterotic LST is the worldvolume theory on a single M5-brane placed between two M9-branes at each end of the interval $ S^1/\mathbb{Z}_2 $.
Under the circle reduction, the M5-brane and the M9-branes reduce to an NS5-brane and two sets of $ \mathrm{O8}^- + 8\mathrm{D8}$-branes located at two ends of the interval as illustrated in Figure~\ref{fig:E8E8-Mtheory} \cite{Gadde:2015tra}. This theory can also be realized in F-theory by two $ -1 $ curves $ \Sigma^1 $ and $ \Sigma^2 $ in the base surface of an elliptic CY3 with the intersection matrix given by \cite{Bhardwaj:2015oru}
\begin{align}
\Omega^{\alpha\beta} = \mqty( -1 & 1 \\ 1 & -1 ) \, .
\end{align}

The partition function of this LST can be factorized into the perturbative part $Z_{\mathrm{pert}}^{\mathrm{HE}}$ for a single tensor multiplet and the contribution from strings $Z_{\mathrm{str}}^{\mathrm{HE}} $ as
\begin{align}
    Z^{\mathrm{HE}}_{\mathrm{GV}} = Z_{\mathrm{pert}}^{\mathrm{HE}} \cdot Z_{\mathrm{str}}^{\mathrm{HE}} = Z_{\mathrm{pert}}^{\mathrm{HE}} \cdot \sum_{k_1, k_2\ge0} Q^{k_1} \qty(\frac{e^{2\pi i w}}{Q})^{k_2} Z_{(k_1, k_2)}^{\mathrm{HE}} \, ,
\end{align}
where $ Q \equiv e^{2\pi i (\phi_{1,0} - \phi_{2,0})} $ and $\phi_{1,0}-\phi_{2,0}$ is the scalar VEV for the dynamical tensor multiplet.

\paragraph{GLSM}

The $E_8\times E_8$ LST contains non-perturbative strings arising from the D2-branes stretched between the D8-branes, the O$8^-$-plane and the NS5-brane in Figure~\ref{fig:E8E8-Mtheory}(b). 
The worldvolume theory on the D2-branes at low energy can be described by a 2d $\mathcal{N}=(0,4)$ gauge theory. For $(k_1,k_2)$-strings, the gauge group is $ O(k_1) \times O(k_2) $. There are an $ \mathcal{N}=(0,4) $ vector multiplet and a symmetric hypermultiplet coming from the D2-D2 string modes for each gauge node, the Fermi multiplets in the bifundamental representations of $ O(k_1) \times SO(16) $ and $ O(k_2) \times SO(16) $ coming from the D2-D8  string modes, and the $ O(k_1) \times O(k_2) $ bifundamental Fermi multiplets and twisted hypermultiplets from the strings between two adjacent D2-branes.
These multiplets form representations of the $ SU(2)_l \times SU(2)_r $ global symmetry which rotates $ 2345 $ directions, and those of the $ SO(3) \sim SU(2)_R $ rotational symmetry for $ 678 $ directions.
In the strong coupling limit, the $ 678 $ directions and the M-theory circle become an $ \mathbb{R}^4 $, so we expect the $ SO(3) $ symmetry enhances to $ SO(4) = SU(2)_R \times SU(2)_{m_0} $.
We summarize the matter content of the 2d theory in Figure~\ref{fig:E8E8-quiver}.
When $k_1=0$ or $k_2=0$, this 2d gauge theory reduces to that for self-dual strings in the 6d E-string theory studied in \cite{Kim:2014dza, Gadde:2015tra}.
\begin{figure}
	\centering
	\begin{subfigure}{1\textwidth}
		\centering
		\begin{tikzpicture}
		\draw (0, 0) node {$ O(k_1) $};
		\draw [thick] (0, 0) circle (0.7);
		\draw [thick, densely dashed] (-0.7, 0) -- (-2.2, 0);
		\draw (-3, 0) node {$ SO(16) $};
		\draw [thick] (-3.8, 0.5) rectangle (-2.2, -0.5);
		\draw (3, 0) node {$ O(k_2) $};
		\draw [thick] (3, 0) circle (0.7);
		\draw (6, 0) node {$ SO(16) $};
		\draw [thick] (6.8, 0.5) rectangle (5.2, -0.5);
		\draw [thick, densely dashed] (3.7, 0) -- (5.2, 0);
		\draw [thick] (-0.3, -0.65) .. controls (-0.9, -1.8) and (0.9, -1.8) .. (0.3, -0.65);
		\draw [thick] (2.7, -0.65) .. controls (2.1, -1.8) and (3.9, -1.8) .. (3.3, -0.65);
		\draw [thick, densely dashed] (0.7, -0.10) -- (2.3, -0.10);
		\draw [thick, decorate, decoration={snake, segment length=5pt, amplitude=1pt}] (0.7, 0.10) -- (2.3, 0.10);
		\draw (-1, -1.2) node {$ \mathbf{sym} $};
		\draw (4, -1.2) node {$ \mathbf{sym} $};
		\end{tikzpicture}
		\subcaption{}
	\end{subfigure}
\phantom{a}
	\begin{subfigure}{1\textwidth}
		\centering
		\begin{tabular}{c|c|cc}
			Field & Type & $ O(k_1) \times O(k_2) $ & $ SO(16) \times SO(16) $ \\ \hline
			$ (A_\mu^{(i)}, \lambda_{+(i)}^{\dot{\alpha}A}) $ & vector & $ (\mathbf{adj}, \mathbf{1}), (\mathbf{1}, \mathbf{adj}) $ \\
			$ (a_{\alpha\dot{\beta}}^{(i)}, \lambda_{-(i)}^{\alpha A}) $ & hyper & $ (\mathbf{sym}, \mathbf{1}), (\mathbf{1}, \mathbf{sym}) $ \\
			$ (\varphi_{A}, \chi_-^{\dot{\alpha}}) $ & twisted hyper & $ (\mathbf{k}_1, \mathbf{k}_2) $ \\
			$ (\chi_+^{\alpha}) $ & Fermi & $ (\mathbf{k}_1, \mathbf{k}_2) $ \\
			$ (\Psi_l^{(i)}) $ & Fermi & $ (\mathbf{k_1}, \mathbf{1}), (\mathbf{1}, \mathbf{k}_2) $ & $ (\mathbf{16}, \mathbf{1}), (\mathbf{1}, \mathbf{16}) $
		\end{tabular}
		\subcaption{}
	\end{subfigure}
	\caption{Quiver diagram (a) and matter content (b) of 2d $ \mathcal{N}=(0,4) $ gauge theory for $ E_8 \times E_8 $ LST. Here, solid, dashed and zigzag lines denote hypermultiplets, Fermi multiplets and twisted hypermultiplets, repectively and $ i=1,2 $ labels each gauge node.} \label{fig:E8E8-quiver}
\end{figure}

We can compute $ Z_{(k_1,k_2)} $ of the 2d gauge theory using the localization method \cite{Benini:2013nda, Benini:2013xpa}. Here, we give explicit expressions of the elliptic genera up to $ (k_1, k_2)=(k_2,k_1)=(2,1) $ order. 
The contour integral expressions for the elliptic genera and the detailed computations are presented in Appendix~\ref{app:E8E8}. When $ k_2=0 $, the elliptic genera reduce to those for the E-strings obtained in \cite{Kim:2014dza}:
\begin{align}
    Z_{(1,0)}^{\mathrm{HE}} &= -\frac{1}{2} \sum_{I=1}^4 \frac{\prod_{l=1}^8 \theta_I(m_l)}{\eta^6 \theta_1(\epsilon_1) \theta_1(\epsilon_2)} \, , \label{Estr-1str}\\
    Z_{(2,0)}^{\mathrm{HE}} &= \frac{1}{4\eta^{12} \theta_1(\epsilon_1) \theta_1(\epsilon_2)} \sum_{I=1}^4 \qty( \frac{\prod_{l=1}^8 \theta_I(m_l \pm \frac{\epsilon_1}{2})}{\theta_1(2\epsilon_1) \theta_1(\epsilon_2-\epsilon_1)} + \frac{\prod_{l=1}^8 \theta_I(m_l \pm \frac{\epsilon_2}{2})}{\theta_1(2\epsilon_2) \theta_1(\epsilon_1-\epsilon_2)} ) \nonumber \\
    &\quad + \sum_{I=1}^4 \sum_{J=I+1}^4 \frac{\theta_{\sigma(I,J)}(0) \theta_{\sigma(I,J)}(2\epsilon_+) \prod_{l=1}^8 \theta_I(m_l) \theta_J(m_l)}{4\eta^{12} \theta_1(\epsilon_{1,2})^2 \theta_{\sigma(I,J)}(\epsilon_1) \theta_{\sigma(I,J)}(\epsilon_2)} \, \label{Estr-2str} ,
\end{align}
where $ \theta_I $ are the Jacobi theta functions defined in \eqref{Jacobi-theta}, $ m_{1,\cdots,8} $ are chemical potentials for the $ SO(16) $ global symmetry, and $\sigma(I, J)$ is defined as
\begin{alignat}{3}
\begin{aligned}
&\sigma(I, J) = \sigma(J, I) \, , \
&&\sigma(I, I) = 0 \, , \
&&\sigma(1, I) = I \, , \\
&\sigma(2, 3) = 4 \, , \
&&\sigma(2, 4) = 3 \, , \
&&\sigma(3, 4) = 2 \, .
\end{aligned}
\end{alignat}
Here we use a shorthand notation $ \theta_I(x\pm y) = \theta_I(x+y) \theta_I(x-y) $. 
For $ (k_1, k_2)=(1,1) $,
\begin{align}\label{E8E8-1,1-str}
Z_{(1,1)}^{\mathrm{HE}} = \frac{1}{4} \sum_{I,J=1}^4 \frac{\prod_{l=1}^8 \theta_I(m_l) \cdot \prod_{l=9}^{16} \theta_J(m_l)}{\eta^{12} \theta_1(\epsilon_1)^2 \theta_1(\epsilon_2)^2} \frac{\theta_{\sigma(I,J)}(\pm m_0 + \epsilon_-)}{\theta_{\sigma(I,J)}(\pm m_0 - \epsilon_+)} \, ,
\end{align}
where $ m_{l=9,\cdots,16} $ are chemical potentials for the other $ SO(16) $ symmetry and $ m_0 $ is a chemical potential for $ SU(2)_{m_0} $. The $ (k_1,k_2)=(2,1) $-string elliptic genus is
\begin{align}\label{E8E8-2,1-str}
Z_{(2,1)}^{\mathrm{HE}} = \frac{1}{4} Z_{(2,1)}^{(0)} + \frac{1}{8} \sum_{K=1}^4 \sum_{I<J}^4 Z_{(2,1)}^{(I,J,K)} \, ,
\end{align}
where
\begin{align}
Z_{(2,1)}^{(0)} &= \sum_{I,J=1}^4 \frac{-\prod_{l=1}^8 \theta_I(m_l \pm \frac{\epsilon_1}{2}) \cdot \prod_{l=9}^{16} \theta_J(m_l)}{2\eta^{18} \theta_1(\epsilon_{1,2})^2 \theta_1(2\epsilon_1) \theta_1(\epsilon_2-\epsilon_1)} \frac{\theta_{\sigma(I,J)}(\pm m_0 + \epsilon_1 - \frac{\epsilon_2}{2})}{\theta_{\sigma(I,J)}(\pm m_0 - \epsilon_1 - \frac{\epsilon_2}{2})} + (\epsilon_1 \leftrightarrow \epsilon_2) \nonumber \\
&\quad + \sum_{I=1}^4 \frac{-\prod_{l=1}^8 \theta_I(m_l \pm (m_0 + \epsilon_+)) \cdot \prod_{l=9}^{16} \theta_I(m_l)}{\eta^{18} \theta_1(\epsilon_{1,2}) \theta_1(2m_0) \theta_1(2m_0+2\epsilon_+) \theta_1(2m_0+2\epsilon_++\epsilon_{1,2})} + (m_0 \to -m_0),
\end{align}
and
\begin{align}
Z_{(2,1)}^{(I,J,K)}
&= -\frac{\theta_{\sigma(I,J)}(0) \theta_{\sigma(I,J)}(2\epsilon_+)}{\eta^{18} \theta_1(\epsilon_{1,2})^3 \theta_{\sigma(I,J)}(\epsilon_{1,2})} \frac{\theta_{\sigma(I,K)}(\pm m_0 + \epsilon_-) \theta_{\sigma(J,K)}(\pm m_0 + \epsilon_-)}{\theta_{\sigma(I,K)}(\pm m_0 - \epsilon_+) \theta_{\sigma(J,K)}(\pm m_0 - \epsilon_+)} \nonumber \\
& \quad \cdot \prod_{l=1}^8 \theta_I(m_l) \theta_J(m_l) \cdot \prod_{l=9}^{16} \theta_K(m_l) \, .
\end{align}

A few remarks are in order. First, we expect that two $ SO(16) $ flavor symmetries which we can see in Figure~\ref{fig:E8E8-quiver} and also from the matter content to be enhanced to the $ E_8 \times E_8 $ symmetry at low energy. One can check this enhancement from the elliptic genera by expanding them in terms of $q$.
Second, although the worldsheet theory seems to have $SU(2)_{m_0}$ flavor symmetry, the bulk 6d LST does not have any matter fields charged under this symmetry.  In fact, this symmetry only acts on the string modes stretched between two O8-planes which correspond to 10d bulk modes moving along the direction parallel to the orientifold planes in Figure~\ref{fig:E8E8-Mtheory}(b).
These modes are decoupled from the 6D LST. Therefore the BPS spectrum of strings in the LST should not depends on $ m_0 $. 
Let us check these expectations.

First, the $ (1,0) $-string and $ (2,0) $-string elliptic genera \eqref{Estr-1str} and \eqref{Estr-2str} do not contain $ m_0 $.
Second, $ (1,1) $-string elliptic genus \eqref{E8E8-1,1-str} has $ m_0 $ dependence. However $ (1,1) $-string order is independent of the dynamical parameter $ Q$, and we can therefore consider it as a contribution from the bulk modes not involved in the LST spectrum. Lastly, $ (2,1) $-string elliptic genus \eqref{E8E8-2,1-str} does contain $ m_0 $ which seems to be a problem. Quite surprisingly, however if we express the partition function in the GV-invariant form given in \eqref{GV} and extract the BPS spectrum, $ m_0 $ dependence in $ (2,1) $-string order disappears completely. We find that the single letter index at $ (2,1) $-string order is
\begin{align}\label{E8E8-2,1-index}
    f_{(2,1)} &= q^{1/2} \big[ \chi_{1,1}(\epsilon_\pm) + \chi_{1/2,1/2}(\epsilon_\pm) (\chi_{\mathbf{248}}^{(1)}+1) + (\chi_{\mathbf{3875}}^{(1)} + \chi_{\mathbf{248}}^{(1)}+2) \big] \\
    & \ + q^{3/2} \big[ \chi_{2,2}(\epsilon_\pm) + \chi_{3/2,3/2}(\epsilon_\pm) (\chi_{\mathbf{248}}^{(1)} + 3) + 2\chi_{3/2,1/2}(\epsilon_\pm) \nonumber \\
    & \qquad + \chi_{1,1}(\epsilon_\pm) (\chi_{\mathbf{3875}}^{(1)} + 4\chi_{\mathbf{248}}^{(1)}+\chi_{\mathbf{248}}^{(2)}+5) + \chi_{1,0}(\epsilon_\pm) (2\chi_{\mathbf{248}}^{(1)}+3) + 2\chi_{1/2,3/2}(\epsilon_\pm) \nonumber  \\
    &\qquad + \chi_{1/2,1/2}(\epsilon_\pm) (\chi_{\mathbf{30380}}^{(1)} + \chi_{\mathbf{248}}^{(1)}\chi_{\mathbf{248}}^{(2)} + \chi_{\mathbf{248}}^{(2)} + 4\chi_{\mathbf{3875}}^{(1)} + 7\chi_{\mathbf{248}}^{(1)} + 10) \nonumber \\
    &\qquad + \chi_{0,1}(\epsilon_\pm) (2\chi_{\mathbf{248}}^{(1)}+3) + (\chi_{\mathbf{147250}}^{(1)} + 2\chi_{\mathbf{30380}}^{(1)} + \chi_{\mathbf{3875}}^{(1)}\chi_{\mathbf{248}}^{(2)} + \chi_{\mathbf{248}}^{(1)}\chi_{\mathbf{248}}^{(2)} \nonumber \\
    & \hspace{30ex} + 4\chi_{\mathbf{3875}}^{(1)} + 8\chi_{\mathbf{248}}^{(1)} + 2\chi_{\mathbf{248}}^{(2)} + 7) \big] + \cdots \, , \nonumber
\end{align}
where $f_{(k_1,k_2)}$ is defined via $ Z_{\mathrm{str}}^{\mathrm{HE}} = \PE[\frac{\sqrt{p_1p_2}}{(1-p_1)(1-p_2)} \sum Q^{k_1} (w/Q)^{k_2} f_{(k_1,k_2)}] $, $ \chi_{j_l,j_r}(\epsilon_\pm) = \chi_{j_l}(\epsilon_-) \chi_{j_r}(\epsilon_+) $ represents spin $ (j_l, j_r) $ state, and $ \chi_{\mathbf{R}}^{(i)} $ is the character of representation $ \mathbf{R} $ in the $ i $-th $ E_8 $ symmetry algebra. This is indeed independent of $m_0$ showing that the dynamical BPS states in the spectrum of the LST are independent of $U(1)_{m_0}$ symmetry. We expect this to hold for higher order computations.

\paragraph{Modularity}

The chiral fermions in the 2d $ \mathcal{N}=(0,4) $ gauge theory given in Figure~\ref{fig:E8E8-quiver}(b) contribute to the 2d anomaly polynomial $ I_4 $ as
\begin{align}
\begin{aligned}
\lambda_{+(1)}^{\dot{\alpha} A} + \lambda_{+(2)}^{\dot{\alpha} A} &\to \sum_{i=1}^2 k_i(k_i-1) \qty(\frac{c_2(r)+c_2(R)}{2} + \frac{p_1(T_2)}{24}) \, , \\
\lambda_{-(1)}^{\alpha A} + \lambda_{-(2)}^{\alpha A} &\to -\sum_{i=1}^2 k_i(k_i+1) \qty(\frac{c_2(l)+c_2(R)}{2} + \frac{p_1(T_2)}{24}) \, , \\
\chi_-^{\dot{\alpha}} + \chi_+^\alpha &\to 2k_1 k_2 \qty(\frac{c_2(l)-c_2(r)}{2}) \, , \\
\Psi_l^{(1)} + \Psi_l^{(2)} &\to \qty( \frac{k_1}{4}\Tr F_1^2 + \frac{k_2}{4}\Tr F_2^2 +(k_1+k_2) \frac{p_1(T_2)}{3} ) \, ,
\end{aligned}
\end{align}
where $ F_1 $ and $ F_2 $ are the 2-form field strengths for two $ SO(16) $ global symmetries. 
This agrees with the anomaly polynomial computed from the anomaly inflow using \eqref{eq:I4-LST}:
\begin{align}
I_4 &= -\frac{(k_1-k_2)^2}{2}(c_2(l)-c_2(r)) - \frac{k_1+k_2}{2}\qty(c_2(l)+c_2(r)+2c_2(R)-\frac{1}{2}p_1(T_2)) \nonumber \\
&\quad + \frac{k_1}{4} \Tr F_1^2 + \frac{k_2}{4} \Tr F_2^2 \, .
\end{align}

We notice that the elliptic genera above for $ k_1, k_2 \geq 1 $ have additional poles at $ m_0 = \epsilon_+ $ beside the center of mass contributions.
These poles come from the bulk modes decoupled from the 6d LST. Based this observation, we write the modular ansatz as
\begin{align}\label{E8-ansatz}
    Z_{(k_1,k_2)}^{\mathrm{HE}} = \frac{1}{\eta^{12(k_1+k_2)}}  \frac{\Phi_{(k_1,k_2)}(\tau, \phi, m_0, m_{l=1,\cdots,16})}{\mathcal{D}_{(k_1,k_2)}^{\mathrm{cm}} \cdot \prod_{s=1}^{\kappa} \varphi_{-1,1/2}(\pm s \lambda( m_0 - \epsilon_+))} \, ,
\end{align}
with $\kappa=\min(k_1,k_2)$.
The $ (1,1) $-string elliptic genus \eqref{E8E8-1,1-str} from ADHM computation and the identity $ \prod_{I=1}^4 \theta_I(\pm m_0 - \epsilon_+) = \eta^6 \theta_1(\pm 2m_0 - 2\epsilon_+) $ suggest $ \lambda = 2 $ here. 

Let us first consider the case where the flavor chemical potentials are switched off $ m_{l=1,\cdots,16}=0 $. In this case the numerator $ \Phi_{(k_1,k_2)} $ can be written in terms of the Eisenstein series and the $ SU(2) $ Jacobi forms for $ \epsilon_\pm $ and $ m_0 $, and we find that this ansatz is compatible with the elliptic genera from the ADHM construction. We explicitly check this up to $ (2,1) $-string order.

However, the cases with generic flavor chemical potentials turn out to be rather subtle. 
It has been shown that the $Z_{(k_1,0)}^{\mathrm{HE}}$ for the E-strings can be expressed in terms of $E_8$ Weyl invariant Jacobi forms  \cite{Kim:2014dza, Gu:2017ccq}, which demonstrates the enhancement of symmetry from $SO(16)$ to $E_8$ at low energy.
Similarly, we expect the symmetry enhancement $SO(16)\times SO(16)\rightarrow E_8\times E_8$ in the 2d CFTs for the strings in the LST. This can be verified by checking if the spectrum of the BPS strings forms $E_8\times E_8$ representations. Indeed, we explicitly checked in \eqref{E8E8-2,1-index} that the single letter index at $(k_1,k_2)=(2,1)$ can be written in terms of $E_8\times E_8$ characters. So it seems that one can formulate a consistent ansatz of the form \eqref{E8-ansatz} with generic flavor chemical poentials.

However, our analysis revealed that this is not the case due to the presence of additional bulk states that do not carry dynamical tensor charge. As these states are decoupled from the LST, we cannot expect them to form representations of the $E_8\times E_8$ symmetry. For instance, the single letter index at $ (k_1, k_2)=(1,1) $, which we compute from the $ (1,1) $-string elliptic genus in \eqref{E8E8-1,1-str}, cannot be written in terms of the $E_8\times E_8$ characters, as demonstrated below:
\begin{align}
    f_{(1,1)}
    &= \frac{e^{4\pi i m_0}}{1-e^{4\pi i(m_0 \pm \epsilon_+)}} \bigg[ \frac{\chi_{0,1/2}(\epsilon_+)}{q} + \Big( 2\chi_{1/2,1}(\epsilon_\pm) - \chi_{1/2,0}(\epsilon_\pm) (\chi_1(m_0)-1) \\
    &\hspace{9ex} + \chi_{0,1/2}(\epsilon_\pm) (\chi_1(m_0) + \chi_{\mathbf{120}}^{(1)} + \chi_{\mathbf{120}}^{(2)} + 1 ) + \chi_{1/2}(m_0) \chi_{\mathbf{16}}^{(1)} \chi_{\mathbf{16}}^{(2)} \Big) + \mathcal{O}(q)\bigg] \,, \nonumber
\end{align}
where the notations are same as those in \eqref{E8E8-2,1-index}, except that $ \chi_{\mathbf{R}}^{(i)} $ is now the $ i $-th $ SO(16) $ character. We believe that this is due to the presence of additional bulk states in the spectrum at this order. For this reason, the ADHM computations above when $k_1,k_2\ge 1$ do not give elliptic genera that exhibit the symmetry enhancement to $E_8\times E_8$. Therefore we are unable to write ansatzes that reproduce the ADHM results in terms of $ E_8 $ Jacobi forms.

Even though the elliptic genera obtained from the ADHM computation do not exhibit manifest $ E_8 \times E_8 $ symmetry, we can still attempt to construct a modular ansatz in terms of $ E_8 $ Weyl invariant Jacobi forms that accurately reproduces the BPS spectrum of the LST up to extra decoupled string states. There are nine fundamental $ E_8 $ Jacobi forms $ A_{1,2,3,4,5} $ and $ B_{2,3,4,6} $ given in \eqref{E8-Jacobi}, where $ A_n $ and $ B_n $ have index $ n $ and weight $ 4 $ and $ 6 $, repectively. We first write an ansatz for $ \Phi_{(k_1, k_2)} $ in \eqref{E8-ansatz} using the $ E_8 $ Jacobi forms for $ m_{1,\cdots,8} $ and $ m_{9,\cdots,16} $, together with the Eisenstein series and $ SU(2) $ Jacobi forms for $ \epsilon_\pm $ and $ m_0 $. We then fix the unknown coefficients in the ansatz by using the dynamical BPS spectrum from the ADHM computation.
To our surprise, we find that there are two values of $\lambda$, namely 1 and 2, that are consistent with both the $E_8\times E_8$ symmetry and the spectrum of the LST obtained through ADHM computation. We check this up to $ (2,1) $-string order and $ q^5 $ order in $ q $-expansion, and report the results from these two choices in Table~\ref{table:E8E8-modular} and \ref{table:E8E8-modular2}, respectively, where the coefficients are listed in ascending order with respect to $ \{\epsilon_+, \epsilon_-, m_0, m_{1,\cdots,8}, m_{9,\cdots,16}\} $.

\begin{table}
	\centering
	\begin{Tabular}{|c|L{70ex}|} \hline
        $ (k_1, k_2) $ & \multicolumn{1}{c|}{$ \big\{C_i^{(k_1, k_2)} \big\} $} \\ \hline \hline
        $ (1, 0) $ & $ \{ 1 \} $ \\ \hline
        $ (2, 0) $ & \scriptsize $ \frac{1}{2^{13} \cdot 3^6} \{4 , \allowbreak -3 , \allowbreak 5 , \allowbreak 3 , \allowbreak 10 , \allowbreak 32 , \allowbreak -15 , \allowbreak 96 , \allowbreak 32 , \allowbreak 36 , \allowbreak -24 , \allowbreak 40 , \allowbreak -12 , \allowbreak -40 , \allowbreak -128 , \allowbreak 0 , \allowbreak -5 , \allowbreak -4 , \allowbreak 5 , \allowbreak -32 , \allowbreak -24 , \allowbreak -9 , \allowbreak 0 , \allowbreak 15 , \allowbreak 9 , \allowbreak -10 , \allowbreak  64 , \allowbreak -5 , \allowbreak 32 , \allowbreak 0 , \allowbreak 3 , \allowbreak 6 , \allowbreak -15 , \allowbreak -9 , \allowbreak 0 , \allowbreak -72 , \allowbreak 15 , \allowbreak -96 , \allowbreak -12 , \allowbreak -3 , \allowbreak 3 , \allowbreak 0 , \allowbreak -27 , \allowbreak 18 , \allowbreak -45 , \allowbreak 9 , \allowbreak 45 , \allowbreak 108 , \allowbreak 0\} $ \\ \hline
        $ (1, 1) $ & $ \frac{1}{2^2 \cdot 3} \{ -1, 1 \} $ \\ \hline
        $ (2, 1) $ & \scriptsize $ \frac{1}{2^{15} \cdot 3^7} \{0, \allowbreak -4, \allowbreak 8, \allowbreak -4, \allowbreak 3, \allowbreak 0, \allowbreak -15, \allowbreak 0, \allowbreak 16, \allowbreak 16, \allowbreak -3, \allowbreak -3, \allowbreak 10, \allowbreak 15, \allowbreak -48, \allowbreak 112, \allowbreak 3, \allowbreak 5, \allowbreak -10, \allowbreak -112, \allowbreak 48, \allowbreak -5, \allowbreak -16, \allowbreak -16, \allowbreak -12, \allowbreak -24, \allowbreak 12, \allowbreak -40, \allowbreak 36, \allowbreak 24, \allowbreak 40, \allowbreak 40, \allowbreak -128, \allowbreak 0, \allowbreak -36, \allowbreak 0, \allowbreak -40, \allowbreak 256, \allowbreak -128, \allowbreak 0, \allowbreak 0, \allowbreak 0, \allowbreak 5, \allowbreak 0, \allowbreak 0, \allowbreak -24, \allowbreak 0, \allowbreak -5, \allowbreak -5, \allowbreak 40, \allowbreak -48, \allowbreak 5, \allowbreak 20, \allowbreak 8, \allowbreak 4, \allowbreak 9, \allowbreak 0, \allowbreak -5, \allowbreak 0, \allowbreak -9, \allowbreak -10, \allowbreak 5, \allowbreak 32, \allowbreak -9, \allowbreak 0, \allowbreak 15, \allowbreak 10, \allowbreak -32, \allowbreak 64, \allowbreak 0, \allowbreak 9, \allowbreak 0, \allowbreak -15, \allowbreak -96, \allowbreak 32, \allowbreak 0, \allowbreak 0, \allowbreak 0, \allowbreak 0, \allowbreak -9, \allowbreak 0, \allowbreak 15, \allowbreak 0, \allowbreak -12, \allowbreak 6, \allowbreak 9, \allowbreak 0, \allowbreak -15, \allowbreak -24, \allowbreak -60, \allowbreak 3, \allowbreak -6, \allowbreak -15, \allowbreak 0, \allowbreak 144, \allowbreak -120, \allowbreak -3, \allowbreak 0, \allowbreak 15, \allowbreak 72, \allowbreak 0, \allowbreak 3, \allowbreak -3, \allowbreak -3, \allowbreak 0, \allowbreak 0, \allowbreak 3, \allowbreak 0, \allowbreak 0, \allowbreak 0, \allowbreak 9, \allowbreak 18, \allowbreak -9, \allowbreak 45, \allowbreak -27, \allowbreak -18, \allowbreak -45, \allowbreak -45, \allowbreak 108, \allowbreak 0, \allowbreak 27, \allowbreak 0, \allowbreak 45, \allowbreak -216, \allowbreak 108, \allowbreak 0, \allowbreak 0, \allowbreak 0, \allowbreak 0, \allowbreak 0\} $\\ \hline
	\end{Tabular}
    \caption{Coefficients $ C_i^{(k_1,k_2)} $ in the modular ansatz of rank 1 $ E_8 \times E_8 $ heterotic LST, written in terms of $ E_8 $ Jacobi forms and $ \lambda=1 $.} \label{table:E8E8-modular}
\end{table}

At $ (k_1, k_2)=(1,0) $ and $ (2,0) $, there are 1 and 49 constants, repectively, and they can be fixed by comparing the ansatz with the E-string elliptic genera as done in \cite{Gu:2017ccq}. At $ (k_1,k_2)=(1,1) $, the comparison does not yield any constraints due to the presence of decoupled states in the elliptic genus that are not part of the LST.  However, the requirement from the GV-invariant form given in \eqref{GV} still impose additional constraints. This allows us to fix all coefficients for $ \lambda=1 $ and 23 coefficients among 37 for $ \lambda=2 $. At $ (k_1,k_2)=(2,1) $, there are 130 coefficients for $ \lambda=1 $ and 831 coefficients for $\lambda=2$, and all of these coefficients are determined by comparison with the LST spectrum computed from the ADHM computation.  

It is worth noting that the spectra of the decoupled states, which do not carry dynamical tensor charge, differ between the ADHM result and the results from the two values of $ \lambda $. We also note that 14 coefficients at $ (1,1) $-string order for $ \lambda=2 $ are not fixed and they appear in the coefficients in $ (2,1) $-string ansatz, as shown in Table~\ref{table:E8E8-modular2}. However, these coefficients do not affect the BPS spectrum for states carrying non-zero dynamical tensor charge.

\begin{table}
	\centering
	\begin{Tabular}{|c|L{70ex}|} \hline
        $ (k_1, k_2) $ & \multicolumn{1}{c|}{$ \big\{C_i^{(k_1, k_2)} \big\} $} \\ \hline \hline
        $ (1, 1) $ & \tiny $ \frac{1}{2^{12} \cdot 3^7} \{-1, \allowbreak 8957952 c_2, \allowbreak 1-8957952 c_2, \allowbreak -2, \allowbreak 8957952 c_5, \allowbreak -16-8957952 c_5, \allowbreak 8957952 c_7, \allowbreak -8957952 c_7, \allowbreak 8957952 c_9, \allowbreak 16-8957952 c_9, \allowbreak 2, \allowbreak 8957952 c_{12}, \allowbreak 32-8957952 c_{12}, \allowbreak -32, \allowbreak 3, \allowbreak 8957952 c_{16}, \allowbreak 6-8957952 c_{16}, \allowbreak 8957952 c_{18}, \allowbreak -6-8957952 c_{18}, \allowbreak 8957952 c_{20}, \allowbreak -3-8957952 c_{20}, \allowbreak 8957952 c_{22}, \allowbreak -12-8957952 c_{22}, \allowbreak 8957952 c_{24}, \allowbreak -8957952 c_{24}, \allowbreak 12, \allowbreak 8957952 c_{27}, \allowbreak 9-8957952 c_{27}, \allowbreak 8957952 c_{29}, \allowbreak 18-8957952 c_{29}, \allowbreak 8957952 c_{31}, \allowbreak -18-8957952 c_{31}, \allowbreak -9, \allowbreak 0, \allowbreak 8957952 c_{35}, \allowbreak -27-8957952 c_{35}, \allowbreak 27\} $ \\ \hline
        $ (2, 1) $ & \scalebox{0.4}{\parbox{175ex}{\raggedright $ \frac{1}{2^{25}\cdot 3^{13}} \{0, \allowbreak -4, \allowbreak 71663616 c_2, \allowbreak 4-71663616 c_2, \allowbreak 0, \allowbreak 0, \allowbreak -8, \allowbreak 3, \allowbreak 0, \allowbreak 71663616 c_5, \allowbreak 0, \allowbreak -15, \allowbreak 0, \allowbreak 0, \allowbreak 16, \allowbreak 143327232 c_2, \allowbreak -3, \allowbreak 0, \allowbreak -64-71663616 c_5, \allowbreak -3, \allowbreak 10, \allowbreak 71663616 c_7, \allowbreak 15, \allowbreak -48, \allowbreak -16+1146617856 c_2, \allowbreak 3, \allowbreak 5, \allowbreak -71663616 c_7, \allowbreak -10, \allowbreak -112+71663616 c_9, \allowbreak 48, \allowbreak -5, \allowbreak 48-71663616 c_9, \allowbreak 128-1146617856 c_2, \allowbreak 8-143327232 c_2, \allowbreak 6, \allowbreak 0, \allowbreak -30, \allowbreak 0, \allowbreak 0, \allowbreak 32, \allowbreak -12, \allowbreak 0, \allowbreak 143327232 c_5, \allowbreak -6, \allowbreak 48, \allowbreak 20, \allowbreak 0, \allowbreak -240, \allowbreak -96, \allowbreak -24, \allowbreak 0, \allowbreak 256+1146617856 c_5, \allowbreak 12, \allowbreak -40, \allowbreak 143327232 c_7, \allowbreak -48, \allowbreak 10, \allowbreak 0, \allowbreak 160, \allowbreak -224+71663616 c_{12}, \allowbreak 36, \allowbreak 0, \allowbreak -768, \allowbreak 24, \allowbreak 40, \allowbreak 1146617856 c_7, \allowbreak 40, \allowbreak -128+143327232 c_9, \allowbreak 0, \allowbreak 80, \allowbreak 96-71663616 c_{12}, \allowbreak -48, \allowbreak 0, \allowbreak -2048-1146617856 c_5, \allowbreak -36, \allowbreak 240, \allowbreak 0, \allowbreak -40, \allowbreak 1146617856 c_9, \allowbreak 160-2293235712 c_2, \allowbreak 48, \allowbreak 0, \allowbreak -128-143327232 c_5, \allowbreak -6, \allowbreak -160, \allowbreak -1146617856 c_7, \allowbreak 30, \allowbreak 768, \allowbreak -32+2293235712 c_2, \allowbreak 6, \allowbreak -80, \allowbreak -143327232 c_7, \allowbreak -20, \allowbreak 1792-1146617856 c_9, \allowbreak 96, \allowbreak -10, \allowbreak 256-143327232 c_9, \allowbreak -32, \allowbreak -24, \allowbreak -48, \allowbreak -192, \allowbreak -80, \allowbreak 72, \allowbreak -384, \allowbreak 80, \allowbreak 0, \allowbreak -640, \allowbreak -256+143327232 c_{12}, \allowbreak -96, \allowbreak 576, \allowbreak 480, \allowbreak 0, \allowbreak 640, \allowbreak 1146617856 c_{12}, \allowbreak 192, \allowbreak 0, \allowbreak -2560-2293235712 c_5, \allowbreak 96, \allowbreak 96, \allowbreak -320, \allowbreak 0, \allowbreak -480, \allowbreak 1536, \allowbreak 384, \allowbreak 0, \allowbreak 512+2293235712 c_5, \allowbreak 24, \allowbreak 640, \allowbreak -2293235712 c_7, \allowbreak -96, \allowbreak -160, \allowbreak 0, \allowbreak 320, \allowbreak 3584-1146617856 c_{12}, \allowbreak -576, \allowbreak 0, \allowbreak -1536, \allowbreak 48, \allowbreak -640, \allowbreak 2293235712 c_7, \allowbreak 80, \allowbreak 2048-2293235712 c_9, \allowbreak 0, \allowbreak 160, \allowbreak 512-143327232 c_{12}, \allowbreak 0, \allowbreak 0, \allowbreak 512, \allowbreak -72, \allowbreak 0, \allowbreak 0, \allowbreak -80, \allowbreak -4096+2293235712 c_9, \allowbreak -256, \allowbreak 384, \allowbreak 768, \allowbreak -384, \allowbreak 1280, \allowbreak -1152, \allowbreak -768, \allowbreak -1280, \allowbreak 0, \allowbreak -1280, \allowbreak 4096-2293235712 c_{12}, \allowbreak 0, \allowbreak 1152, \allowbreak 0, \allowbreak 0, \allowbreak 1280, \allowbreak -8192+2293235712 c_{12}, \allowbreak 0, \allowbreak 0, \allowbreak 4096, \allowbreak 0, \allowbreak 0, \allowbreak 0, \allowbreak 12, \allowbreak 0, \allowbreak 5, \allowbreak 71663616 c_{16}, \allowbreak 0, \allowbreak 0, \allowbreak -214990848 c_2, \allowbreak 0, \allowbreak -5, \allowbreak 24-71663616 c_{16}, \allowbreak -5, \allowbreak 40+71663616 c_{18}, \allowbreak -429981696 c_2, \allowbreak 5, \allowbreak -4-71663616 c_{18}, \allowbreak -40+429981696 c_2+71663616 c_{20}, \allowbreak -32+214990848 c_2-71663616 c_{20}, \allowbreak 0, \allowbreak -9, \allowbreak 10, \allowbreak 0, \allowbreak 45, \allowbreak 0, \allowbreak 9, \allowbreak 0, \allowbreak -48-214990848 c_5, \allowbreak 0, \allowbreak -5, \allowbreak 143327232 c_{16}, \allowbreak 9, \allowbreak -10, \allowbreak -18, \allowbreak 50, \allowbreak 80+71663616 c_{22}, \allowbreak 0, \allowbreak 90, \allowbreak 144-429981696 c_5, \allowbreak -9, \allowbreak -10, \allowbreak -96-214990848 c_7+1146617856 c_{16}, \allowbreak 5, \allowbreak 32+143327232 c_{18}, \allowbreak 18, \allowbreak -95, \allowbreak -8-71663616 c_{22}, \allowbreak 9, \allowbreak -60, \allowbreak 976+429981696 c_5+71663616 c_{24}, \allowbreak 0, \allowbreak -75, \allowbreak 288-429981696 c_7, \allowbreak 10, \allowbreak 64-214990848 c_9+1146617856 c_{18}, \allowbreak -32+859963392 c_2+143327232 c_{20}, \allowbreak -18, \allowbreak -30, \allowbreak 368+214990848 c_5-71663616 c_{24}, \allowbreak 18, \allowbreak -20, \allowbreak 768+429981696 c_7-1146617856 c_{16}, \allowbreak -60, \allowbreak -384-429981696 c_9, \allowbreak 32+1146617856 c_{20}, \allowbreak -9, \allowbreak 110, \allowbreak 48+214990848 c_7-143327232 c_{16}, \allowbreak 20, \allowbreak -1312+429981696 c_9-1146617856 c_{18}, \allowbreak -144-859963392 c_2, \allowbreak 25, \allowbreak -416+214990848 c_9-143327232 c_{18}, \allowbreak -320-1146617856 c_{20}, \allowbreak -40-143327232 c_{20}, \allowbreak 18, \allowbreak 36, \allowbreak -10, \allowbreak 0, \allowbreak 216, \allowbreak -20, \allowbreak 72, \allowbreak 40, \allowbreak 64+143327232 c_{22}, \allowbreak 18, \allowbreak -108, \allowbreak -150, \allowbreak 144, \allowbreak -280, \allowbreak 128-214990848 c_{12}+1146617856 c_{22}, \allowbreak -72, \allowbreak 240, \allowbreak 896+859963392 c_5+143327232 c_{24}, \allowbreak -36, \allowbreak -144, \allowbreak -40, \allowbreak -216, \allowbreak 240, \allowbreak -768-429981696 c_{12}, \allowbreak -288, \allowbreak -240, \allowbreak -512+1146617856 c_{24}, \allowbreak -36, \allowbreak -160, \allowbreak 960+859963392 c_7-2293235712 c_{16}, \allowbreak 0, \allowbreak 220, \allowbreak -36, \allowbreak 160, \allowbreak -2624+429981696 c_{12}-1146617856 c_{22}, \allowbreak 216, \allowbreak 180, \allowbreak -859963392 c_5, \allowbreak -90, \allowbreak 400, \allowbreak -192+2293235712 c_{16}, \allowbreak -110, \allowbreak -1280+859963392 c_9-2293235712 c_{18}, \allowbreak 36, \allowbreak -160, \allowbreak -832+214990848 c_{12}-143327232 c_{22}, \allowbreak 144, \allowbreak -120, \allowbreak -256-1146617856 c_{24}, \allowbreak 108, \allowbreak -240, \allowbreak 576-859963392 c_7, \allowbreak 140, \allowbreak 2048+2293235712 c_{18}, \allowbreak -256-2293235712 c_{20}, \allowbreak 0, \allowbreak -60, \allowbreak -128-143327232 c_{24}, \allowbreak 18, \allowbreak 0, \allowbreak -192, \allowbreak -30, \allowbreak 1536-859963392 c_9, \allowbreak 832+2293235712 c_{20}, \allowbreak -144, \allowbreak -576, \allowbreak 0, \allowbreak -320, \allowbreak 432, \allowbreak 288, \allowbreak 800, \allowbreak 144, \allowbreak -160, \allowbreak -2560+859963392 c_{12}-2293235712 c_{22}, \allowbreak 288, \allowbreak 0, \allowbreak -480, \allowbreak 288, \allowbreak -320, \allowbreak 4096+2293235712 c_{22}, \allowbreak 0, \allowbreak 480, \allowbreak -2048-2293235712 c_{24}, \allowbreak 0, \allowbreak -288, \allowbreak 0, \allowbreak -432, \allowbreak 480, \allowbreak 3072-859963392 c_{12}, \allowbreak 0, \allowbreak -480, \allowbreak -1024+2293235712 c_{24}, \allowbreak 0, \allowbreak 0, \allowbreak -1536, \allowbreak 0, \allowbreak 0, \allowbreak 0, \allowbreak 0, \allowbreak 0, \allowbreak 0, \allowbreak 0, \allowbreak -15, \allowbreak 71663616 c_{27}, \allowbreak -9, \allowbreak 0, \allowbreak 0, \allowbreak 0, \allowbreak 15, \allowbreak -214990848 c_{16}, \allowbreak 0, \allowbreak -12, \allowbreak 0, \allowbreak 15, \allowbreak 36-71663616 c_{27}, \allowbreak 6, \allowbreak -30, \allowbreak -120+71663616 c_{29}, \allowbreak 9, \allowbreak 0, \allowbreak -429981696 c_{16}, \allowbreak -15, \allowbreak -24-214990848 c_{18}, \allowbreak 12-644972544 c_2, \allowbreak 3, \allowbreak 30, \allowbreak 12-71663616 c_{29}, \allowbreak -6, \allowbreak 15, \allowbreak -240+429981696 c_{16}+71663616 c_{31}, \allowbreak 0, \allowbreak 144-429981696 c_{18}, \allowbreak 24-1289945088 c_2-214990848 c_{20}, \allowbreak -3, \allowbreak -30, \allowbreak -192+214990848 c_{16}-71663616 c_{31}, \allowbreak 30, \allowbreak 240+429981696 c_{18}, \allowbreak -144+1289945088 c_2-429981696 c_{20}, \allowbreak -15, \allowbreak 156+214990848 c_{18}, \allowbreak 48+644972544 c_2+429981696 c_{20}, \allowbreak 60+214990848 c_{20}, \allowbreak -18, \allowbreak -27, \allowbreak 30, \allowbreak 0, \allowbreak 15, \allowbreak -24+143327232 c_{27}, \allowbreak -15, \allowbreak -144, \allowbreak 135, \allowbreak -54, \allowbreak 270, \allowbreak -192-214990848 c_{22}+1146617856 c_{27}, \allowbreak 3, \allowbreak 30, \allowbreak -288-644972544 c_5+143327232 c_{29}, \allowbreak 33, \allowbreak 69, \allowbreak -60, \allowbreak 0, \allowbreak 225, \allowbreak 720-429981696 c_{22}, \allowbreak 51, \allowbreak 60, \allowbreak -576-1289945088 c_5-214990848 c_{24}+1146617856 c_{29}, \allowbreak -3, \allowbreak -30, \allowbreak -192-644972544 c_7+859963392 c_{16}+143327232 c_{31}, \allowbreak 102, \allowbreak -105, \allowbreak 108, \allowbreak -420, \allowbreak 1632+429981696 c_{22}-1146617856 c_{27}, \allowbreak 144, \allowbreak -360, \allowbreak 3456+1289945088 c_5-429981696 c_{24}, \allowbreak 30, \allowbreak -300, \allowbreak 192-1289945088 c_7+1146617856 c_{31}, \allowbreak -15, \allowbreak 240-644972544 c_9+859963392 c_{18}, \allowbreak -54, \allowbreak -90, \allowbreak 384+214990848 c_{22}-143327232 c_{27}, \allowbreak -123, \allowbreak 120, \allowbreak 3168+644972544 c_5+429981696 c_{24}-1146617856 c_{29}, \allowbreak 18, \allowbreak -45, \allowbreak -864+1289945088 c_7-859963392 c_{16}, \allowbreak -60, \allowbreak -816-1289945088 c_9, \allowbreak -24+859963392 c_{20}, \allowbreak -75, \allowbreak 150, \allowbreak 288+214990848 c_{24}-143327232 c_{29}, \allowbreak -39, \allowbreak 330, \allowbreak -1920+644972544 c_7-1146617856 c_{31}, \allowbreak 45, \allowbreak -3312+1289945088 c_9-859963392 c_{18}, \allowbreak -336, \allowbreak -6, \allowbreak 45, \allowbreak -240-143327232 c_{31}, \allowbreak 30, \allowbreak -1008+644972544 c_9, \allowbreak -288-859963392 c_{20}, \allowbreak 114, \allowbreak 318, \allowbreak 264, \allowbreak 300, \allowbreak -36, \allowbreak 384, \allowbreak -960, \allowbreak -216, \allowbreak 720, \allowbreak 1920-644972544 c_{12}+859963392 c_{22}-2293235712 c_{27}, \allowbreak -300, \allowbreak -936, \allowbreak 180, \allowbreak -540, \allowbreak -240, \allowbreak -1920-1289945088 c_{12}+2293235712 c_{27}, \allowbreak -156, \allowbreak -660, \allowbreak 4224+859963392 c_{24}-2293235712 c_{29}, \allowbreak -96, \allowbreak 192, \allowbreak 480, \allowbreak 648, \allowbreak 0, \allowbreak -5760+1289945088 c_{12}-859963392 c_{22}, \allowbreak -168, \allowbreak 840, \allowbreak -768+2293235712 c_{29}, \allowbreak -6, \allowbreak -360, \allowbreak -1536-2293235712 c_{31}, \allowbreak 96, \allowbreak 0, \allowbreak 108, \allowbreak -480, \allowbreak -2304+644972544 c_{12}, \allowbreak 324, \allowbreak -180, \allowbreak -859963392 c_{24}, \allowbreak 6, \allowbreak 360, \allowbreak 4992+2293235712 c_{31}, \allowbreak 0, \allowbreak 1152, \allowbreak -96, \allowbreak 96, \allowbreak 96, \allowbreak 0, \allowbreak 0, \allowbreak -96, \allowbreak 0, \allowbreak 0, \allowbreak 0, \allowbreak 0, \allowbreak 0, \allowbreak 27, \allowbreak 0, \allowbreak 0, \allowbreak -45, \allowbreak -214990848 c_{27}, \allowbreak 9, \allowbreak 0, \allowbreak 36, \allowbreak 0, \allowbreak -18, \allowbreak -45, \allowbreak 54, \allowbreak 0, \allowbreak -429981696 c_{27}, \allowbreak 18, \allowbreak -90, \allowbreak 72-214990848 c_{29}, \allowbreak -9, \allowbreak 45, \allowbreak 72-644972544 c_{16}, \allowbreak -9, \allowbreak 45, \allowbreak -36, \allowbreak -45, \allowbreak -360+429981696 c_{27}+71663616 c_{35}, \allowbreak -81, \allowbreak 0, \allowbreak -432-429981696 c_{29}, \allowbreak -18, \allowbreak 45, \allowbreak 144-1289945088 c_{16}-214990848 c_{31}, \allowbreak -45, \allowbreak 36-644972544 c_{18}, \allowbreak -18, \allowbreak 90, \allowbreak -288+214990848 c_{27}-71663616 c_{35}, \allowbreak 36, \allowbreak 180, \allowbreak -720+429981696 c_{29}, \allowbreak 0, \allowbreak 0, \allowbreak -864+1289945088 c_{16}-429981696 c_{31}, \allowbreak 90, \allowbreak -144-1289945088 c_{18}, \allowbreak -144+1934917632 c_2-644972544 c_{20}, \allowbreak 18, \allowbreak -90, \allowbreak -468+214990848 c_{29}, \allowbreak 18, \allowbreak -45, \allowbreak 288+644972544 c_{16}+429981696 c_{31}, \allowbreak 0, \allowbreak 1296+1289945088 c_{18}, \allowbreak 144-1934917632 c_2-1289945088 c_{20}, \allowbreak 9, \allowbreak -45, \allowbreak 360+214990848 c_{31}, \allowbreak -45, \allowbreak 360+644972544 c_{18}, \allowbreak 432+1289945088 c_{20}, \allowbreak 0, \allowbreak -36, \allowbreak 216+644972544 c_{20}, \allowbreak 18, \allowbreak -45, \allowbreak 135, \allowbreak 135, \allowbreak -162, \allowbreak 135, \allowbreak 180, \allowbreak -18, \allowbreak 810, \allowbreak -216-644972544 c_{22}+859963392 c_{27}+143327232 c_{35}, \allowbreak 234, \allowbreak -432, \allowbreak -540, \allowbreak 180, \allowbreak -540, \allowbreak -1289945088 c_{22}+1146617856 c_{35}, \allowbreak -126, \allowbreak -90, \allowbreak 1584+1934917632 c_5-644972544 c_{24}+859963392 c_{29}, \allowbreak -45, \allowbreak 81, \allowbreak 90, \allowbreak 108, \allowbreak 135, \allowbreak 1296+1289945088 c_{22}-859963392 c_{27}, \allowbreak -225, \allowbreak -360, \allowbreak 144-1934917632 c_5-1289945088 c_{24}, \allowbreak -9, \allowbreak -765, \allowbreak -144+1934917632 c_7+859963392 c_{31}, \allowbreak 81, \allowbreak 135, \allowbreak -234, \allowbreak -270, \allowbreak -2160+644972544 c_{22}-1146617856 c_{35}, \allowbreak 432, \allowbreak 270, \allowbreak 3024+1289945088 c_{24}-859963392 c_{29}, \allowbreak -45, \allowbreak 630, \allowbreak -2016-1934917632 c_7, \allowbreak -90, \allowbreak -2304+1934917632 c_9, \allowbreak -36, \allowbreak -135, \allowbreak -432-143327232 c_{35}, \allowbreak -81, \allowbreak 180, \allowbreak -432+644972544 c_{24}, \allowbreak 54, \allowbreak 135, \allowbreak -1728-859963392 c_{31}, \allowbreak 90, \allowbreak 3168-1934917632 c_9, \allowbreak 216, \allowbreak -576, \allowbreak -1260, \allowbreak 612, \allowbreak -2520, \allowbreak 1836, \allowbreak 1224, \allowbreak 2520, \allowbreak -36, \allowbreak 2520, \allowbreak -6912+1934917632 c_{12}-2293235712 c_{35}, \allowbreak 0, \allowbreak -1836, \allowbreak 0, \allowbreak 36, \allowbreak -2520, \allowbreak 13824-1934917632 c_{12}+2293235712 c_{35}, \allowbreak 0, \allowbreak 0, \allowbreak -6912, \allowbreak 0, \allowbreak 0, \allowbreak -27, \allowbreak 0, \allowbreak 81, \allowbreak -54, \allowbreak -135, \allowbreak -54, \allowbreak -135, \allowbreak 108-644972544 c_{27}, \allowbreak -54, \allowbreak 243, \allowbreak 0, \allowbreak -108, \allowbreak -135, \allowbreak 216-1289945088 c_{27}-214990848 c_{35}, \allowbreak 54, \allowbreak -270, \allowbreak -108-644972544 c_{29}, \allowbreak -27, \allowbreak -108, \allowbreak 270, \allowbreak 0, \allowbreak 0, \allowbreak -1296+1289945088 c_{27}-429981696 c_{35}, \allowbreak 108, \allowbreak 540, \allowbreak 432-1289945088 c_{29}, \allowbreak 27, \allowbreak 270, \allowbreak -864+1934917632 c_{16}-644972544 c_{31}, \allowbreak -54, \allowbreak -135, \allowbreak 108, \allowbreak 135, \allowbreak 432+644972544 c_{27}+429981696 c_{35}, \allowbreak -243, \allowbreak 0, \allowbreak -3888+1289945088 c_{29}, \allowbreak 54, \allowbreak -135, \allowbreak 864-1934917632 c_{16}-1289945088 c_{31}, \allowbreak 135, \allowbreak 1188+1934917632 c_{18}, \allowbreak 54, \allowbreak 135, \allowbreak 540+214990848 c_{35}, \allowbreak 54, \allowbreak -270, \allowbreak -1080+644972544 c_{29}, \allowbreak -81, \allowbreak 0, \allowbreak 2592+1289945088 c_{31}, \allowbreak -135, \allowbreak -216-1934917632 c_{18}, \allowbreak 324+1934917632 c_{20}, \allowbreak 27, \allowbreak 0, \allowbreak 108, \allowbreak 0, \allowbreak -135, \allowbreak 1296+644972544 c_{31}, \allowbreak 0, \allowbreak 0, \allowbreak -648-1934917632 c_{20}, \allowbreak 81, \allowbreak 486, \allowbreak -54, \allowbreak 405, \allowbreak -324, \allowbreak -189, \allowbreak -810, \allowbreak -54, \allowbreak 135, \allowbreak 2160+1934917632 c_{22}+859963392 c_{35}, \allowbreak -243, \allowbreak 0, \allowbreak 405, \allowbreak -270, \allowbreak 270, \allowbreak -3456-1934917632 c_{22}, \allowbreak 27, \allowbreak -540, \allowbreak 1728+1934917632 c_{24}, \allowbreak 0, \allowbreak 243, \allowbreak 0, \allowbreak 324, \allowbreak -405, \allowbreak -2592-859963392 c_{35}, \allowbreak -27, \allowbreak 540, \allowbreak 864-1934917632 c_{24}, \allowbreak 0, \allowbreak 0, \allowbreak 1296, \allowbreak 0, \allowbreak 0, \allowbreak 0, \allowbreak 0, \allowbreak -81, \allowbreak -162, \allowbreak -162, \allowbreak -405, \allowbreak 0, \allowbreak -324, \allowbreak 810, \allowbreak 162, \allowbreak -810, \allowbreak -1296+1934917632 c_{27}-644972544 c_{35}, \allowbreak 162, \allowbreak 729, \allowbreak 0, \allowbreak 324, \allowbreak 405, \allowbreak 1296-1934917632 c_{27}-1289945088 c_{35}, \allowbreak 81, \allowbreak 810, \allowbreak -3564+1934917632 c_{29}, \allowbreak 81, \allowbreak -162, \allowbreak -405, \allowbreak -486, \allowbreak 0, \allowbreak 3888+1289945088 c_{35}, \allowbreak 162, \allowbreak -810, \allowbreak 648-1934917632 c_{29}, \allowbreak 0, \allowbreak 405, \allowbreak 1944+1934917632 c_{31}, \allowbreak -81, \allowbreak 0, \allowbreak 0, \allowbreak 405, \allowbreak 1944+644972544 c_{35}, \allowbreak -243, \allowbreak 0, \allowbreak 0, \allowbreak 0, \allowbreak -405, \allowbreak -3888-1934917632 c_{31}, \allowbreak 0, \allowbreak -972, \allowbreak 81, \allowbreak -81, \allowbreak -81, \allowbreak 0, \allowbreak 0, \allowbreak 81, \allowbreak 0, \allowbreak 0, \allowbreak 0, \allowbreak 0, \allowbreak 243, \allowbreak 486, \allowbreak -243, \allowbreak 1215, \allowbreak -729, \allowbreak -486, \allowbreak -1215, \allowbreak 0, \allowbreak -12 15, \allowbreak 2916+1934917632 c_{35}, \allowbreak 0, \allowbreak 729, \allowbreak 0, \allowbreak 0, \allowbreak 1215, \allowbreak -5832-1934917632 c_{35}, \allowbreak 0, \allowbreak 0, \allowbreak 2916, \allowbreak 0, \allowbreak 0, \allowbreak 0, \allowbreak 0, \allowbreak 0\} $}} \\ \hline
	\end{Tabular}
    \caption{Coefficients $ C_i^{(k_1,k_2)} $ in the modular ansatz of rank 1 $ E_8 \times E_8 $ heterotic LST, written in terms of $ E_8 $ Jacobi forms and $ \lambda=2 $. There are 14 undetermined conatants $ c_i \equiv C_i^{(1,1)} $ but they do not related with the LST spectrum. $ (k_1,k_2)=(1,0) $ and $ (2,0) $ cases are same with Table~\ref{table:E8E8-modular}.} \label{table:E8E8-modular2}
\end{table}

\paragraph{Blowup equation}
The tree-level and one-loop contributions to the effective prepotential of the $E_8\times E_8$ LST are identical to those for the E-string theory, but there are additional contributions from the auxiliary 2-form  field. Collecting these contributions yields the full effective prepotential which is given by
\begin{align}\label{E8E8-E}
    \mathcal{E} &= \frac{1}{\epsilon_1 \epsilon_2} \qty( \frac{\tau}{2}(\phi_{1,0}-\phi_{2,0})^2 + \qty( \frac{\epsilon_1^2 + \epsilon_2^2}{4} - \frac{1}{2}\sum_{i=1}^{8} m_i^2 + \epsilon_+^2 ) (\phi_{1,0}-\phi_{2,0}) ) + \mathcal{E}_{\mathrm{tree}}^{(0)} \, , \nonumber \\
    \mathcal{E}_{\mathrm{tree}}^{(0)}
    &= \frac{1}{\epsilon_1 \epsilon_2} \qty( \frac{\epsilon_1^2 + \epsilon_2^2}{2} - \frac{1}{2} \sum_{i=1}^{16} m_i^2 + 2\epsilon_+^2 ) \phi_{0,0} \, ,
\end{align}
with an auxiliary scalar VEV $ \phi_{0,0} $.

For a blowup equation, we consider a set of consistent magnetic fluxes for the dynamical tensor and the auxiliary 2-form field such that
\begin{align}
	\phi_{1,0} - \phi_{2,0}\ \rightarrow  \ \phi_{1,0} - \phi_{2,0}+n_{1,0}\epsilon_{1,2} \ , \quad \phi_{0,0} \ \rightarrow \  \phi_{0,0} +n_{0,0}\epsilon_{1,2} \ ,
\end{align}
where the fluxes are quantized as
\begin{align}
    n_1  \equiv n_{1,0}+n_{0,0} \in \mathbb{Z} + 1/2 \, , \quad n_2 \equiv n_{0,0} \in \mathbb{Z} \, .
\end{align}
We also choose background magnetic flxlues for the masses, winding number and KK momentum as 
\begin{align}
	B_{m_i} = \left\{
        \begin{array}{ll}
            1/2 & \quad (0 \leq i \leq 8) \\
            -1/2 & \quad (9 \leq i \leq 16)
    \end{array}\right. \, , \quad
    B_\tau = B_w = 0 \, .
\end{align}
We propose that the partition function of the $E_8\times E_8$ LST satisfies the blowup equation given by
\begin{align}\label{eq:HE-blowup}
    \Lambda \hat{Z}^{\mathrm{HE}}_{\mathrm{str}} = \sum_{n_1, n_2} (-1)^{n_1+n_2} q^{\frac{1}{2}(n_1-n_2)^2} e^{2\pi i (-\frac{n_1}{2} \sum_{i=1}^8 m_i + \frac{n_2}{2}\sum_{i=9}^{16} m_i - 2(n_1 + n_2)\epsilon_+)} \hat{Z}^{\mathrm{HE}(N)}_{\mathrm{str}}\hat{Z}^{\mathrm{HE}(S)}_{\mathrm{str}} \, ,
\end{align}
where $ \Lambda = \Lambda(w, \tau, m_i) $. We checked the the elliptic genera computed from the 2d gauge theory description for the strings satisfy this blowup equation, with $ m_1=\cdots=m_8 $ and $ m_9=\cdots=m_{16} $, up to $ (k_1, k_2)=(2, 1) $ order and to second order in the $ q $-expansion.

The elliptic genera can also be computed  by solving the blowup equation with a modular ansatz written in terms of $ E_8 $ Jacobi forms in the following way. The $ (1,0) $- and $ (2,0) $-string elliptic genera have already been calculated in this way for the 6d E-strings in previous work \cite{Gu:2019pqj, Kim:2020hhh}. At the $ (1,1) $- and $(2,1)$-string order, the modular ansatzes are constrained by the requirement of the GV-invariant expression in \eqref{GV}, which we use to fix several coefficients in the ansatzes. Finally, we solve the blowup equation, which completely fixes all the coefficients in $ (2,1) $-string modular ansatz for both $ \lambda=1 $ and $ 2 $. These results are in agreement with those presented in Table~\ref{table:E8E8-modular} and \ref{table:E8E8-modular2}. We expect that the elliptic genera at higher orders can be calculated using the blowup equation in a similar manner.

\subsubsection{\texorpdfstring{$ SO(32) $}{SO(32)} picture} \label{subsubsec:SO32LST}

The $ SO(32) $ heterotic LST is the worldvolume theory on $N$ NS5-branes in the $SO(32)$ heterotic string theory. At low energies, it is described by an $ Sp(N) $ gauge theory with 16 fundamental hypermultiplets. In the F-theory construction \cite{Bhardwaj:2015oru}, this theory is engineered by a rational curve $ \Sigma $ in the base surface with $ \Sigma^2=0 $. It is also T-dual to the $ E_8 \times E_8 $ heterotic LST upon compactification on a circle.

  The partition function when $N=1$ can be written as
\begin{align}
Z^{\mathrm{HO}}_{\mathrm{GV}} = Z_{\mathrm{pert}}^{\mathrm{HO}} \cdot Z_{\mathrm{str}}^{\mathrm{HO}} = Z_{\mathrm{pert}}^{\mathrm{HO}} \cdot \sum_{k=0}^\infty e^{2\pi i k w} Z_k^{\mathrm{HO}} \, ,
\end{align}
where $ w \sim 1/g_{\mathrm{YM}}^2 $ is intrepreted as the inverse gauge coupling and $k$ is the little string number.
The 1-loop contributions coming from the $ Sp(1) $ vector and the fundamental hypermultiplets are
\begin{align}\label{so(32)-pert}
\begin{aligned}
Z_{\mathrm{pert}}^{\mathrm{HO}}
= \PE\bigg[& -\frac{1+p_1p_2}{(1-p_1)(1-p_2)} \qty(Q^2 + q Q^{-2}) \frac{1}{1-q} \\
& + \frac{\sqrt{p_1 p_2}}{(1-p_1)(1-p_2)} \qty(Q + q Q^{-1}) \sum_{l=1}^{16} \qty(e^{2\pi i m_l} + e^{-2\pi i m_l}) \frac{1}{1-q} \bigg],
\end{aligned}
\end{align}
where $ m_l $ are the chemical potentials for the $ SO(32) $ flavor symmetry and $ Q = e^{2\pi i \phi_1} $ is the $ Sp(1) $ fugacity.

\paragraph{GLSM}

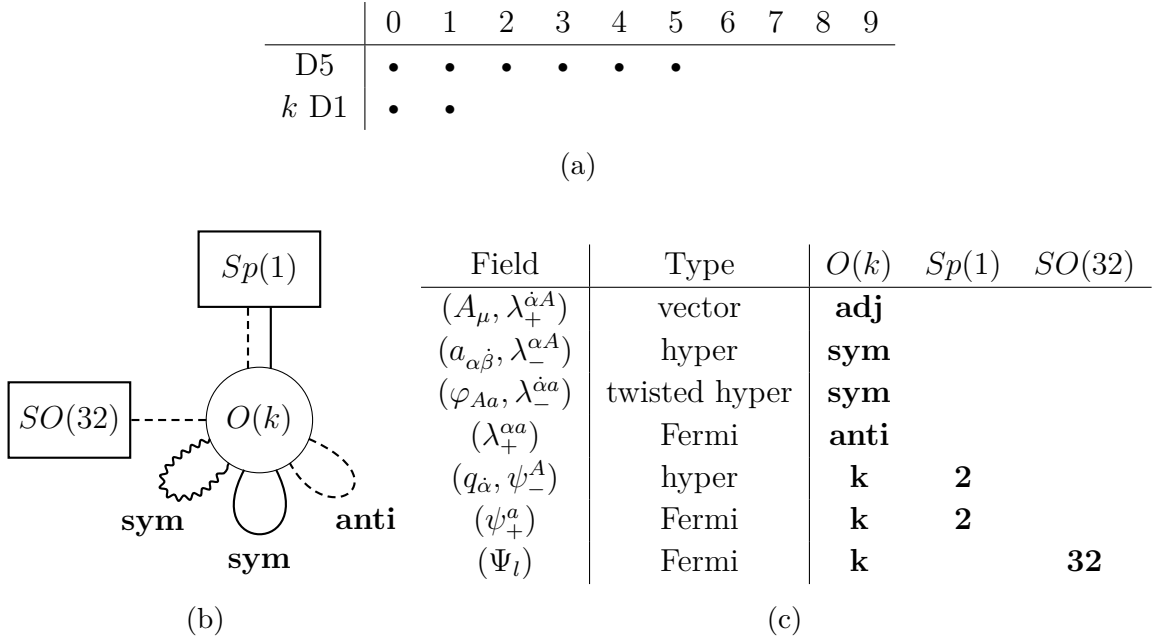
\begin{figure}
	\centering
    \begin{subfigure}[b]{1\textwidth}
        \centering
        \begin{tabular}{c|cccccccccc}
            & 0 & 1 & 2 & 3 & 4 & 5 & 6 & 7 & 8 & 9 \\ \hline
            D5 & \textbullet & \textbullet & \textbullet & \textbullet & \textbullet & \textbullet \\
            $k$ D1 & \textbullet & \textbullet
        \end{tabular}
        \subcaption{}
    \end{subfigure}
    \phantom{a}

	\begin{subfigure}[b]{0.35\textwidth}
		\centering
		\begin{tikzpicture}
		\draw (0, 0) node {$ O(k) $};
		\draw (0, 0) circle (0.7);
		\draw [thick, densely dashed] (-0.7, 0) -- (-1.7, 0);
		\draw (-2.5, 0) node {$ SO(32) $};
		\draw [thick] (-3.3, 0.5) rectangle (-1.7, -0.5);
		\draw (0, 2) node {$ Sp(1) $};
		\draw [thick] (-0.8, 2.5) rectangle (0.8, 1.5);
		\draw [thick, densely dashed] (-0.15, 0.68) -- (-0.15, 1.5);
		\draw [thick] (0.15, 0.68) -- (0.15, 1.5);
		\draw [thick] (-0.2, -0.66) .. controls (-0.8, -1.8) and (0.8, -1.8) .. (0.22, -0.66);
		\draw [thick, decorate, decoration={snake, segment length=4pt, amplitude=1pt}] (-0.65, -0.25) .. controls (-1.9, -0.8) and (-0.9, -1.5) .. (-0.4, -0.57);
		\draw [thick, densely dashed] (0.65, -0.25) .. controls (1.9, -0.8) and (0.9, -1.5) .. (0.4, -0.57);
		\draw (0, -1.9) node {\textbf{sym}};
		\draw (-1.4, -1.4) node {\textbf{sym}};
		\draw (1.4, -1.3) node {\textbf{anti}};
		\end{tikzpicture}
		\subcaption{}
	\end{subfigure}
	\begin{subfigure}[b]{0.64\textwidth}
		\centering
		\begin{tabular}{c|c|ccc}
			Field & Type & $ O(k) $ & $ Sp(1) $ & $ SO(32) $ \\ \hline
			$ (A_\mu, \lambda_+^{\dot{\alpha}A}) $ & vector & $ \mathbf{adj} $ \\
			$ (a_{\alpha\dot{\beta}}, \lambda_-^{\alpha A}) $ & hyper & $ \mathbf{sym} $ \\
			$ (\varphi_{Aa}, \lambda_-^{\dot{\alpha}a}) $ & twisted hyper & $ \mathbf{sym} $ \\
			$ (\lambda_+^{\alpha a}) $ & Fermi & $ \mathbf{anti} $ \\
			$ (q_{\dot{\alpha}}, \psi_-^A) $ & hyper & $ \mathbf{k} $ & $ \mathbf{2} $ \\
			$ (\psi_+^a) $ & Fermi & $ \mathbf{k} $ & $ \mathbf{2} $ \\
			$ (\Psi_l) $ & Fermi & $ \mathbf{k} $ & & $ \mathbf{32} $
		\end{tabular}
		\subcaption{}
	\end{subfigure}
	\caption{(a) Brane configuration, (b) quiver description, and (c) matter content for the rank 1 $ SO(32) $ heterotic LST.} \label{fig:SO32}
\end{figure}

Under S-duality, the $SO(32)$ LST can be mapped into a system of a D5-brane in type I string theory.
In this system, the little strings are $ k $ D1-branes bound to the D5-brane, as shown in Figure~\ref{fig:SO32}(a) \cite{Johnson:1998yw, Kim:2018gak}. The partition function for the little strings can be computed using the 2d gauge theory description for the worldvolume theory on the D1-branes. 

The 2d gauge theory is an $\mathcal{N}=(0,4)$ $ O(k) $ gauge theory with a symmetric hypermultiplet, twisted hypermultiplet, and an antisymmetric Fermi multiplet describing the motion of the D1-branes on O9-plane. In addition, there are $ O(k) \times Sp(1) $ bifundamental matters coming from 
the D1-D5 strings, and the D1-D9 string modes give rise to Fermi multiplets in bifundamental representation of $ O(k) \times SO(32) $. This theory has an $ SO(4) = SU(2)_l \times SU(2)_r $ global symmetry which rotates 2345 directions and another $ SO(4) = SU(2)_R \times SU(2)_{m_0} $ rotation symmetry corresponds to $ 6789 $ directions. Essentially, the 2d gauge theory agrees with the ADHM data for $k$-instantons in the $Sp(1)$ gauge theory with 16 fundamentals.
The 2d gauge theory description and its matter content are summarized in Figure~\ref{fig:SO32}(b) and (c).

The elliptic genera of the 2d gauge theory for $k$ little strings can be calculated using localization. 
The computational details will be explained in Appendix~\ref{app:SO32}. The 1-string elliptic genus is given by
\begin{align}\label{SO32-1str}
Z_1^{\mathrm{HO}}
= -\sum_{I=1}^4 \frac{\prod_{l=1}^{16} \theta_I(m_l)}{2\eta^{12} \theta_1(\epsilon_1) \theta_1(\epsilon_2) \theta_1(\pm m_0 - \epsilon_+)} \frac{\theta_I(m_0 \pm \phi_1)}{\theta_I(\epsilon_+ \pm \phi_1)} \, ,
\end{align}
where $ m_0 $ is the chemical potential for $ SU(2)_{m_0} $. Also, the explicit expression of the 2-string elliptic genus is presented in \eqref{SO32-2str}. We note that although the elliptic genera seem to depend on the mass parameter $m_0$, which plays no role in the 6d worldvolume theory, the BPS states carrying $Sp(1)$ gauge charge are all independent of $m_0$. One can see this by checking that all BPS states captured by the elliptic genera depending on $ \phi_1 $ are independent of $ m_0 $, which we checked up to 2-string and $ q^2 $ order in the $q$-expansion.

The $ E_8 \times E_8 $ and $ SO(32) $ heterotic LSTs are related via the T-duality \cite{Narain:1986am, Ginsparg:1986bx, Polchinski:1995df, Horava:1995qa, Ganor:1996mu}. This implies that the partition functions of these two theories are related each other up to appropriate reparametrizations of fugacities. To compare two elliptic genera, we first turn on Wilson lines for the flavor symmetries along the T-dual circle such that they break $ E_8 \times E_8 $ and $ SO(32) $ symmetries to their common subgroup $ SO(16) \times SO(16) $. In the $ E_8 \times E_8 $ picture, the Wilson lines shift some of the chemical potentials for $ E_8 \times E_8 $ symmetry as
\begin{align}\label{E8E8-massshift}
    \tilde{m}_8^{\mathrm{HE}} = m_8^{\mathrm{HE}} + \tau^{\mathrm{HE}} \, , \quad
    \tilde{m}_{16}^{\mathrm{HE}} = m_{16}^{\mathrm{HE}} + \tau^{\mathrm{HE}} \, , \quad
    \tilde{m}_l^{\mathrm{HE}} = m_l^{\mathrm{HE}} \ (l \neq 8, 16) \, ,
\end{align}
and we also redefine
\begin{align}\label{E8E8-redefine}
    \begin{aligned}
        \phi_{1,0}^{\mathrm{HE}} - \phi_{2,0}^{\mathrm{HE}} &\to \phi_{1,0}^{\mathrm{HE}} - \phi_{2,0}^{\mathrm{HE}} + \tilde{m}_8^{\mathrm{HE}} - \frac{\tau^{\mathrm{HE}}}{2} \, , \\
        w^{\mathrm{HE}} &\to w^{\mathrm{HE}} + \tilde{m}_8^{\mathrm{HE}} + \tilde{m}_{16}^{\mathrm{HE}} - \tau^{\mathrm{HE}} \, .
    \end{aligned}
\end{align}
To distinguish chemical potentials in two LSTs, we add a superscript `HE' for the chemical potentials in $ E_8 \times E_8 $ LST. We list some leading BPS states in Table~\ref{table:E8E8-bps}, where we only show the states carrying nonzero charge for $ \phi_{1,0}-\phi_{2,0} $.
\begin{table}
	\centering
	\begin{Tabular}{|c|C{27ex}||c|C{27ex}|} \hline
		$ \mathbf{d} $ & $ \oplus N_{j_l,j_r}^{\mathbf{d}} (j_l, j_r) $ & $ \mathbf{d} $ & $ \oplus N_{j_l,j_r}^{\mathbf{d}} (j_l, j_r) $ \\ \hline
		$ (1, 0, 0) $ & $ \mathbf{16}_1(0, 0) $ & $ (1, 0, \frac{1}{2}) $ & $ \overline{\mathbf{128}}_1 (0, 0) $ \\ \hline
		$ (1, 0, 1) $ & $ [\mathbf{560}_1 + \mathbf{16}_1](0, 0) \oplus \mathbf{16}_1 (\frac{1}{2}, \frac{1}{2}) $ & $ (2, 0, 0) $ & $ (0, \frac{1}{2}) $ \\ \hline
		$ (2, 0, \frac{1}{2}) $ & $ \mathbf{128}_1 (0, \frac{1}{2}) $ & $ (2, 0, 1) $ & $ [\mathbf{1820}_1 + \mathbf{120}_1 + 2](0, \frac{1}{2}) \oplus (\frac{1}{2}, 0) \oplus [\mathbf{120}_1 + 1](\frac{1}{2}, 1) \oplus (1, \frac{3}{2}) $ \\ \hline
		$ (2, 1, 0) $ & $ \mathbf{16}_2(0, 0) $ & $ (2, 1, \frac{1}{2}) $ & $ [\mathbf{128}_1 \cdot \mathbf{16}_2 + \overline{\mathbf{128}}_2] (0, 0) $ \\ \hline
		$ (2, 1, 1) $ & \multicolumn{3}{C{68ex}|}{$ [\mathbf{128}_1 \cdot \overline{\mathbf{128}}_2 + (\mathbf{1820}_1 + 2 \cdot \mathbf{120}_1 + 4) \mathbf{16}_2 + \mathbf{560}_2] (0, 0) \allowbreak \oplus [(\mathbf{120}_1 + 3) \cdot \mathbf{16}_2] (\frac{1}{2}, \frac{1}{2}) + \mathbf{16}_2 (1, 1) $} \\ \hline
	\end{Tabular}
    \caption{BPS spectrum of the rank 1 $ E_8 \times E_8 $ heterotic LST after introducing the Wilson lines, up to $ d_1 \leq 2 $, $ d_2 \leq 1 $ and $ d_3 \leq 1 $. Here, $ \mathbf{d}=(d_1, d_2, d_3) $ labels the BPS states with charge $ d_1(\phi_{1,0}^{\mathrm{HE}}-\phi_{2,0}^{\mathrm{HE}}) + d_2 (w^{\mathrm{HE}} - \phi_{1,0}^{\mathrm{HE}} + \phi_{2,0}^{\mathrm{HE}}) + d_3 \tau^{\mathrm{HE}} $ after the redefinition as in \eqref{E8E8-redefine}. $ \mathbf{R}_{1,2} $ labels representation of $ SO(16)_{1,2} $ whose chemical potentials are $ \{\tilde{m}_1^{\mathrm{HE}},\cdots,\tilde{m}_8^{\mathrm{HE}}\} $, and $ \{\tilde{m}_9^{\mathrm{HE}},\cdots,\tilde{m}_{16}^{\mathrm{HE}}\} $ given in \eqref{E8E8-massshift}. The states related by the symmetry $ d_1 \leftrightarrow d_2 $ and $ SO(16)_1 \leftrightarrow SO(16)_2 $ are omitted in the table. We only show the LST BPS states which have nonzero charge for $ \phi_{1,0}-\phi_{2,0} $.} \label{table:E8E8-bps}
\end{table}

In the $ SO(32) $ picture, the Wilson lines shift
\begin{align}\label{SO32-massshift}
    \tilde{m}_l^{\mathrm{HO}} = m_l^{\mathrm{HO}} \ (1 \leq l \leq 8) \, , \quad
    \tilde{m}_l^{\mathrm{HO}} = m_l^{\mathrm{HO}} + \frac{\tau^{\mathrm{HO}}}{2} \ (9 \leq l \leq 16) \, ,
\end{align}
and we redefine
\begin{align}\label{SO32-redefine}
    w^{\mathrm{HO}} \to w^{\mathrm{HO}} + \frac{1}{2} \sum_{l=9}^{16} \tilde{m}_l^{\mathrm{HO}} - \tau^{\mathrm{HO}} \, ,
\end{align}
where we put a superscript `HO' to denote the $ SO(32) $ chemical potentials.
The perturbative partition function in \eqref{so(32)-pert} now becomes
\begin{align}\label{so32-pert-shift}
    Z_{\mathrm{pert}}^{\mathrm{HO}}
    &= \PE\bigg[ -\frac{1+p_1p_2}{(1-p_1)(1-p_2)} \qty(e^{4\pi i \phi_1^{\mathrm{HO}}} + e^{2\pi i (\tau^{\mathrm{HO}} - 2\pi i \phi_1^{\mathrm{HO}})}) \frac{1}{1-e^{2\pi i \tau^{\mathrm{HO}}}} \\
    &\: + \frac{\sqrt{p_1 p_2}}{(1-p_1)(1-p_2)} \qty(e^{2\pi i \phi_1^{\mathrm{HO}}} + e^{2\pi i (\tau^{\mathrm{HO}}-\phi_1^{\mathrm{HO}})}) \sum_{l=1}^{16} \qty(e^{2\pi i \tilde{m}_l} + e^{-2\pi i \tilde{m}_l}) \frac{e^{2\pi i r_l \tau^{\mathrm{HO}}}}{1-e^{2\pi i \tau^{\mathrm{HO}}}} \bigg] , \nonumber
\end{align}
where $ r_l=0 $ for $ 1 \leq l \leq 8 $ and $ r_l = 1/2 $ for $ 9 \leq l \leq 16 $. The first few BPS states of the $ SO(32) $ LST from the elliptic genera are listed in Table~\ref{table:SO32-bps}.
Again, we show  only the BPS states carrying nonzero charge for $ \phi_1^{\mathrm{HO}} $. 

\begin{table}
	\centering
	\begin{Tabular}{|c|C{27ex}||c|C{27ex}|} \hline
		$ \mathbf{d} $ & $ \oplus N_{j_l,j_r}^{\mathbf{d}} (j_l, j_r) $ & $ \mathbf{d} $ & $ \oplus N_{j_l,j_r}^{\mathbf{d}} (j_l, j_r) $ \\ \hline
		$ (1, 0, 1) $ & $ \overline{\mathbf{128}}_1 (0, 0) $ & $ (1, 0, 2) $ & $ \mathbf{128}_1 (0, \frac{1}{2}) $ \\ \hline
		$ (1, \frac{1}{2}, -1) $ & $ \overline{\mathbf{128}}_2 (0, 0) $ & $ (1, \frac{1}{2}, 1) $ & $ [\mathbf{128}_1 \cdot \mathbf{16}_2 + \overline{\mathbf{128}}_2] (0, 0) $ \\ \hline
		$ (1, 1, -2) $ & $ \mathbf{128}_2 (0, \frac{1}{2}) $ & $ (1, 1, -1) $ & $ [\mathbf{16}_1 \cdot \mathbf{128}_2 + \overline{\mathbf{128}}_1] (0, 0) $ \\ \hline
		$ (2, 0, 1) $ & $ [\mathbf{560}_1 + \mathbf{16}_1](0, 0) \oplus \mathbf{16}_1 (\frac{1}{2}, \frac{1}{2}) $ & $ (2, 0, 2) $ & $ [\mathbf{1820}_1 + \mathbf{120}_1 + 2](0, \frac{1}{2}) \oplus (\frac{1}{2}, 0) \oplus [\mathbf{120}_1 + 1](\frac{1}{2}, 1) \oplus (1, \frac{3}{2}) $ \\ \hline
		$ (2, \frac{1}{2}, -1) $ & $ [\mathbf{560}_2 + \mathbf{16}_2](0, 0) \oplus \mathbf{16}_2 (\frac{1}{2}, \frac{1}{2}) $ & $ (2, \frac{1}{2}, 1) $ & $ [\mathbf{128}_1 \cdot \overline{\mathbf{128}}_2 + (\mathbf{1820}_1 + 2 \cdot \mathbf{120}_1 + 4) \cdot \mathbf{16}_2 + \mathbf{560}_2] (0, 0) \allowbreak \oplus [(\mathbf{120}_1 + 3) \cdot \mathbf{16}_2] (\frac{1}{2}, \frac{1}{2}) \oplus \mathbf{16}_2 (1, 1) $ \\ \hline
		$ (2, 1, -2) $ & $ [\mathbf{1820}_2 + \mathbf{120}_2 + 2](0, \frac{1}{2}) \oplus (\frac{1}{2}, 0) \oplus [\mathbf{120}_2 + 1](\frac{1}{2}, 1) \oplus (1, \frac{3}{2}) $ & $ (2, 1, -1) $ & $ [\overline{\mathbf{128}}_1 \cdot \mathbf{128}_2 + \mathbf{16}_1 \cdot  (\mathbf{1820}_2 + 2 \cdot \mathbf{120}_2 + 4) + \mathbf{560}_1] (0, 0) \oplus [\mathbf{16}_1  \cdot (\mathbf{120}_2+3) ] (\frac{1}{2}, \frac{1}{2}) \oplus \mathbf{16}_1 (1, 1) $ \\ \hline
	\end{Tabular}
    \caption{BPS spectrum of rank 1 $ SO(32) $ heterotic LST with the Wilson lines for $ 1 \leq d_1 \leq 2 $, $ d_2 \leq 1 $ and first two orders of $ d_3 $. Here, $ \mathbf{d}=(d_1, d_2, d_3) $ labels the BPS states with charge $ d_1 w^{\mathrm{HO}} + d_2 \tau^{\mathrm{HO}} + d_3 \phi_1^{\mathrm{HO}} $ after redefining \eqref{SO32-redefine}, and $ \mathbf{R}_{1,2} $ labels the representation of $ SO(16)_{1,2} $ whose corresponding chemical potentials are $ \{\tilde{m}_1^{\mathrm{HO}},\cdots,\tilde{m}_8^{\mathrm{HO}}\} $ and $ \{\tilde{m}_9^{\mathrm{HO}},\cdots,\tilde{m}_{16}^{\mathrm{HO}}\} $ in \eqref{SO32-massshift}. Note that $ d_1=0 $ sector is given in \eqref{so32-pert-shift}.} \label{table:SO32-bps}
\end{table}

Now one can verify that the BPS spectra of two LSTs are the same under the exchange of winding number and KK-momentum as $ w^{\mathrm{HE}} \leftrightarrow \tau^{\mathrm{HO}}/2 $ and $ \tau^{\mathrm{HE}}/2 \leftrightarrow w^{\mathrm{HO}} $, as well as the exchange of $ \phi_{1,0}^{\mathrm{HE}} - \phi_{2,0}^{\mathrm{HE}} \leftrightarrow \phi_1^{\mathrm{HO}} $ and $ \tilde{m}_l^{\mathrm{HE}} \leftrightarrow \tilde{m}_l^{\mathrm{HO}} $. However, due to the presence of extra decoupled states at $ k_1 = k_2 $ sectors in the $ E_8 \times E_8 $ heterotic LST mentioned above, the spectra at $ k_1 = k_2 $ do not match each other.

\paragraph{Modularity}

Each chiral fermion in Figure~\ref{fig:SO32}(c) contributes to the 2d anomaly polynomial as
\begin{align}
\begin{aligned}
\lambda_+^{\dot{\alpha}A} + \lambda_+^{\alpha a} &\to k(k-1) \qty( \frac{c_2(r) + c_2(R)}{2} + \frac{c_2(l)+c_2(m_0)}{2} + \frac{p_1(T_2)}{12} ) \, ,  \\
\lambda_-^{\alpha A} + \lambda_-^{\dot{\alpha} a} &\to -k(k+1) \qty( \frac{c_2(l) + c_2(R)}{2} + \frac{c_2(r)+c_2(m_0)}{2}+\frac{p_1(T_2)}{12} ) \, , \\
\psi_-^A + \psi_+^a + \Psi_l &\to 2k \qty(\frac{-c_2(R)+c_2(m_0)}{2}) + k \qty( \frac{1}{4}\Tr F_m^2 + \frac{2}{3}p_1(T_2) ) \, ,
\end{aligned}
\end{align}
where $ F_m $ is the 2-form field strength for the $ SO(32) $ global symmetry. The anomaly polynomial is the sum of these contributions.
This can also be derived from the anomaly inflow presented in \eqref{eq:I4-LST} as
\begin{align}
I_4 = k X_{4,0} =  k \qty( -c_2(l) - c_2(r) - 2c_2(R) + \frac{1}{2}p_1(T_2) + \frac{1}{4} \Tr F_m^2 ) \, .
\end{align}
Here, $X_{4,0}$ is the 4-form appearing in the mixed gauge anomalies in the 6d $Sp(1)$ gauge theory
\begin{align}
    I_8^{\rm mixed} = Y_4\wedge X_{4,0} =  \frac{1}{4} \Tr F_{Sp(1)}^2 \wedge \qty( \frac{1}{2} p_1(T_6) - 2c_2(R) + \frac{1}{4} \Tr F_m^2 ) \, .
\end{align}

Then, the modular ansatz for the $ k $-string elliptic genus can be taken as
\begin{align}
Z_k = \frac{1}{\eta^{24k}} \frac{\Phi_k(\tau, \epsilon_\pm, \phi_1, m_0, m_l)}{\mathcal{D}_k^{\mathrm{cm}} \cdot \mathcal{D}_k^{A_1} \cdot \prod_{s=1}^k \varphi_{-1,1/2}(\pm s m_0 - s \epsilon_+)} \, ,
\end{align}
where $ \Phi_k $ is written in terms of the $ SU(2) $ Weyl invariant Jacobi forms for $ \epsilon_\pm, \phi_1, m_0 $ and the $ SO(32) $ Weyl invariant Jacobi forms for $ m_{l=1,\cdots,16} $ given in Appendix~\ref{app:jacobi}. 
We have found that the 1-string elliptic genus \eqref{SO32-1str} can be reproduced by the modular ansatz with coefficients given in Table~\ref{table:SO32-modular}. The coefficients in this ansatz are listed in ascending order with respect to $ \{\epsilon_+, \epsilon_-, \phi_1, m_0, m_{l=1,\cdots,16}\} $ that we defined in the footnote~\ref{ascending}.
We expect that this ansatz is consistent with the elliptic genera from the ADHM computation for any value of $k$, since the 2D quiver theory possesses the SO(32) flavor symmetry explicitly.

\begin{table}
	\centering
	\begin{Tabular}{|c|L{70ex}|} \hline
        $ k $ & \multicolumn{1}{c|}{$ \big\{ C_i^{(k)} \big\} $} \\ \hline \hline
		$ 1 $ & \scriptsize $ \frac{1}{2^{15} \cdot 3^{13}} \{0,\allowbreak 0,\allowbreak -27648,\allowbreak 13824,\allowbreak 0,\allowbreak -248832,\allowbreak 0,\allowbreak 27648,\allowbreak -13824,\allowbreak -13824,\allowbreak 165888,\allowbreak 248832,\allowbreak 13824,\allowbreak 82944,\allowbreak -165888,\allowbreak -82944,\allowbreak -165888,\allowbreak -55296,\allowbreak -55296,\allowbreak -165888,\allowbreak -276480,\allowbreak -110592,\allowbreak 55296,\allowbreak -663552,\allowbreak 276480,\allowbreak 165888,\allowbreak 165888,\allowbreak 110592,\allowbreak 663552,\allowbreak 663552,\allowbreak 55296,\allowbreak 165888,\allowbreak 0,\allowbreak -165888,\allowbreak 0,\allowbreak -663552,\allowbreak -1,\allowbreak 1,\allowbreak 0,\allowbreak 884736,\allowbreak -884736,\allowbreak 0,\allowbreak 0,\allowbreak -2,\allowbreak 2,\allowbreak -18,\allowbreak 18,\allowbreak -18,\allowbreak 18,\allowbreak -2,\allowbreak 2,\allowbreak 0,\allowbreak 32,\allowbreak -32,\allowbreak 0,\allowbreak 0,\allowbreak -20736,\allowbreak -124416,\allowbreak 0,\allowbreak 20736,\allowbreak 124416,\allowbreak 124416,\allowbreak -124416,\allowbreak 0,\allowbreak 82944,\allowbreak -82944,\allowbreak -41472,\allowbreak 746496,\allowbreak 27648,\allowbreak 165888,\allowbreak -124416,\allowbreak -248832,\allowbreak 1492992,\allowbreak -746496,\allowbreak -27648,\allowbreak -193536,\allowbreak 165888,\allowbreak 124416,\allowbreak -2239488,\allowbreak -1492992,\allowbreak -55296,\allowbreak 41472,\allowbreak 165888,\allowbreak 0,\allowbreak 2239488,\allowbreak 0,\allowbreak -442368,\allowbreak 110592,\allowbreak 110592,\allowbreak 663552,\allowbreak 331776,\allowbreak -110592,\allowbreak -663552,\allowbreak 0,\allowbreak 0,\allowbreak 3,\allowbreak -3,\allowbreak 9,\allowbreak -9,\allowbreak 3,\allowbreak -3,\allowbreak 0,\allowbreak 0,\allowbreak -12,\allowbreak 12,\allowbreak -12,\allowbreak 12,\allowbreak 0,\allowbreak 0,\allowbreak 124416,\allowbreak -124416,\allowbreak 82944,\allowbreak 0,\allowbreak 0,\allowbreak 62208,\allowbreak 373248,\allowbreak 248832,\allowbreak -82944,\allowbreak -82944,\allowbreak 20736,\allowbreak 124416,\allowbreak 746496,\allowbreak 1617408,\allowbreak 82944,\allowbreak -20736,\allowbreak -145152,\allowbreak -870912,\allowbreak -746496,\allowbreak -41472,\allowbreak -248832,\allowbreak -1119744,\allowbreak -165888,\allowbreak -82944,\allowbreak 414720,\allowbreak -207360,\allowbreak 497664,\allowbreak 580608,\allowbreak -41472,\allowbreak -41472,\allowbreak -3981312,\allowbreak -248832,\allowbreak -995328,\allowbreak 248832,\allowbreak 41472,\allowbreak 3981312,\allowbreak 0,\allowbreak 0,\allowbreak 0,\allowbreak 0,\allowbreak 0,\allowbreak 0,\allowbreak 0,\allowbreak 0,\allowbreak 9,\allowbreak -9,\allowbreak 27,\allowbreak -27,\allowbreak 9,\allowbreak -9,\allowbreak 0,\allowbreak 0,\allowbreak -55296,\allowbreak 6912,\allowbreak 0,\allowbreak 248832,\allowbreak -373248,\allowbreak 55296,\allowbreak -6912,\allowbreak -6912,\allowbreak -290304,\allowbreak -995328,\allowbreak 6912,\allowbreak 41472,\allowbreak 1036800,\allowbreak 331776,\allowbreak -82944,\allowbreak -110592,\allowbreak 96768,\allowbreak -331776,\allowbreak -801792,\allowbreak 380160,\allowbreak 89856,\allowbreak 1534464,\allowbreak 304128,\allowbreak 331776,\allowbreak -290304,\allowbreak -6912,\allowbreak -41472,\allowbreak -41472,\allowbreak 110592,\allowbreak 580608,\allowbreak -186624,\allowbreak -82944,\allowbreak -1492992,\allowbreak 41472,\allowbreak 2,\allowbreak -2,\allowbreak 0,\allowbreak 1022976,\allowbreak -1022976,\allowbreak 0,\allowbreak 0,\allowbreak 4,\allowbreak -4,\allowbreak 36,\allowbreak -36,\allowbreak 36,\allowbreak -36,\allowbreak 4,\allowbreak -4,\allowbreak 0,\allowbreak -91,\allowbreak 91,\allowbreak 0,\allowbreak 165888,\allowbreak 20736,\allowbreak 0,\allowbreak -995328,\allowbreak 55296,\allowbreak 331776,\allowbreak 0,\allowbreak -20736,\allowbreak -3110400,\allowbreak -124416,\allowbreak -55296,\allowbreak -387072,\allowbreak -20736,\allowbreak 0,\allowbreak 2985984,\allowbreak 870912,\allowbreak -110592,\allowbreak 0,\allowbreak 20736,\allowbreak 1119744,\allowbreak -746496,\allowbreak 0,\allowbreak -138240,\allowbreak -525312,\allowbreak -525312,\allowbreak -559872,\allowbreak 663552,\allowbreak 525312,\allowbreak 559872,\allowbreak 0,\allowbreak 0,\allowbreak -6,\allowbreak 6,\allowbreak -18,\allowbreak 18,\allowbreak -6,\allowbreak 6,\allowbreak 0,\allowbreak 0,\allowbreak 24,\allowbreak -24,\allowbreak 24,\allowbreak -24,\allowbreak 331776,\allowbreak 1161216,\allowbreak 165888,\allowbreak 62208,\allowbreak -995328,\allowbreak -165888,\allowbreak -62208,\allowbreak -62208,\allowbreak 3359232,\allowbreak -497664,\allowbreak 0,\allowbreak 0,\allowbreak 62208,\allowbreak -3359232,\allowbreak 0,\allowbreak 0,\allowbreak 0,\allowbreak 0,\allowbreak 0,\allowbreak 0,\allowbreak 0,\allowbreak 0,\allowbreak -18,\allowbreak 18,\allowbreak -54,\allowbreak 54,\allowbreak -18,\allowbreak 18,\allowbreak 0,\allowbreak -1,\allowbreak 1,\allowbreak 0,\allowbreak -1492992,\allowbreak 1492992,\allowbreak 0,\allowbreak 0,\allowbreak -2,\allowbreak 2,\allowbreak -18,\allowbreak 18,\allowbreak -18,\allowbreak 18,\allowbreak -2,\allowbreak 2,\allowbreak 0,\allowbreak 86,\allowbreak -86,\allowbreak 3,\allowbreak -3,\allowbreak 9,\allowbreak -9,\allowbreak 3,\allowbreak -3,\allowbreak 0,\allowbreak 0,\allowbreak -12,\allowbreak 12,\allowbreak -12,\allowbreak 12,\allowbreak 0,\allowbreak 9,\allowbreak -9,\allowbreak 27,\allowbreak -27,\allowbreak 9,\allowbreak -9,\allowbreak 0,\allowbreak 0,\allowbreak -27,\allowbreak 27\} $ \\ \hline
	\end{Tabular}
	\caption{Coefficients in the modular ansatz for the rank 1 $ SO(32) $ heterotic LST.} \label{table:SO32-modular}
\end{table}

\paragraph{Blowup equation}

Since the partition functions of the $E_8\times E_8$ LST and the $SO(32)$ LST are the same, the partition function of the $SO(32)$ LST should satisfy the same blowup equation for the $E_8\times E_8$ LST in \eqref{eq:HE-blowup}. More precisely, two partition functions are the same, after the identification of fugacities of two theories, up to decoupled states which are independent of the dynamical K\"ahler parameter $\phi_{1,0}^{\rm HE}-\phi_{2,0}^{\rm HE}$ or $\phi_1^{\rm HO}$. In the blowup equation, all the difference from the decoupled states can be absorbed by the prefactor $\Lambda$. Therefore, the partition function of the $SO(32)$ LST satisfies the blowup equation in \eqref{eq:HE-blowup} with a different prefactor $\Lambda$ for this theory.

As usual, the blowup equation can be solved iteratively starting from the effective prepotential and the perturbative partition function in \eqref{so32-pert-shift} with a choice of magnetic fluxes on $\mathbb{P}^1$. The effective prepotential of the $SO(32)$ LST receives tree level contributions from the gauge kinetic term and the counterterm with an auxiliary 2-form field, and 1-loop contributions from the $ Sp(1) $ vector multiplet and the 16 fundamental hypermultiplets. With the parametrization given in \eqref{SO32-massshift} and \eqref{SO32-redefine}, we compute the 1-loop prepotential as
\begin{align}
\mathcal{F} = \frac{1}{12} \sum_{n \in \mathbb{Z}} \qty( \abs{n\tau \pm 2\phi_1}^3 - \sum_{i=1}^{16} \abs{(n+r_i) \tau \pm \phi_1 + m_i}^3 ) = -\frac{1}{2} \sum_{i=1}^8 m_i^2 \phi_1 \, ,
\end{align}
where $ r_i=0 $ for $ 1 \leq i \leq 8 $, $ r_i=1/2 $ for $ 9 \leq i \leq 16 $, and we used the zeta function regularization to compute the infinite KK momenta summations.
The mixed Chern-Simons coefficients can be computed in the same manner.
Collecting all the contributions yields the effective prepotential
\begin{align}
    \mathcal{E} = \frac{1}{\epsilon_1 \epsilon_2} \qty( -\frac{1}{2} \sum_{i=1}^8 m_i^2 \phi_1 + \frac{\epsilon_1^2 + \epsilon_2^2}{4} \phi_1 + \epsilon_+^2 \phi_1 ) + \mathcal{E}^{(0)}_{\mathrm{tree}} \, ,
\end{align}
where
\begin{align}
    \mathcal{E}_{\mathrm{tree}}^{(0)} = \frac{1}{\epsilon_1 \epsilon_2} \qty[w \phi_1^2 + \qty(  \frac{\epsilon_1^2 + \epsilon_2^2}{2} - \frac{1}{2} \sum_{i=1}^{16} m_i^2 + 2\epsilon_+^2 ) \phi_{0} ] \, ,
\end{align}
with the auxiliary scalar VEV $ \phi_0 \equiv \phi_{0,0} $. Indeed this under the reparametrization $ w \to \tau/2 $ and $ \phi_1 \to \phi_{1,0} - \phi_{2,0} $ coincides with the effective prepotential of the $ E_8\times E_8 $ LST in \eqref{E8E8-E}, as expected from the T-duality.

Then we turn on the magnetic fluxes on the blowup background such as
\begin{align}
    n_1 = n'_1+n'_0 \in \mathbb{Z} \, , \
    n_2 = n_0' \in \mathbb{Z} \, , \
    B_{m_i} = \left\{
        \begin{array}{ll}
            1/2 & \quad (0 \leq i \leq 8) \\
            -1/2 & \quad (9 \leq i \leq 16)
    \end{array}\right. \, , \
    B_\tau = B_w = 0 \, ,
\end{align}
where $ n_0',n_1' $ denote the fluxes for $ \phi_0 $ and those for $ \phi_1 $ respectively. 

Using these ingredients, it is now possible to construct the blowup equation for the $SO(32)$ LST. To compute the elliptic genera for the little strings, we first expand the blowup equation in terms of the K\"ahler parameter $e^{2\pi i w}$ for the string number and substitute the modular ansatz into the $k$-string elliptic genera that appear at each order in the expansion. The coefficients in the modular ansatz are then determined by solving the blowup equation. We have carried out this calculation for $k=1$ and 
reproduced the result in Table~\ref{table:SO32-modular}. We expect that the higher order elliptic genera can also be computed in this manner.

\subsection{\texorpdfstring{$ SU(3) + 1\mathbf{sym} + 1\mathbf{\Lambda}^2 $}{SU(3)+1Sym+1Anti}} \label{subsec:su3-symm}

As a last example, we consider the $\mathcal{N}=(1,0)$ LST whose low energy theory is given by an $ SU(N) $ gauge theory with a symmetric hypermultiplet and an antisymmetric hypermultiplet, as introduced in \cite{Bhardwaj:2015oru}. This theory can be realized by $ N $ D6-branes stretched between two half NS5-branes in the type IIA string theory on an interval $ S^1/\mathbb{Z}_2 $ with an $ \mathrm{O8}^- $- and an $ \mathrm{O8}^+ $-plane at each end, which is called the O$8^{\pm}$ background \cite{Witten:1997bs,Aharony:2007du}, as depicted in Figure \ref{fig:SU3-brane}(b). 
Here, the half NS5-branes are located at each orientifold plane.

The index part of partition function is factorized as
\begin{align}
    Z_{\mathrm{GV}} = Z_{\mathrm{pert}} \cdot Z_{\mathrm{str}} = Z_{\mathrm{pert}} \cdot \sum_{k=0}^\infty e^{2\pi i k w} Z_k \, ,
\end{align}
where $ w = 1/g_{\mathrm{YM}}^2 $, and $ Z_k $ is the elliptic genus of $k$-strings. The 1-loop contribution $ Z_{\mathrm{pert}} $ from the $ SU(N) $ vector multiplet and the hypermultiplets is given by
\begin{align}\label{SU3-pert}
    Z_{\mathrm{pert}} &= \PE\bigg[-\frac{1+p_1 p_2}{(1-p_1)(1-p_2)(1-q)}\sum_{\rho \in \mathbf{R}^+} \Big( e^{2\pi i \rho \cdot \phi} + q e^{-2\pi i \rho \cdot \phi}\Big) \\
    &\qquad + \frac{\sqrt{p_1p_2}}{(1-p_1)(1-p_2)} \sum_{n \in \mathbb{Z}} \Big( \sum_{w \in \mathbf{sym}} e^{2\pi i |n \tau + w \cdot \phi + m_1| } + \sum_{w \in \mathbf{\Lambda}^2} e^{2\pi i |n \tau + w \cdot \phi + m_2| } \Big) \bigg] \nonumber
\end{align}
from \eqref{Ivec} and \eqref{Ihyp}, where $ \mathbf{R}^+ $ is the set of positive roots of $ SU(N) $, $ \mathbf{sym} $ and $ \mathbf{\Lambda}^2 $ are weight vectors of symmetric and antisymmetric representations, repectively.

\paragraph{GLSM}

\begin{figure}
	\centering
	\begin{subfigure}[b]{0.59\textwidth}
		\centering
		\begin{tabular}{c|cccccccccc}
			& 0 & 1 & 2 & 3 & 4 & 5 & 6 & 7 & 8 & 9 \\ \hline
			$ \mathrm{O8}^\pm $ & \textbullet & \textbullet & \textbullet & \textbullet & \textbullet & \textbullet & \textbullet & \textbullet & \textbullet \\ 
			NS5 & \textbullet & \textbullet & \textbullet & \textbullet & \textbullet & \textbullet \\
			D6 & \textbullet & \textbullet & \textbullet & \textbullet & \textbullet & \textbullet & & & & \textbullet \\
			D2 & \textbullet & \textbullet & & & & & & & & \textbullet
		\end{tabular}
		\subcaption{}
	\end{subfigure}
	\begin{subfigure}[b]{0.4\textwidth}
		\centering
		\begin{tikzpicture}
		\draw [thick] (0, 1.5) -- (0, -1.5);
		\draw (-0.6, 1.2) node {$ \mathrm{O8}^+ $};
		\draw [thick] (3, 1.5) -- (3, -1.5);
		\draw (3.6, 1.2) node {$ \mathrm{O8}^- $};
		\draw (-0.7, 0) node {NS5};
		\draw (3.7, 0) node {NS5};
		
		\draw [thick] (0, -0.18) -- (3, -0.18);
		\draw [thick] (0, -0.09) -- (3, -0.09);
		\draw [thick] (0, 0) -- (3, 0);
		\draw (1.5, -0.55) node {$ N $ D6's};
		
		\draw [thick, color=orange] (0, 0.09) -- (3, 0.09);
		\draw [thick, color=orange] (0, 0.18) -- (3, 0.18);
		\draw (1.5, 0.55) node {$ k $ D2's};
		
		\filldraw [color=black!70] (0, 0) circle (0.25);
		\filldraw [color=black!70] (3, 0) circle (0.25);
		\end{tikzpicture}
        \subcaption{}
    \end{subfigure}
    \phantom{a}
    \begin{subfigure}{1\textwidth}
        \centering
        \begin{tabular}{c|c|cccc}
            Field & Type & $ U(k) $ & $ U(N) $ & $ U(1)_S $ & $ U(1)_A $ \\ \hline
            $ (A_\mu, \lambda_+^{\dot{\alpha} A}) $ & vec & $ \mathbf{adj} $ \\
            $ (a_{\alpha\dot{\beta}}, \lambda_-^{\alpha A}) $ & hyp & $ \mathbf{adj} $ \\
            $ (q_{\dot{\alpha}}, \psi_-^A) $ & hyp & $ \mathbf{k} $ & $ \overline{\mathbf{N}} $ \\
            $ (\varphi_A, \Phi_-^{\dot{\alpha}}) $ & twisted hyp & $ \mathbf{anti} $ & & $ 1 $  \\
            $ (\Psi_+^\alpha) $ & Fermi & $ \mathbf{sym} $ & & $ 1 $ \\
            $ (\psi_+) $ & Fermi & $ \mathbf{k} $ & $ \mathbf{N} $ & $ 1 $ & \\
            $ (\tilde{\varphi}_A, \tilde{\Phi}_-^{\dot{\alpha}}) $ & twisted hyp & $ \mathbf{sym} $ & & & $ 1 $ \\
            $ (\tilde{\Psi}_+^\alpha) $ & Fermi & $ \mathbf{anti} $ & & & $ 1 $ \\
            $ (\tilde{\psi}_+) $ & Fermi & $ \mathbf{k} $ & $ \mathbf{N} $ & & $ 1 $
        \end{tabular}
        \subcaption{}
    \end{subfigure}
    \caption{Brane configuration (a), (b) and the matter content for the $k$-strings in the  $ SU(N) + 1\mathbf{sym} + 1\mathbf{\Lambda}^2 $ LST.} \label{fig:SU3-brane}
\end{figure}

In this theory, the little strings are $SU(N)$ instanton strings realized by $k$ D2-branes on top of the D6-branes.
By examining the brane configuration, it is possible to deduce the 2d $\mathcal{N}=(0,4)$ gauge theory description, which has a $U(k)$ gauge symmetry and matter content as summarized in Figure~\ref{fig:SU3-brane}(c). The vector multiplet, adjoint hypermultiplet, and hypermultiplets in the bifundamental representation of $U(k) \times U(N)$ agree with the ADHM data for the $SU(N)$ instanton moduli space, while the remaining fields charged under $U(1)_S$ and $U(1)_A$ arise from zero modes of the 6d symmetric and antisymmetric hypermultiplets, respectively, at $k$-instantons \cite{Shadchin:2005mx}. Note that the gauge anomaly of the 2d theory is cancelled as
\begin{align}
\begin{aligned}
&-4 \times k + 4 \times k + 2N \times \frac{1}{2} + 2 \times \frac{k-2}{2} - 2 \times \frac{k+2}{2} - N \times \frac{1}{2}  \\
&\qquad + 2 \times \frac{k+2}{2} - 2 \times \frac{k-2}{2} - N \times \frac{1}{2} = 0 \, ,
\end{aligned}
\end{align}
where each term comes from the charged chiral fermions given in Figure~\ref{fig:SU3-brane}(c). 

There are mixed anomalies between the gauge and global $U(1)$ symmetries. Let $ T_{U(1)} $, $ S $, $ A $, $ G $ be generators of $ U(1) \subset U(k) $, $ U(1)_S $, $ U(1)_A $ and $ U(1)_G \subset U(N) $, respectively. Then mixed anomalies are
\begin{align}\label{SU3-mixed-anomaly}
\Tr \gamma_3 T_{U(1)} S = -4 - N \, , \quad
\Tr \gamma_3 T_{U(1)} A = 4 - N \, , \quad
\Tr \gamma_3 T_{U(1)} G = -4N \, .
\end{align}
Thus, the anomaly free $ U(1) $ global symmetry in the 2d gauge theory is the subgroup of $ U(1)_S \times U(1)_A \times U(1)_G $ generated by $ 2S + 2A - G $. There is a decoupled $ U(1) $ symmetry generated by $ T_{U(1)} - 2S - 2A  + G  $ which acts trivially on the 2d fields.

We compute the elliptic genera of the little strings using the 2d gauge theory description and the localization technique. 
The 1-string elliptic genus when $N=3$ is
\begin{align}\label{SU3-Z1}
Z_1 &= -\sum_{j=1}^3 \frac{\theta_1(2a_j+m_1-\epsilon_+) \theta_1(2a_j + m_1 - \epsilon_+ - \epsilon_{1,2})}{\theta_1(\epsilon_{1,2}) \theta_1(2a_j+m_2-3\epsilon_+)} \prod_{k \neq j}^3  \frac{\theta_1(a_{jk} + m_{1,2} - \epsilon_+)}{\theta_1(a_{jk}) \theta_1(2\epsilon_+ - a_{jk})} \nonumber \\
& \quad + \sum_{I=1}^4 \frac{\theta_1(m_1-m_2+\epsilon_{1,2})}{\theta_1(\epsilon_{1,2})} \prod_{j=1}^3 \frac{\theta_I(a_j + m_1 - \frac{m_2}{2} + \frac{\epsilon_+}{2})}{\theta_I(a_j-\frac{3\epsilon_+-m_2}{2})} \, ,
\end{align}
where $ a_1,a_2,a_3 $ are the chemical potentials for $ U(3) $, and $ m_1 $ and $ m_2 $ are the $ U(1)_S $ and $ U(1)_A $ chemical potentials, respectively.
Here we use a shorthand notation, $ a_{jk} = a_j - a_k $. We present the computational details and the 2-string elliptic genus in Appendix~\ref{app:AOA}. 

We have checked as expected that the leading order of the elliptic genera in $ q $-expansion correctly reproduces the BPS spectrum of the 5d $ SU(3) + 1\mathbf{sym} + 1\mathbf{\Lambda}^2 $ theory \cite{Kim:2020hhh}, where $ m_1 $ and $ m_2 $ are identified as the mass parameters of the symmetric and antisymmetric hypermultiplets in the 5d SCFT. However, when considering the BPS states in the 6d LST, the chemical potentials appearing in the elliptic genera are further constrained by the mixed anomalies given in \eqref{SU3-mixed-anomaly}, and for $N=3$, the chemical potential for the anomaly-free $U(1)$ is determined by the condition
\begin{align}\label{SU3-2d-mixed-anomaly}
    -7m_1 + m_2 - 4\sum_{i=1}^3 a_i = 0 \, .
\end{align}
We can also set $\sum_i a_i=0$ using the fact that the $U(1)$ symmetry generated by $T_{U(1)} - 2S - 2A  + G$ decouples from the 2d CFT.
By imposing these conditions, we can rewrite the elliptic genera such that they only depend on a chemical potential $ m_1-m_2 $ as well as $SU(3)$ chemical potentials $a_i$ with $\sum_ia_i=0$.
This is consistent with the fact that the 6d LST has only one anomaly-free $U(1)$ symmetry, as we will show below.

\paragraph{Modularity}

The modular property of the elliptic genus can be read from the anomaly polynomial of the 2d theory. Each chiral fermion in Figure~\ref{fig:SU3-brane}(c) contributes to the 2d anomaly polynomial as
\begin{align}\label{SU3-LST-2d-anomaly}
    \begin{aligned}
        \lambda_+^{\dot{\alpha}A} + \lambda_-^{\alpha A} &\to 2k^2 \qty( \frac{c_2(r)+c_2(R)}{2} - \frac{c_2(l)+c_2(R)}{2} ) \, , \\
        \Psi_+^\alpha + \tilde{\Phi}_-^{\dot{\alpha}} &\to k(k+1) \qty( \frac{c_2(l) - c_2(r)}{2} + \frac{1}{2} F_1^2 - \frac{1}{2}  F_2^2 ) \, , \\
        \Phi_-^{\dot{\alpha}} + \tilde{\Psi}_+^\alpha &\to k(k-1) \qty(\frac{c_2(l)-c_2(r)}{2} - \frac{1}{2} F_1^2 + \frac{1}{2}  F_2^2) \, , \\
        \psi_-^A + \psi_+ + \tilde{\psi}_+ &\to N k \qty( -c_2(R) + \frac{1}{2} F_1^2 + \frac{1}{2} F_2^2 ) \, ,
    \end{aligned}
\end{align}
where $ F_1 $ and $ F_2 $ are the field strengths for $ U(1)_S $ and $ U(1)_A $, respectively. Thus the full anomaly polynomial of the 2d theory for $k$-strings is given by
\begin{align}\label{SU3-LST-anomaly-polynomial}
    I_4 = k \qty( -N c_2(R) + \frac{N+2}{2} F_1^2 + \frac{N-2}{2} F_2^2 ) \, .
\end{align}

The same result can be decuced using the anomaly inflow from the 6d LST in the presence of $k$-strings. The 1-loop anomalies from the chiral fields in the 6d $SU(N)$ gauge theory contain the mixed gauge anomalies
\begin{align}\label{SU3-LST-6d-anomaly}
    \begin{aligned}
        I_8 &\supset \frac{1}{4} \Tr F_{SU(N)}^2 \wedge \qty( -N c_2(R) + \frac{N+2}{2} F_1^2 + \frac{N-2}{2} F_2^2 ) \\
        &\quad + \frac{1}{6} \tr F_{SU(N)}^3 \wedge \qty( (N+4) F_1 + (N-4) F_2 ) \, ,
    \end{aligned}
\end{align}
where `$ \tr $' is the trace in fundamental representation.
To obtain this, we used following relations,
\begin{align}
    \tr_{\mathbf{sym}} F^3 = (N+4) \tr F^3 \, , \quad
    \tr_{\mathbf{\Lambda}^2} F^3 = (N-4) \tr F^3,
\end{align}
for the $ SU(N) $ representations. The gauge anomaly in the first line is cancelled by adding the counterterm as \eqref{B0-S}. 
The second line is the ABJ anomaly and it imposes a constraint on the flavor symmetries as
\begin{align}\label{SU3-ABJ}
    F_2 = -\frac{N+4}{N-4} F_1 \, .
\end{align}
Thus, there is only one anomaly-free global symmetry, given by $ U(1) \subset U(1)_S \times U(1)_A $. Then, the anomaly inflow from the 6d LST on the $k$-string background leads to the same anomaly polynomial in \eqref{SU3-LST-anomaly-polynomial} for the worldsheet CFT.

Now we make a modular ansatz for the elliptic genus of $k$-strings in the $ SU(3) $ LST based on the anomaly polynomial. The elliptic genus $ Z_k $ has a modular anomaly $ \int I_4 = -3k\epsilon_+^2 $. Thus the modular ansatz we propose is
\begin{align}\label{su3-ansatz}
Z_k = \frac{\Phi_k(\tau, \epsilon_\pm, \phi_1, \phi_2)}{\mathcal{D}_k^{\mathrm{cm}} \cdot \mathcal{D}_k^{A_2}} \, .
\end{align}
The $SU(3)$ chemical potentials $ \phi_{1,2} $ are related to $ a_{1,2,3} $ in the elliptic genus given in \eqref{SU3-Z1} by
\begin{align}
    a_1 = \phi_1 \, , \quad
    a_2 = -\phi_1 + \phi_2 \, , \quad
    a_3 = -\phi_2 \, .
\end{align}
We turn off the $ U(1) $ flavor chemical potential because $ U(1) $ has trivial Weyl group and does not fit into the standard theory of Weyl invariant Jacobi forms.

At 1-string order, the ansatz has 514 unknown coefficients $C_i^{(k)}$, and we check that this ansatz with the coefficients in Table~\ref{table:su3-modular}, which are listed  in ascending order with respect to $ \{\epsilon_+, \epsilon_-, \phi_{1,2}\} $, reproduces the 1-string elliptic genus obtained from the ADHM construction in \eqref{SU3-Z1}.

\begin{table}
	\centering
	\begin{Tabular}{|c|L{76ex}|} \hline
		$ k $ & \multicolumn{1}{c|}{$C_i^{(k)}$} \\ \hline \hline
		$ 1 $ & \tiny $ \frac{1}{2^{30} \cdot 3^{13}} \{0, \allowbreak -256, \allowbreak -512, \allowbreak -960, \allowbreak -1024, \allowbreak -512, \allowbreak -10752, \allowbreak -12288, \allowbreak -64512, \allowbreak -36864, \allowbreak -90112, \allowbreak -32768, \allowbreak 64, \allowbreak 32, \allowbreak 128, \allowbreak 192, \allowbreak 16, \allowbreak 256, \allowbreak -1024, \allowbreak -224, \allowbreak 1536, \allowbreak -3840, \allowbreak -1536, \allowbreak -1152, \allowbreak -13824, \allowbreak 5376, \allowbreak -36864, \allowbreak -12288, \allowbreak -163840, \allowbreak 1536, \allowbreak -32768, \allowbreak -29184, \allowbreak -98304, \allowbreak 208896, \allowbreak 368640, \allowbreak 221184, \allowbreak 81920, \allowbreak -131072, \allowbreak 1, \allowbreak -4, \allowbreak 8, \allowbreak 240, \allowbreak -2496, \allowbreak 912, \allowbreak 768, \allowbreak -1728, \allowbreak -4096, \allowbreak 12288, \allowbreak -2304, \allowbreak -8192, \allowbreak -30720, \allowbreak -384, \allowbreak -12288, \allowbreak -18432, \allowbreak 12288, \allowbreak -101376, \allowbreak 368640, \allowbreak -43008, \allowbreak -73728, \allowbreak 18432, \allowbreak 32768, \allowbreak -104448, \allowbreak -524288, \allowbreak 184320, \allowbreak -1572864, \allowbreak -196608, \allowbreak 0, \allowbreak 589824, \allowbreak 524288, \allowbreak -131072, \allowbreak 24, \allowbreak -168, \allowbreak 3840, \allowbreak 960, \allowbreak -21504, \allowbreak -4992, \allowbreak -27648, \allowbreak -7296, \allowbreak 32768, \allowbreak 86016, \allowbreak 13056, \allowbreak 65536, \allowbreak -73728, \allowbreak -9216, \allowbreak -196608, \allowbreak -221184, \allowbreak 196608, \allowbreak -73728, \allowbreak 294912, \allowbreak -688128, \allowbreak 0, \allowbreak -81920, \allowbreak 1048576, \allowbreak -98304, \allowbreak -262144, \allowbreak -16384, \allowbreak -1179648, \allowbreak -672, \allowbreak -192, \allowbreak 9216, \allowbreak 6144, \allowbreak -12288, \allowbreak 12288, \allowbreak -16384, \allowbreak 13312, \allowbreak 0, \allowbreak 0, \allowbreak -16384, \allowbreak 0, \allowbreak -32768, \allowbreak 2048, \allowbreak 0, \allowbreak 98304, \allowbreak 512, \allowbreak 256, \allowbreak -2048, \allowbreak 0, \allowbreak 192, \allowbreak 32, \allowbreak 384, \allowbreak 672, \allowbreak 768, \allowbreak -768, \allowbreak 6912, \allowbreak -6144, \allowbreak 4608, \allowbreak 1536, \allowbreak 12288, \allowbreak 4032, \allowbreak -30720, \allowbreak 7680, \allowbreak -190464, \allowbreak -104448, \allowbreak -221184, \allowbreak 36864, \allowbreak 319488, \allowbreak 196608, \allowbreak 0, \allowbreak 4, \allowbreak -56, \allowbreak 576, \allowbreak -256, \allowbreak -672, \allowbreak 576, \allowbreak 1536, \allowbreak -13824, \allowbreak 1584, \allowbreak 3072, \allowbreak 19968, \allowbreak -384, \allowbreak -12288, \allowbreak 43776, \allowbreak -24576, \allowbreak 20736, \allowbreak -239616, \allowbreak 9216, \allowbreak 92160, \allowbreak -122880, \allowbreak 921600, \allowbreak 69120, \allowbreak 405504, \allowbreak -158208, \allowbreak 1892352, \allowbreak -417792, \allowbreak -1916928, \allowbreak -1032192, \allowbreak -688128, \allowbreak 393216, \allowbreak -9, \allowbreak 72, \allowbreak -1920, \allowbreak -1920, \allowbreak 13824, \allowbreak 7680, \allowbreak 26112, \allowbreak 8640, \allowbreak -24576, \allowbreak -142848, \allowbreak -13824, \allowbreak -49152, \allowbreak 38400, \allowbreak 2688, \allowbreak 319488, \allowbreak 460800, \allowbreak -24576, \allowbreak 36864, \allowbreak -1032192, \allowbreak 700416, \allowbreak 442368, \allowbreak 135168, \allowbreak -1081344, \allowbreak -258048, \allowbreak 491520, \allowbreak 12288, \allowbreak 4325376, \allowbreak -196608, \allowbreak -1179648, \allowbreak 504, \allowbreak 504, \allowbreak -11904, \allowbreak -11904, \allowbreak 18432, \allowbreak -5376, \allowbreak 30720, \allowbreak -16128, \allowbreak 0, \allowbreak -36864, \allowbreak 23040, \allowbreak 0, \allowbreak 12288, \allowbreak 7168, \allowbreak 0, \allowbreak -98304, \allowbreak 0, \allowbreak -98304, \allowbreak 0, \allowbreak 98304, \allowbreak -576, \allowbreak -576, \allowbreak 2048, \allowbreak -1024, \allowbreak 0, \allowbreak -2048, \allowbreak 8, \allowbreak -12, \allowbreak 16, \allowbreak -288, \allowbreak -224, \allowbreak -288, \allowbreak 768, \allowbreak -256, \allowbreak -1152, \allowbreak -2304, \allowbreak 576, \allowbreak -2880, \allowbreak -11808, \allowbreak -2304, \allowbreak 6912, \allowbreak -23040, \allowbreak 24576, \allowbreak 9216, \allowbreak 59904, \allowbreak -267264, \allowbreak -14208, \allowbreak 110592, \allowbreak 78336, \allowbreak 414720, \allowbreak 276480, \allowbreak 1566720, \allowbreak 36864, \allowbreak -368640, \allowbreak -294912, \allowbreak 0, \allowbreak 0, \allowbreak 396, \allowbreak -96, \allowbreak 5088, \allowbreak -5376, \allowbreak -15264, \allowbreak -1344, \allowbreak 16128, \allowbreak 53376, \allowbreak 11424, \allowbreak 29952, \allowbreak 23040, \allowbreak 4032, \allowbreak -4608, \allowbreak -347904, \allowbreak -105984, \allowbreak 165888, \allowbreak 442368, \allowbreak -109056, \allowbreak -442368, \allowbreak 36864, \allowbreak -1548288, \allowbreak 496128, \allowbreak 368640, \allowbreak -161280, \allowbreak -3096576, \allowbreak 995328, \allowbreak 3907584, \allowbreak -368640, \allowbreak -442368, \allowbreak -126, \allowbreak 0, \allowbreak -2736, \allowbreak 6432, \allowbreak 53760, \allowbreak 10272, \allowbreak 0, \allowbreak 9408, \allowbreak -110592, \allowbreak -113664, \allowbreak -23808, \allowbreak -36864, \allowbreak 82944, \allowbreak -1152, \allowbreak 258048, \allowbreak -202752, \allowbreak -147456, \allowbreak 138240, \allowbreak 368640, \allowbreak 374784, \allowbreak -147456, \allowbreak -55296, \allowbreak -1179648, \allowbreak 67584, \allowbreak 1224, \allowbreak 648, \allowbreak -9216, \allowbreak 1216, \allowbreak 15360, \allowbreak -18304, \allowbreak 0, \allowbreak -5248, \allowbreak 0, \allowbreak 49152, \allowbreak 4864, \allowbreak -288, \allowbreak -1, \allowbreak 4, \allowbreak -72, \allowbreak -96, \allowbreak -1044, \allowbreak 672, \allowbreak 5472, \allowbreak 1344, \allowbreak -3456, \allowbreak 13824, \allowbreak -3072, \allowbreak -3456, \allowbreak -8064, \allowbreak -960, \allowbreak -6912, \allowbreak 85824, \allowbreak 89856, \allowbreak -6912, \allowbreak 165888, \allowbreak -55296, \allowbreak 165888, \allowbreak 73728, \allowbreak 663552, \allowbreak -162432, \allowbreak -387072, \allowbreak 23040, \allowbreak -608256, \allowbreak -387072, \allowbreak -2211840, \allowbreak 552960, \allowbreak 663552, \allowbreak 0, \allowbreak -24, \allowbreak 168, \allowbreak -984, \allowbreak 48, \allowbreak -27936, \allowbreak -11136, \allowbreak 16416, \allowbreak 1920, \allowbreak 69120, \allowbreak 228096, \allowbreak 2400, \allowbreak -41472, \allowbreak 41472, \allowbreak 13440, \allowbreak -525312, \allowbreak 260352, \allowbreak -138240, \allowbreak 89856, \allowbreak -110592, \allowbreak -27648, \allowbreak -55296, \allowbreak 147456, \allowbreak 2101248, \allowbreak 317952, \allowbreak 110592, \allowbreak 4608, \allowbreak -663552, \allowbreak 294, \allowbreak -456, \allowbreak -1344, \allowbreak -7296, \allowbreak -11520, \allowbreak -4224, \allowbreak 16128, \allowbreak -10176, \allowbreak 0, \allowbreak -13824, \allowbreak 15360, \allowbreak 0, \allowbreak 32256, \allowbreak -7296, \allowbreak 0, \allowbreak -92160, \allowbreak -728, \allowbreak -472, \allowbreak 4992, \allowbreak 9, \allowbreak -72, \allowbreak 576, \allowbreak 1584, \allowbreak -1944, \allowbreak 3744, \allowbreak -7776, \allowbreak -4608, \allowbreak -15552, \allowbreak -96768, \allowbreak -2304, \allowbreak 10368, \allowbreak -20736, \allowbreak -8064, \allowbreak 72576, \allowbreak -95040, \allowbreak 20736, \allowbreak -145152, \allowbreak -290304, \allowbreak -110592, \allowbreak -82944, \allowbreak -241920, \allowbreak -539136, \allowbreak -209088, \allowbreak -248832, \allowbreak 55296, \allowbreak 787968, \allowbreak -165888, \allowbreak -248832, \allowbreak -504, \allowbreak -504, \allowbreak 15768, \allowbreak 12960, \allowbreak -63072, \allowbreak -4608, \allowbreak -30240, \allowbreak 14112, \allowbreak 93312, \allowbreak 15552, \allowbreak -18864, \allowbreak 0, \allowbreak -13824, \allowbreak -6624, \allowbreak 0, \allowbreak 317952, \allowbreak 0, \allowbreak 93312, \allowbreak -373248, \allowbreak -172800, \allowbreak 333, \allowbreak 900, \allowbreak -4248, \allowbreak 1008, \allowbreak 1728, \allowbreak 9936, \allowbreak 126, \allowbreak 0, \allowbreak -1296, \allowbreak -7776, \allowbreak 14904, \allowbreak 5184, \allowbreak 2592, \allowbreak -5184, \allowbreak -31104, \allowbreak -41472, \allowbreak 13824, \allowbreak 31104, \allowbreak -67392, \allowbreak -864, \allowbreak 155520, \allowbreak -121824, \allowbreak 124416, \allowbreak -124416, \allowbreak 435456, \allowbreak 0, \allowbreak 124416, \allowbreak 41472, \allowbreak 0, \allowbreak -108864, \allowbreak -1224, \allowbreak -648, \allowbreak 10260, \allowbreak -1080, \allowbreak -12960, \allowbreak 17280, \allowbreak 0, \allowbreak 5184, \allowbreak 0, \allowbreak -62208, \allowbreak -4752, \allowbreak 531, \allowbreak 378, \allowbreak 648, \allowbreak -10368, \allowbreak 648, \allowbreak 34020, \allowbreak -3888, \allowbreak 0, \allowbreak -2592, \allowbreak -23328, \allowbreak 62208, \allowbreak 0, \allowbreak 0, \allowbreak 0, \allowbreak 5184, \allowbreak -46656, \allowbreak -54432, \allowbreak 216, \allowbreak 216, \allowbreak -2916, \allowbreak 243, \allowbreak -324, \allowbreak 1944, \allowbreak 0, \allowbreak -2916, \allowbreak -7776, \allowbreak -243\} $ \\ \hline
	\end{Tabular}
	\caption{Coefficients in the modular ansatz for the 1-string elliptic genus of the $ SU(3) + 1\mathbf{sym} + 1\mathbf{\Lambda}^2 $ LST.} \label{table:su3-modular}
\end{table}

\paragraph{Blowup equation}

Lastly, let us consider the blowup equation for the $ SU(3) + 1\mathbf{sym} + 1\mathbf{\Lambda}^2 $ LST. 
We first compute the effective prepotential. The 1-loop prepotential from the $ SU(3) $ vector and hypermultiplets is
\begin{align}\label{SU3-prepotential}
6\mathcal{F}
&= \frac{1}{2} \sum_{n\in\mathbb{Z}} \qty( \sum_{e\in\mathbf{R}} \abs{n\tau + e \cdot \phi}^3 - \! \sum_{w \in \mathbf{sym}} \abs{n\tau + w \cdot \phi + m_1}^3 - \! \sum_{w \in \mathbf{\Lambda}^2} \abs{n\tau + w \cdot \phi + m_2}^3) \nonumber \\
&= \qty(8\phi_1^3 - 3\phi_1^2 \phi_2 - 3\phi_1 \phi_2^2 + 8\phi_2^3) - \frac{1}{2} \qty((\phi_2  + m_2)^3 \! + \! (\phi_2 - \phi_1 - m_2)^3 \! + \! (\phi_1 - m_2)^3 ) \nonumber \\
&\quad - \frac{1}{2} \big( (2\phi_1+m_1)^3 + (\phi_2+m_1)^3 + (-2\phi_1 + 2\phi_2+m_1)^3 + (-\phi_1+\phi_2-m_1)^3  \nonumber \\
&\qquad \quad + (\phi_1-m_1)^3 + (2\phi_2-m_1)^3 ) \, ,
\end{align}
in the chamber $ \phi_2 \geq \phi_1 > 0 $. In the last expression, we keep only the terms dependent on the dynamical K\"ahler parameter $\phi_i$'s. We also evaluate the perturbative partition function \eqref{SU3-pert} in this chamber. With the tree level contributions and the contributions from the mixed Chern-Simons terms, the full effective prepotential is given by
\begin{align}\label{SU3-E}
    \mathcal{E} = \frac{1}{\epsilon_1 \epsilon_2} \qty(\mathcal{F} - \frac{\epsilon_1^2 + \epsilon_2^2}{48}(4\phi_1 - 4\phi_2) + \epsilon_+^2 (\phi_1 + \phi_2) ) + \mathcal{E}_{\mathrm{tree}}^{(0)} \, ,
\end{align}
where
\begin{align}
    \mathcal{E}_{\mathrm{tree}}^{(0)}
= \frac{1}{\epsilon_1 \epsilon_2} \qty[ w(\phi_1^2 - \phi_1 \phi_2 + \phi_2^2) + \phi_0 \qty( 3\epsilon_+^2 - \frac{5}{2} m_1^2 - \frac{1}{2} m_2^2 ) ] \, ,
\end{align}
with $ w \sim 1/g_{\mathrm{YM}}^2 $ and an auxiliary scalar VEV $ \phi_0 \equiv \phi_{0,0} $. Also, because of the 6d mixed anomaly free condition given in \eqref{SU3-ABJ}, we impose the condition
\begin{align}\label{SU3-anomaly-free}
m_2 = 7m_1
\end{align}
in the effective prepotential \eqref{SU3-E}. This is compatible with the 2d mixed anomaly free condition \eqref{SU3-2d-mixed-anomaly}.

The blowup equation can be constructed with a set of magnetic fluxes
\begin{align}
n_0 \in \mathbb{Z} \, , \quad
n_1 \in \mathbb{Z} + 2/3 \, , \quad
n_2 \in \mathbb{Z} + 1/3 \, , \quad
B_{m_1} = 1/6 \, , \quad
B_\tau = B_w = 0 \, .
\end{align}
We propose that the blowup equation of the form given in \eqref{eq:blowup-GV} with these inputs is satisfied for the partition function of the $SU(3)$ LST.
We have checked that the elliptic genera computed using the 2d gauge theory satisfy this blowup equation order by order in the expansion with respect to the fugacities $ e^{2\pi i w} $, $ q=e^{2\pi i \tau} $, $ t = e^{2\pi i (2\phi_1-\phi_2)} $, and $ u = e^{2\pi i(-\phi_1+2\phi_2)} $, up to 2-strings, $ q^1 $, $ t^1 $ and $ u^1 $ orders. 

We also attempted to solve the blowup equation with the help of the modular ansatz in \eqref{su3-ansatz}.  By utilizing the blowup equation and modular ansatz, we were able to find a BPS spectrum up to $ q^1 $, $ t^{-1} $ and $ u^1 $ order which mathches with the ADHM computation given in \eqref{SU3-Z1}. This fixes 419 unknown coefficients in the modular ansatz, which are compatible with those listed in Table~\ref{table:su3-modular}. We expect that higher order computations of blowup equation will fix the remaining unknowns in the modular ansatz.

\section{Conclusion}\label{sec:conclusion}

In this paper, we have proposed the blowup equations for six-dimensional little string theories (LSTs), and demonstrated how our proposal works in some cases.
In order to formulate the blowup equations, we have found that we need to introduce an auxiliary 2-form field to cancel the mixed gauge-global anomalies and also take into account the summation over its magnetic fluxes on the blown-up $\mathbb{P}^1$ as well as the fluxes for the dynamical tensor and gauge symmetries.
Although the flux sum for the auxiliary 2-form field in the blowup equation is divergent, which is essentially because the auxiliary 2-form field has no quadratic kinetic term and it is thus non-dynamical, we have found that the blowup equation still makes sense as a Laurant series expansion in terms of K\"ahler parameters, and we can even use it to determine the BPS spectra of strings in the LSTs with the help of modular ansatz.
As concrete examples, we have computed the elliptic genera of strings in $\hat{A}_1$ type LSTs in IIA/IIB string theories, LSTs in $E_8 \times E_8$ and $SO(32)$ heterotic string theories, and an rank-2 LST with $SU(3)$ guage symmetry and $1 \mathbf{sym} + 1\mathbf{\Lambda}^2$ hypermultiplets. We then checked that these elliptic genera satisfy the blowup equations, and conversely, that the unknown coefficients of their modular ansatz can be fixed by solving the blowup equations.

There are some interesting extensions of the results in this paper. First, it would be quite interesting to generalize the blowup formalism into supergravity theories. The blowup equations for little string theories may suggest the possibility of this generalization because elliptic genera of the string worldsheet theories in some supergravity theories such as 9d/10d heterotic string theories are related with the elliptic genera of LSTs through RG-flows by higgsings. A key difference between supergravity theories and LSTs or SCFTs is that all symmetries in supergravity theories are gauged. As a result, we need to turn on dynamical magnetic fluxes for all the symmetries in the theory, and the $ \Lambda $ factor in the blowup equation can only depend on the $\Omega$-deformation parameters. We leave this generalization as a future work.

Another extension of the current work is the consideration of twisted compactifications of little string theories. The blowup formalism for 5d Kaluza-Klein theories resulting from 6d SCFTs compactified on a circle with automorphism twists has been previously explored in \cite{Kim:2020hhh}. It is straightforward to extend this approach to derive the blowup equations for twisted compactifications of LSTs by simply replacing the intersection form $\Omega^{\alpha\beta}$ of the tensor nodes and the Killing forms $K_{ij}$ for the gauge algebras in the blowup equation for untwisted LSTs with their twisted counterparts. 
One potential use of this formulation is to confirm T-dualities between LSTs including twists along the T-dual circle. For example, the $SU(3)$ gauge theory with $1\mathbf{sym}+1\mathbf{\Lambda}^2$ we discussed in section \ref{subsec:su3-symm} is expected to be T-dual to another little string theory with a twist \cite{Bhardwaj:2022ekc}, which is due to the presence of the symmetric hypermultiplet. The blowup equations for twisted LSTs may provide a more rigorous method for identifying and verifying such dualities.

As another generalization, one can also study little string theories with supersymmetric defects. Various BPS defects in superconformal field theories have been widely studied.
For instance, the partition functions of 5d/6d field theories in the presence of the codimension 4 defects were investigated in \cite{Kim:2021gyj} in the context of the blowup formalism. It should be straightforward to extend this approach to the study of LSTs coupled to codimension 4 defects, offering a concrete method for analyzing the dynamics of these defects within the LSTs.

Recently, a systematic method for calculating the partition functions of LSTs engineered by NS5-branes on D- and E-type singularities using the topological vertex formalism was proposed in \cite{Kim:2022xlf}.
The resulting partition functions for D-type LSTs were found to be consistent with those obtained using the elliptic genus computation in \cite{Kim:2017xan}, while the partition functions for E-type LSTs represent new results. It would be valuable to verify these proposed partition functions using the blowup equations.

\acknowledgments

We are grateful to Sung-Soo Kim and Kimyeong Lee for valuable discussions. HK, MK and YS thank APCTP for its hospitality during completion of this work. HK also thanks the Simons Center for Geometry and Physics, Stony Brook University for the hospitality and partial support during the final stage of this work at the workshops “2022 Simons Summer Workshop” and “Geometry of (S)QFT”.
The research of HK, MK and YS is supported by Samsung Science and Technology Foundation under Project Number SSTF-BA2002-05 and by the National Research Foundation of Korea (NRF) grant funded by the Korea government (MSIT) (No. 2018R1D1A1B07042934).
Some of the computations that were conducted using {\sf mathematica} were carried out on the computer {\it sushiki} at Yukawa Institute for Theoretical Physics in Kyoto University.

\appendix

\section{Elliptic functions}\label{app:elliptic}

In this appendix, we summarize definintions and properties of the modular forms and Jacobi forms used in this paper.

\subsection{Modular forms}

Let $ \mathcal{H} = \{ z \in \mathbb{C} \mid \imaginary z > 0 \} $ be the upper half plane of the complex plane, $ \tau \in \mathcal{H} $ be the complex structure of the torus and $ q = e^{2\pi i \tau} $. A \emph{modular form of weight} $ k $ is a function $ f : \mathcal{H} \to \mathbb{C} $ satisfying
\begin{align}
f\qty(\frac{a\tau+b}{c\tau+d}) = (c\tau+d)^k f(\tau) \, , \quad
\mqty(a & b \\ c & d) \in \mathrm{SL}(2,\mathbb{Z}) \, .
\end{align}
An example of the modular form is the Eisenstein series defined by
\begin{align}
E_{2k}(\tau) = \frac{1}{2\zeta(2k)} \sum_{(m,n) \neq (0, 0)} \frac{1}{(m+n\tau)^{2k}}
= 1 + \frac{(2\pi i)^{2k}}{\zeta(2k) (2k-1)!} \sum_{n=1}^\infty \sigma_{2k-1}(n) q^n \, ,
\end{align}
where $ \zeta(s) $ is the Riemann zeta function and $ \sigma_k(n) = \sum_{d | n} d^k $ is the divisor function. $ E_{2k}(\tau) $ with $ k>1 $ are the holomorphic modular forms of weight $ 2k $, while $ E_2(\tau) $ is only quasi-modular:
\begin{align}
E_2\qty(\frac{a\tau+b}{c\tau+d}) = (c\tau+d)^2 E_2(\tau) - \frac{6i}{\pi} c(c\tau + d) \, .
\end{align}
Two Eisenstein series $ E_4(\tau) $ and $ E_6(\tau) $ generate the ring of holomorphic modular forms $ \mathcal{M}_*(\mathrm{SL}(2,\mathbb{Z})) = \bigoplus_{k \geq 0} \mathcal{M}_{2k}(\mathrm{SL}(2,\mathbb{Z})) $, where $ \mathcal{M}_{2k}(\mathrm{SL}(2,\mathbb{Z})) $ is the space of weight $ 2k $ modular forms. In other words, $ \mathcal{M}_{2k}(\mathrm{SL}(2,\mathbb{Z})) $ can be written as
\begin{align}
\mathcal{M}_{2k}(\mathrm{SL}(2,\mathbb{Z})) = \bigoplus_{4a+6b=2k} \mathbb{C} E_4(\tau)^a E_6(\tau)^b \, .
\end{align}

As a related function with the Eisenstein series, we define the Dedekind eta function as
\begin{align}\label{dedekind-eta}
\eta(\tau) = q^{1/24} \prod_{n=1}^\infty (1-q^n) \, .
\end{align}
Its 24th power $ \Delta(\tau) = \eta(\tau)^{24} = (E_4(\tau)^3 - E_6(\tau)^2)/1728 $ is a weight 12 modular form called \emph{modular discriminant} and $ \eta(\tau) $ itself has following modular transformation properties:
\begin{align}
\eta(\tau+1) = e^{\pi i/12} \eta(\tau) \, , \quad
\eta(-1/\tau) = \sqrt{-i\tau} \eta(\tau) \, .
\end{align}

\subsection{Jacobi forms} \label{app:jacobi}

There is a generalization of the modular forms including additional fugacities. 
A function $ \varphi_{k,m} : \mathcal{H} \times \mathbb{C} \to \mathbb{C} $ is called a \emph{Jacobi form} \cite{Eicher:Zagier} if it has two transformation properties
\begin{alignat}{2}
    \varphi_{k,m}\qty(\frac{a\tau+b}{c\tau+d}, \frac{z}{c\tau+d}) &= (c\tau+d)^k e^{\frac{2\pi i m c z^2}{c\tau+d}} \varphi_{k,m}(\tau, z) \ && \text{ for }\ \mqty(a & b \\ c & d) \in \mathrm{SL}(2, \mathbb{Z}) \, , \label{Jacobi-modular} \\
    \varphi_{k,m}(\tau, z + \lambda \tau + \mu) &= e^{-2\pi i m (\lambda^2\tau + 2\lambda z)} \varphi_{k,m}(\tau, z) \ && \text{ for }\ \lambda, \mu \in \mathbb{Z} \, ,
\end{alignat}
and a Fourier expansion of the form
\begin{align}
    \varphi_{k,m}(\tau, z) = \sum_{n,r} c(n, r) q^n e^{2\pi i r z} \, ,
\end{align}
where $ k\in \mathbb{Z} $ is called the \emph{weight} and $ m \in \mathbb{Z}_{\geq 0} $ is called the \emph{index} or \emph{level} of the Jacobi form. When $ m=0 $, $ \varphi_{k,m} $ is independent of $ z $ and reduces to a modular form of weight $ k $. $ \varphi_{k,m} $ is called a holomorphic Jacobi form if $ c(n, r) = 0 $ unless $ 4mn \geq r^2 $, a cusp Jacobi form if $ c(n, r) = 0 $ unless $ 4mn > r^2 $, and a weak Jacobi form if $ c(n, r) = 0 $ unless $ n \geq 0 $.

Let $ J_{k,m} $ be a space of weak Jacobi forms of weight $ k $ and level $ m $. The ring of weak Jacobi form $ J_{*,*} = \bigoplus_{k,m} J_{k,m} $ is freely generated over the ring of modular forms $ \mathcal{M}_*(\mathrm{SL}(2,\mathbb{Z})) $, whose generators are
\begin{align}
\varphi_{-2,1}(\tau, z) = -\frac{\theta_1(\tau, z)^2}{\eta(\tau)^6} \, , \quad
\varphi_{0,1}(\tau, z) = 4 \sum_{i=2}^4 \frac{\theta_i(\tau, z)^2}{\theta_i(\tau, 0)^2} \, ,
\end{align}
where $ \theta_i(\tau, x) $ are Jacobi theta functions defined by
\begin{align}
\begin{alignedat}{2}\label{Jacobi-theta}
&\theta_1(\tau, x) = -i \sum_{n \in \mathbb{Z}} (-1)^{n} q^{\frac{1}{2}(n+1/2)^2} y^{n+1/2} \, , \quad
&&\theta_2(\tau, x) = \sum_{n \in \mathbb{Z}} q^{\frac{1}{2}(n+1/2)^2} y^{n+1/2} \, , \\
&\theta_3(\tau, x) = \sum_{n \in \mathbb{Z}} q^{\frac{n^2}{2}} y^n \, , \quad
&&\theta_4(\tau, x) = \sum_{n \in \mathbb{Z}} (-1)^n q^{\frac{n^2}{2}} y^n \, ,
\end{alignedat}
\end{align}
for $ y = e^{2\pi i x} $. In other words, any weak Jacobi form $ \varphi_{k,m} $ can be written as
\begin{align}
J_{k,m} \ni \varphi_{k,m}(\tau, z) = \underset{a_3 + a_4 = m, a_i \in \mathbb{Z}_{\geq 0}}{\sum_{4a_1+6a_2-2a_3 = k}} C_{a_i} E_4(\tau)^{a_1} E_6(\tau)^{a_2} \varphi_{-2,1}(\tau, z)^{a_3} \varphi_{0,1}(\tau, z)^{a_4}
\end{align}
for some $ C_{a_i} \in \mathbb{C} $. We also frequently use
\begin{align}
\varphi_{-1,1/2}(\tau, z) = i \frac{\theta_1(\tau, z)}{\eta(\tau)^3} \, ,
\end{align}
which satisfies $ \varphi_{-1,1/2}(\tau, z)^2 = \varphi_{-2,1}(\tau, z) $.

The notion of weak Jacobi forms is further generalized to Weyl invariant Jacobi forms \cite{Wirthmuller}. Let $ \mathfrak{g} $ be a Lie algebra of rank $ l $, $ \mathfrak{h}_{\mathbb{C}} \cong \mathbb{C}^l $ be the complexification of the Cartan subalgebra, $ W $ be its Weyl group, $ Q^\vee $ be the coroot lattice, and $ P $ be its weight lattice. Denote $ \langle \cdot, \cdot \rangle $ a Killing form on $ \mathfrak{h}_\mathbb{C} $ normalized to $ 2 $ for the shortest coroot. A \emph{Weyl invariant Jacobi form of weight $ k $ and index $ m $} is a function $ \varphi_{k,m} : \mathcal{H} \times \mathfrak{h}_\mathbb{C} \to \mathbb{C} $ satisfying following conditions.
\begin{itemize}
	\item[(i)] Weyl-invariance: for $ w \in W $,
	\begin{align}
	\varphi_{k,m}(\tau, wz) = \varphi_{k,m}(\tau, z) \, .
	\end{align}
	
	\item[(ii)] Modularity: for $ \smqty(a & b \\ c & d) \in \mathrm{SL}(2,\mathbb{Z}) $,
	\begin{align}
	\varphi_{k,m}\qty(\frac{a\tau+b}{c\tau+d}, \frac{z}{c\tau+d}) = (c\tau+d)^k \exp(\frac{\pi i m c}{c\tau+d} \langle z, z \rangle ) \varphi_{k,m}(\tau, z) \, .
	\end{align}

	\item[(iii)] Quasi-periodicity: for $ \lambda, \mu \in Q^\vee $,
	\begin{align}
	\varphi_{k,m}(\tau, z + \lambda \tau + \mu) = \exp(-\pi i m \qty[\langle \lambda, \lambda \rangle \tau + 2\langle \lambda, z \rangle] ) \varphi_{k,m}(\tau, z) \, .
	\end{align}
	
	\item[(iv)] Fourier expansion:
	\begin{align}
	\varphi_{k,m}(\tau, z) = \sum_{n=0}^\infty \sum_{\ell \in P} c(n, \ell) q^n e^{2\pi i \langle \ell, z \rangle} \, .
	\end{align}
\end{itemize}
The weak Jacobi forms defined above are $ \mathfrak{g} = A_1 $ case.

Let $ J_{k,m}(\mathfrak{g}) $ be the space of the $ \mathfrak{g} $ Weyl invariant Jacobi forms with weight $ k $ and index $ m $. 
Then, for a simple Lie algebra except for $ E_8 $, the bigraded ring,
\begin{align}
J_{*,*}(\mathfrak{g}) = \bigoplus_{k,m \in \mathbb{Z}} J_{k,m}(\mathfrak{g})
\end{align}
is freely generated by $ l+1 $ fundamental Weyl invariant Jacobi forms over the ring of modular forms $ \mathcal{M}_*(\mathrm{SL}(2,\mathbb{Z})) $. The Wirthm\"uller's theorem \cite{Wirthmuller} provides weights and indices for fundamental Weyl invariant Jacobi forms of simple Lie algebras except for $ E_8 $ as we list in Table~\ref{table:Jacobi-data}.
\begin{table}
	\centering
	\begin{Tabular}{c|c}
		$ \mathfrak{g} $ & $ (-k, m) $ \\ \hline
		$ A_l $ & $ (0, 1), (j, 1) $ for $ 2 \leq j \leq l+1 $ \\
		$ B_l $ & $ (2j, 1) $ for $ 0 \leq j \leq l $ \\
		$ C_l $ & $ (0, 1), (2, 1), (4, 1), (2j, 2) $ for $ 3 \leq j \leq l $ \\
		$ D_l $ & $ (0, 1), (2, 1), (4, 1), (l, 1), (2j, 2) $ for $ 3 \leq j \leq l-1 $ \\
		$ E_6 $ & $ (0, 1), (2, 1), (5, 1), (6, 2), (8, 2), (9, 2), (12, 3) $ \\
		$ E_7 $ & $ (0, 1), (2, 1), (6, 2), (8, 2), (10, 2), (12, 3), (14, 3), (18, 4) $ \\
		$ F_4 $ & $ (0, 1), (2, 1), (6, 2), (8, 2), (12, 3) $ \\
		$ G_2 $ & $ (0, 1), (2, 1), (6, 2) $ \\
	\end{Tabular}
	\caption{Weights and indices for the fundamental Weyl invariant Jacobi forms} \label{table:Jacobi-data}
\end{table}
Although the theorem does not give explicit form of the Jacobi forms, generators of Weyl invariant Jacobi forms for each Lie algebra have been studied in many literatures \cite{Bertola:Jacobi:phd, BERTOLA200019, BERTOLA2000213, Adler:D8, Adler:F4, Satake:1993cp}. The $ E_8 $ is exceptional case for Wirthm\"uller's theorem, but its Weyl invariant Jacobi forms are also studied recently \cite{Sakai:2011xg, Sakai:2017ihc, Wang:2018fil, Sun:2021ije}. See also \cite{Duan:2020imo, Kim:2018gak} for a review in physics liturature. Here, we give a construction of Weyl invariant Jacobi forms used in this paper.

Let us consider $ \mathfrak{g} = A_l $. The weight $ -k $ Jacobi form $ \varphi_{k}^{A_l} \in J_{-k,1}(A_l) $ is given by
\begin{align}
\varphi_{k}^{A_l} = \left. \mathcal{Z}^{l+1-k} \prod_{j=1}^{l+1} \frac{i \theta_1(x_j)}{\eta^3} \right|_{\sum x_j=0}, \quad (k = 0, 2, 3, \cdots, l+1)
\end{align}
where
\begin{align}
\mathcal{Z} = \frac{1}{2\pi i} \qty( \sum_{j=1}^{l+1} \frac{\partial}{\partial x_j} + \frac{\pi^2}{3} E_2(\tau) \sum_{j=1}^{l+1} x_j ) \, .
\end{align}
The orthogonal basis $ x_j $'s are related with the Dynkin basis $ \phi_i $'s by $ x_1 = \phi_1 $, $ x_j = -\phi_{j-1} + \phi_j $ for $ 2 \leq j \leq l $ and $ x_{l+1} = -\phi_l $. In particular, we use
\begin{align}
\varphi_3^{A_2} &= \qty(\chi_\mathbf{3} - \chi_{\overline{\mathbf{3}}}) + \qty(\chi_{\overline{\mathbf{6}}} - \chi_\mathbf{6} + 7\chi_\mathbf{3} - 7\chi_{\overline{\mathbf{3}}})q + \mathcal{O}(q^2), \\
\varphi_2^{A_2} &= \frac{1}{2} \qty[ \qty(6 - \chi_\mathbf{3} - \chi_{\overline{\mathbf{3}}} ) + \qty(42 + 6\chi_\mathbf{8} - \chi_\mathbf{6} - \chi_{\overline{\mathbf{6}}} - 13\chi_\mathbf{3} - 13\chi_{\overline{\mathbf{3}}} )q + \mathcal{O}(q^2) ], \nonumber \\
\varphi_0^{A_2} &= \frac{1}{4} \qty[ \qty(18 + \chi_\mathbf{3} + \chi_{\overline{\mathbf{3}}}) + \qty(342 + 18\chi_\mathbf{8} + \chi_\mathbf{6} + \chi_{\overline{\mathbf{6}}} - 83\chi_\mathbf{3} - 83\chi_{\overline{\mathbf{3}}})q + \mathcal{O}(q^2) ], \nonumber
\end{align}
for $ \mathfrak{g}=A_2 $ in section \ref{subsec:su3-symm} to write the modular ansatz for the $ SU(3) + 1\mathbf{sym} + 1\mathbf{\Lambda}^2 $ LST, where $ \chi_{\mathbf{R}} $ denotes character of $ SU(3) $ for representation $ \mathbf{R} $.\footnote{Note that $ \varphi_0^{A_2} $ in our paper is $ -6\varphi_0 $ defined in Appendix~B of \cite{DelZotto:2017mee}.}

Next, to study the $ D_l $ Jacobi forms, we first consider the $ B_l $ Jacobi forms. The generators of $ B_l $ Jacobi forms $ \varphi_{2j}^{B_l} \in J_{-2j,1} $ can be computed from the generating function
\begin{align}
\prod_{j=1}^l \frac{i\theta_1(v - x_i)}{\eta^3} \frac{i\theta_1(v + x_i)}{\eta^3}
= \qty(i \frac{\theta_1(v)}{\eta^3})^{2l} \sum_{j=0}^l \frac{\wp^{(2j-2)}(v)}{(2j-1)!} \varphi_{2j}^{B_l}(x_1,\cdots,x_l) \, ,
\end{align}
where $ j=0$ term in the summation is understood as $ \varphi_0^{B_l}(x_1,\cdots,x_l) $, and $ \wp $ is the Weierstra\ss\ $ \wp $ function defined as
\begin{align}
\wp(z) = \frac{\theta_3(0)^2 \theta_2(0)^2}{4} \frac{\theta_4(z)^2}{\theta_1(z)^2} - \frac{1}{12} \qty( \theta_3(0)^4 + \theta_2(0)^4 ).
\end{align}
Then the $ l-3 $ generators of $ D_l $ Jacobi forms with index $ 2 $ is identified with $ B_l $ Jacobi forms:
\begin{align}
\varphi_{-k,2}^{D_l} = \varphi_{k}^{B_l} \quad (k = 6, 8, \cdots, 2l-2) \, ,
\end{align}
where $ \varphi_{-k,2}^{D_l} \in J_{-k,2}(D_l) $ and $ \varphi_k^{B_l} \in J_{-k,1}(B_l) $. The index $ 1 $ generators are
\begin{align}
\varphi_{-n,1}^{D_l} &= \prod_{j=1}^{l} \frac{\theta_1(x_j)}{\eta^3} \, , \quad
\varphi_{-4,1}^{D_l} = \frac{1}{\eta^{12}} \qty( \frac{\prod_{j=1}^l \theta_3(x_j)}{\theta_3(0)^{l-4}} - \frac{\prod_{j=1}^l \theta_4(x_j)}{\theta_4(0)^{l-4}} - \frac{\prod_{j=1}^l \theta_2(x_j)}{\theta_2(0)^{l-4}} ), \nonumber \\
\varphi_{-2,1}^{D_l} &= \frac{\theta_3(0)^4+\theta_4(0)^4}{\eta^{12}} \qty(\frac{\prod_{j=1}^l \theta_3(x_j)}{\theta_3(0)^{l-4}} - \frac{\prod_{j=1}^l \theta_4(x_j)}{\theta_4(0)^{l-4}} + \frac{2\prod_{j=1}^l \theta_2(x_j)}{\theta_2(0)^{l-4}} ) \nonumber \\
&\quad - \frac{3\theta_2(0)^4}{\eta^{12}} \qty(\frac{\prod_{j=1}^l \theta_3(x_j)}{\theta_3(0)^{l-4}} + \frac{\prod_{j=1}^l \theta_4(x_j)}{\theta_4(0)^{l-4}} ) \, , \nonumber \\
\varphi_{0,1}^{D_l} &= \frac{1}{\eta^{12}} \qty( \frac{\prod_{j=1}^l \theta_3(x_j)}{\theta_3(0)^{l-12}} - \frac{\prod_{j=1}^l \theta_4(x_j)}{\theta_4(0)^{l-12}} - \frac{\prod_{j=1}^l \theta_2(x_j)}{\theta_2(0)^{l-12}} ) \, ,
\end{align}
where $ \varphi_{-k,1}^{D_l} \in J_{-k,1}(D_l) $. These level 1 Jacobi forms are used to construct the 1-string elliptic genus of the $ SO(32) $ heterotic LST in subsection~\ref{subsubsec:SO32LST}.

Lastly, we review the $ E_8 $ Jacobi forms. The bigraded ring $ J_{*,*}(E_8) $ for the $ E_8 $ Weyl invariant Jacobi forms are contained in a polynomial algebra over $ \mathcal{M}_{*}(\mathrm{SL}(2,\mathbb{Z})) $ generated by nine functions \cite{Wang:2018fil}:
\begin{align}
J_{*,*}(E_8) \subsetneq \mathcal{M}_*(\mathrm{SL}(2,\mathbb{Z}))[ A_1, A_2, A_3, A_4, A_5, B_2, B_3, B_4, B_6] \, ,
\end{align}
where \cite{Sakai:2011xg}
\begin{align}\label{E8-Jacobi}
A_1 &= \Theta_{E_8}(\tau, x) = \frac{1}{2}\sum_{k=1}^4 \prod_{j=1}^8 \theta_k(\tau, x_j) \, , \quad
A_4 = \Theta_{E_8}(\tau, 2x) \, , \\
A_n &= \frac{n^3}{n^3+1}\qty(\Theta_{E_8}(n\tau, nx) + \frac{1}{n^4}\sum_{k=0}^{n-1} \Theta_{E_8}(\tfrac{\tau+k}{n}, x) ) \quad (n=2, 3, 5) \, , \nonumber \\
B_2 &= \frac{32}{5}\qty(e_1(\tau) \Theta_{E_8}(2\tau, 2x) + \frac{1}{2^4} e_3(\tau) \Theta_{E_8}(\tfrac{\tau}{2}, x) + \frac{1}{2^4} \Theta_{E_8}(\tfrac{\tau+1}{2}, x) ) \, , \nonumber \\
B_3 &= \frac{81}{80} \qty(h(\tau)^2 \Theta_{E_8}(3\tau, 3x) - \frac{1}{3^5}\sum_{k=0}^2 h(\tfrac{\tau+k}{3})^2 \Theta_{E_8}(\tfrac{\tau+k}{3}, x)) \, , \nonumber \\
B_4 &= \frac{16}{15} \bigg(\theta_4(2\tau,0)^4 \Theta_{E_8}(4\tau, 4x) - \frac{1}{2^4}\theta_4(2\tau,0)^4 \Theta_{E_8}(\tau+\tfrac{1}{2}, 2x) \nonumber \\
&\qquad \quad - \frac{1}{2^{10}}\sum_{k=0}^3 \theta_2(\tfrac{\tau+k}{2},0)^4 \Theta_{E_8}(\tfrac{\tau+k}{4}, x) \bigg) \, , \nonumber \\
B_6 &= \frac{9}{10} \bigg( h(\tau)^2 \Theta_{E_8}(6\tau, 6x) + \frac{1}{2^4} \sum_{k=0}^1 h(\tau+k)^2 \Theta_{E_8}(\tfrac{3\tau+3k}{2}, 3x) \nonumber \\
&\qquad \quad - \frac{1}{3^5}\sum_{k=0}^2 h(\tfrac{\tau+k}{3})^2 \Theta_{E_8}(\tfrac{2\tau+2k}{3}, 2x) - \frac{1}{2^4 \cdot 3^5} \sum_{k=0}^5 h(\tfrac{\tau+k}{3})^2 \Theta_{E_8}(\tfrac{\tau+k}{6}, x) \bigg) \, . \nonumber
\end{align}
Here,
\begin{alignat}{2}
&e_1(\tau) = \frac{1}{12}\qty(\theta_3(\tau,0)^4 + \theta_4(\tau,0)^4) \, , \quad
&&e_2(\tau) = \frac{1}{12}\qty(\theta_2(\tau,0)^4 - \theta_4(\tau,0)^4) \, , \\
&e_2(\tau) = \frac{1}{12}\qty(-\theta_2(\tau,0)^4 - \theta_3(\tau,0)^4) \, , \quad
&&h(\tau) = \theta_3(2\tau, 0) \theta_3(6\tau, 0) + \theta_2(2\tau, 0) \theta_2(6\tau, 0) \, , \nonumber
\end{alignat}
$ A_n $ and $ B_n $ have index $ n $ and weight $ 4 $ and $ 6 $, repectively, and normalized such that $ A_n(\tau, 0) = E_4(\tau) $ and $ B_n(\tau, 0) = E_6(\tau) $. They are used to construct modular ansatz of $ E_8 \times E_8 $ LST in subsection~\ref{subsec:E8xE8}.

\section{Derivation of elliptic genera} \label{app:integral}

In this appendix, we present the details for elliptic genus computations of $ E_8 \times E_8 $ heterotic LST, $ SO(32) $ heterotic LST and $ SU(3) + 1\mathbf{sym} + 1\mathbf{\Lambda}^2 $ LST using the 2d ADHM constructions for the moduli spaces of (instanton) strings.

\subsection{Elliptic genus of \texorpdfstring{$ E_8 \times E_8 $}{E8×E8} heterotic LST} \label{app:E8E8}

We can evaluate the elliptic genera of the rank 1 $ E_8 \times E_8 $ heterotic LST from the 2d $ \mathcal{N}=(0,4) $ gauge theory description given in Figure~\ref{fig:E8E8-quiver}. The elliptic genus is given by the integration of 1-loop determinants of supermultiplets in 2d gauge theory over flat connections of the $ O(k_1) \times O(k_2) $ gauge group. Note that we also have to sum over disconnected sectors of the flat connections corresponding to the disconnected components of the orthogonal gauge group. For a $ O(k) $ group with $ k \geq 3 $, there are at most $ \lfloor k/2 \rfloor $ complex moduli $ u_I $ and in total eight disconnected sectors for flat connections, while $ O(2) $ has seven sectors consist of one continuous complex modulus and six discrete holonomies, and $ O(1) $ has four discrete sectors \cite{Kim:2014dza}. In total, $ (k_1,k_2) $-string elliptic genus is given by
\begin{align}
    Z_{(k_1,k_2)} = \sum_{I_1,I_2} \frac{1}{|W^{(I_1)}| \cdot |W^{(I_2)}|} \frac{1}{(2\pi i)^{r_1+r_2}} \oint Z_{\mathrm{1-loop}} \, ,
\end{align}
where $ I_1 $ and $ I_2 $ represents disconnected sectors of $ O(k_1) $ and $ O(k_2) $ flat conections, $ W^{(I_{1,2})} $ are corresponding Weyl group factors and $ r_{1,2} $ are number of continuous complex moduli. The integration contour is chosen by Jeffery-Kirwan residue (JK-residue for short) prescription as discussed in \cite{Benini:2013nda, Benini:2013xpa}. The 1-loop determinant $ Z_{\mathrm{1-loop}} $ is the collection of following 1-loop determinants
\begin{align}
    \begin{aligned}
        Z_{\mathrm{vec}}^{(j)} &= \qty( \prod_{I=1}^{r_j} \frac{2\pi \eta^2 du_I}{i} \frac{i \theta_1(2\epsilon_+)}{\eta}) \qty( \prod_{e \in \mathbf{R}_j} \frac{i \theta_1(e \cdot u)}{\eta} \frac{i \theta_1(2\epsilon_+ + e \cdot u)}{\eta} ) \, , \\
        Z_{\mathrm{sym,hyp}}^{(j)} &= \prod_{w\in\mathbf{sym}_j} \frac{(i\eta)^2}{\theta_1(\epsilon_{1,2} + w \cdot u)} \, , \quad
        Z_{\mathrm{fund,Fermi}}^{(j)} = \prod_{w \in\mathbf{fund}_j} \prod_{l=l_j}^{l_j+7} \frac{i \theta_1(m_l+w \cdot u)}{\eta} \, , \\
        Z_{\mathrm{bifund}} &= \prod_{w\in\mathbf{bifund}} \frac{\theta_1(\pm m_0 + \epsilon_- + w \cdot u)}{\theta_1(\pm m_0 - \epsilon_+ + w \cdot u)} \, ,
    \end{aligned}
\end{align}
for $ j=1,2 $, where $ \mathbf{R}_j $, $ \mathbf{sym}_j $ and $ \mathbf{fund}_j $ denotes root system, symmetric and fundamental representation of $ SO(k_j) $, repectively, $ \mathbf{bifund} $ is the bifundamental representation of $ SO(k_1) \times SO(k_2) $ and $ (l_1, l_2) = (1, 9) $.  
The details of the contour integral for $O(k)$ gauge group are explained in \cite{Kim:2014dza}, and we will use some of their results.

\paragraph{(1,0)-string}
The $ O(1) $ gauge group consists of four discrete flat connections labelled by $ u^I = 0, \frac{1}{2}, \frac{\tau+1}{2}, \frac{\tau}{2} $. For each sector, the 1-loop determinant is 
\begin{align}
Z_{(1,0)}^I = \frac{(i \eta)^2}{\theta_1(\epsilon_1 + 2u) \theta_1(\epsilon_2 + 2u^I)} \prod_{l=1}^8 \frac{i \theta_1(m_l + u^I)}{\eta} \, .
\end{align}
Thus, the $ (1,0) $-string elliptic genus is 
\begin{align}
Z_{(1,0)} = -\frac{1}{2} \sum_{I=1}^4 \frac{\prod_{l=1}^8 \theta_I(m_l)}{\eta^6 \theta_1(\epsilon_1) \theta_1(\epsilon_2)} \, ,
\end{align}
where $ 1/2 $ is the Weyl group factor.

\paragraph{(2,0)-string}
The $ O(2) $ gauge group has one continuous flat connection and six discrete flat connections. The contribution from the continuous sector is
\begin{align}
Z_{(2,0)}^{(0)}
= \frac{1}{2\pi i} \oint \frac{2\pi \eta^2 du}{i} \frac{i \theta_1(2\epsilon_+)}{\eta} \frac{(i\eta)^6}{\theta_1(\epsilon_{1,2}) \theta_1(\epsilon_{1,2} \pm 2u)} \prod_{l=1}^8 \frac{i \theta_1(m_l \pm u)}{\eta} \, .
\end{align}
The JK-residue comes from $ u = -\frac{\epsilon_{1,2}}{2} + u^I $, where $ u^I = 0, \frac{1}{2}, \frac{\tau+1}{2}, \frac{\tau}{2} $. In total, we compute\footnote{The overall sign is chosen by requiring the GV-invariant structure \eqref{GV}.}
\begin{align}
Z_{(2,0)}^{(0)} = \frac{1}{2\eta^{12} \theta_1(\epsilon_1) \theta_1(\epsilon_2)} \sum_{I=1}^4 \qty( \frac{\prod_{l=1}^8 \theta_I(m_l \pm \frac{\epsilon_1}{2})}{\theta_1(2\epsilon_1) \theta_1(\epsilon_2-\epsilon_1)} + \frac{\prod_{l=1}^8 \theta_I(m_l \pm \frac{\epsilon_2}{2})}{\theta_1(2\epsilon_2) \theta_1(\epsilon_1-\epsilon_2)} ) \, .
\end{align}
The six discrete sectors are
\begin{align}
Z_{(2,0)}^{(I,J)} &= \frac{i\theta_1(u^I+u^J)}{\eta} \frac{i \theta_1(2\epsilon_+ + u^I + u^J)}{\eta} \nonumber \\
&\quad \cdot \frac{(i\eta)^6}{\theta_1(\epsilon_{1,2}+2u^I) \theta_1(\epsilon_{1,2}+2u^J) \theta_1(\epsilon_{1,2}+u^I+u^J)} \prod_{l=1}^8 \frac{i\theta_1(m_l + u^{I,J})}{\eta},
\end{align}
where $ (I,J)=(1,2), (1, 3), (1, 4), (2, 3), (2, 4), (3, 4) $ labels the six sectors of flat connections for $ (u^1, u^2, u^3, u^4) = (0, \frac{1}{2}, \frac{\tau+1}{2}, \frac{\tau}{2}) $.
These sectors can be rewritten as
\begin{align}
Z_{(2,0)}^{(I,J)} & = \frac{\theta_{\sigma(I,J)}(0) \theta_{\sigma(I,J)}(2\epsilon_+) \prod_{l=1}^8 \theta_I(m_l) \theta_J(m_l)}{\eta^{12} \theta_1(\epsilon_{1,2})^2 \theta_{\sigma(I,J)}(\epsilon_1) \theta_{\sigma(I,J)}(\epsilon_2)} \, ,
\end{align}
where
\begin{alignat}{3}\label{sigma}
\begin{aligned}
&\sigma(I, J) = \sigma(J, I) \, , \
&&\sigma(I, I) = 0 \, , \
&&\sigma(1, I) = I \, , \\
&\sigma(2, 3) = 4 \, , \
&&\sigma(2, 4) = 3 \, , \
&&\sigma(3, 4) = 2 \, .
\end{aligned}
\end{alignat}
After dividing it by the Weyl group factor, the $ (2,0) $-string elliptic genus is given by
\begin{align}
Z_{(2,0)} = \frac{1}{2} Z_{(2,0)}^{(0)} + \frac{1}{4} \sum_{I=1}^4 \sum_{J=I+1}^4 Z_{(2,0)}^{(I,J)} \, .
\end{align}

\paragraph{(1,1)-string}
Let $ u_1 $ and $ u_2 $ label the flat connections for two $ O(1) $ gauge groups.
As we explained in the (1,0)-string case above, there are four distinct flat connections labelled by $ u^I = 0, \frac{1}{2}, \frac{\tau+1}{2}, \frac{\tau}{2} $ for each $O(1)$ gauge group.
The 1-loop determinant is
\begin{align}
\begin{aligned}
Z_{(1,1)}^{(I,J)} &=  \frac{(i\eta)^4}{\theta_1(\epsilon_1+2u^{I,J}) \theta_1(\epsilon_2+2u^{I,J})}  \frac{\theta_1(\pm m + \epsilon_- + u^I + u^I)}{\theta_1(\pm m - \epsilon_+ + u^I+u^J)} \\
&\quad \cdot \qty(\prod_{l=1}^8 \frac{i\theta_1(m_l+u^I)}{\eta}) \qty(\prod_{l=9}^{16} \frac{i\theta_1(m_l+u^J)}{\eta}) \, ,
\end{aligned}
\end{align}
where $ I,J = 1,2,3,4 $ when $ u^{I,J} = 0, \frac{1}{2}, \frac{\tau+1}{2}, \frac{\tau}{2} $, repectively. By dividing it by the Weyl group factor, the $ (1,1) $-string elliptic genus is
\begin{align}
Z_{(1,1)} = \frac{1}{4} \sum_{I,J=1}^4 \frac{\prod_{l=1}^8 \theta_I(m_l) \cdot \prod_{l=9}^{16} \theta_J(m_l)}{\eta^{12} \theta_1(\epsilon_1)^2 \theta_1(\epsilon_2)^2} \frac{\theta_{\sigma(I,J)}(\pm m_0 + \epsilon_-)}{\theta_{\sigma(I,J)}(\pm m_0 - \epsilon_+)} \, .
\end{align}

\paragraph{(2,1)-string}
To compute the $ (2,1) $-string elliptic genus, we need to consider all combinations of $ O(2) $ and $ O(1) $ flat connections. First, from the continuous sector of $ O(2) $ and four discrete sectors of $ O(1) $, we have
\begin{align}
Z_{(2,1)}^{(0)} 
&= \frac{1}{2\pi i} \oint \frac{2\pi \eta^2 du_1}{i} \frac{i\theta_1(2\epsilon_+)}{\eta} \frac{(i\eta)^6}{\theta_1(\epsilon_{1,2}) \theta_1(\epsilon_{1,2} \pm 2u_1)} \frac{(i \eta)^2}{\theta_1(\epsilon_1+2u^J) \theta_1(\epsilon_2+2u^J)} \nonumber \\
&\quad \cdot \frac{\theta_1(\pm m + \epsilon_- + u_1 + u^J) \theta_1(\pm m + \epsilon_- - u_1 + u^J)}{\theta_1(\pm m - \epsilon_+ + u_1 + u^J) \theta_1(\pm m - \epsilon_+ - u_1 + u^J)} \\
&\quad \cdot \qty(\prod_{l=1}^8 \frac{i\theta_1(m_l\pm u_1)}{\eta}) \qty(\prod_{l=9}^{16} \frac{i\theta_1(m_l+u^J)}{\eta}) \, , \nonumber
\end{align}
where $ u^J = 0, \frac{1}{2}, \frac{\tau+1}{2}, \frac{\tau}{2} $ labels $ O(1) $ flat connections. Then $ Z_{(2,1)}^{(0)} $ is given by sum of following two JK-residues:
\begin{itemize}
	\item $ u_1 = -\frac{\epsilon_{1,2}}{2} + u^I $ for $ u^I = 0, \frac{1}{2}, \frac{\tau+1}{2}, \frac{\tau}{2} $
	\begin{align}
	\sum_{I,J=1}^4 \frac{-\prod_{l=1}^8 \theta_I(m_l \pm \frac{\epsilon_1}{2}) \cdot \prod_{l=9}^{16} \theta_J(m_l)}{2\eta^{18} \theta_1(\epsilon_{1,2})^2 \theta_1(2\epsilon_1) \theta_1(\epsilon_2-\epsilon_1)} \frac{\theta_{\sigma(I,J)}(\pm m_0 + \epsilon_1 - \frac{\epsilon_2}{2})}{\theta_{\sigma(I,J)}(\pm m_0 - \epsilon_1 - \frac{\epsilon_2}{2})} + (\epsilon_1 \leftrightarrow \epsilon_2)
	\end{align}
	\item $ u_1 = \pm m + \epsilon_+ -u^J $
	\begin{align}
	\sum_{I=1}^4 \frac{-\prod_{l=1}^8 \theta_I(m_l \pm (m_0 + \epsilon_+)) \cdot \prod_{l=9}^{16} \theta_I(m_l)}{\eta^{18} \theta_1(\epsilon_{1,2}) \theta_1(2m_0) \theta_1(2m_0+2\epsilon_+) \theta_1(2m_0+2\epsilon_++\epsilon_{1,2})} + (m_0 \to -m_0)
	\end{align}
\end{itemize}
Next, there are combinations of six discrete sectors for $ O(2) $ and four discrete sectors for $ O(1) $. If we denote $ (u^I, u^J) = (0, \frac{1}{2}), (0, \frac{\tau+1}{2}), (0, \frac{\tau}{2}), (\frac{1}{2}, \frac{\tau+1}{2}), (\frac{1}{2}, \frac{\tau}{2}), (\frac{\tau+1}{2}, \frac{\tau}{2}) $ as the $ O(2) $ discrete flat connections and $ u^K = 0, \frac{1}{2}, \frac{\tau+1}{2}, \frac{\tau}{2} $ as the $ O(1) $ flat connections, the 1-loop determinant is
\begin{align}
Z_{(2,1)}^{(I,J,K)}
&= \frac{i\theta_1(u^I+u^J)}{\eta} \frac{i\theta_1(2\epsilon_++u^I+u^J)}{\eta} \frac{(i\eta)^6}{\theta_1(\epsilon_{1,2}+2u^{I,J}) \theta_1(\epsilon_{1,2}+u^I+u^J)} \nonumber \\
&\quad \cdot \frac{(i\eta)^2}{\theta_1(\epsilon_{1,2}+2u^K)} \frac{\theta_1(\pm m + \epsilon_- + u^I + u^K) \theta_1(\pm m + \epsilon_- + u^J + u^K)}{\theta_1(\pm m - \epsilon_+ + u^I + u^K) \theta_1(\pm m - \epsilon_+ + u^J + u^K)}  \nonumber \\
&\quad \cdot \qty(\prod_{l=1}^8 \frac{i\theta_1(m_l+u^{I,J})}{\eta}) \qty(\prod_{l=9}^{16} \frac{i\theta_1(m_l+u^K)}{\eta}) \, .
\end{align}
Then we get
\begin{align}
Z_{(2,1)}^{(I,J,K)}
&= -\frac{\theta_{\sigma(I,J)}(0) \theta_{\sigma(I,J)}(2\epsilon_+)}{\eta^{18} \theta_1(\epsilon_{1,2})^3 \theta_{\sigma(I,J)}(\epsilon_{1,2})} \frac{\theta_{\sigma(I,K)}(\pm m_0 + \epsilon_-) \theta_{\sigma(J,K)}(\pm m_0 + \epsilon_-)}{\theta_{\sigma(I,K)}(\pm m_0 - \epsilon_+) \theta_{\sigma(J,K)}(\pm m_0 - \epsilon_+)} \nonumber \\
& \quad \cdot \prod_{l=1}^8 \theta_I(m_l) \theta_J(m_l) \cdot \prod_{l=9}^{16} \theta_K(m_l) \, .
\end{align}
By dividing it by the Weyl group factor, the $ (2,1) $-string elliptic genus can be written as
\begin{align}
Z_{(2,1)} = \frac{1}{4} Z_{(2,1)}^{(0)} + \frac{1}{8} \sum_{K=1}^4 \sum_{I<J}^4 Z_{(2,1)}^{(I,J,K)} \, .
\end{align}

\subsection{Elliptic genus of \texorpdfstring{$ SO(32) $}{SO(32)} heterotic LST} \label{app:SO32}

In this appendix, we compute the elliptic genus of the rank 1 $ SO(32) $ heterotic LST based on the 2d gauge theory description given in Figure~\ref{fig:SO32}. The 2d theory has orthogonal gauge group, so $ k $-string elliptic genus can be written as
\begin{align}
    Z_k &= \sum_K \frac{1}{|W^{(K)}|} \frac{1}{(2\pi i)^r} \oint \qty( \prod_{I=1}^{r} \frac{2\pi \eta^2 du_I}{i} \frac{i \theta_1(2\epsilon_+)}{\eta}) \qty( \prod_{e \in \mathbf{R}} \frac{i \theta_1(e \cdot u)}{\eta} \frac{i \theta_1(2\epsilon_+ + e \cdot u)}{\eta} ) \nonumber \\
    & \quad \qty( \prod_{\rho\in\mathbf{sym}} \frac{(i\eta)^2}{\theta_1(\epsilon_{1,2}+\rho(u))} \frac{(i\eta)^2}{\theta_1(\pm m_0 - \epsilon_+ + \rho(u))} ) \qty( \prod_{\rho \in \mathbf{anti}} \frac{i^2 \theta_1(\pm m_0 + \epsilon_- + \rho(u))}{\eta^2} ) \nonumber \\
    &\quad \qty(\prod_{\rho\in\mathbf{bifund}} \frac{\theta_1(m_0 + \rho(a,u))}{\theta_1(\epsilon_+ + \rho(a,u))}) \qty(\prod_{\rho\in\mathbf{fund}} \prod_{l=1}^{16} \frac{i\theta_1(m_l+\rho(u))}{\eta}) \, ,
\end{align}
for a number $ r $ of continuous complex moduli $ u_I $ as explained in \cite{Kim:2014dza} and briefly reviewed in previous subsection. Here, $ K $ denotes the disconnected sectors of $ O(k) $ flat connections, $ W^{(K)} $ is corresponding Weyl group, $ \mathbf{R} $, $ \mathbf{sym} $, $ \mathbf{anti} $ and $ \mathbf{fund} $ are $ SO(k) $ root system, symmetric, antisymmetric (i.e., adjoint) and fundamental representations, repectively, and $ \mathbf{bifund} $ is the bifundamental representation in $ SO(k) \times Sp(1) $.

\paragraph{1-string}
There are 4 discrete flat connections in the $ O(1) $ gauge group, labelled by $ u^I = 0, \frac{1}{2}, \frac{\tau+1}{2}, \frac{\tau}{2} $. 
The 1-string elliptic genus by summing over the contributions for these flat connections is
\begin{align}
Z_1
= -\sum_{I=1}^4 \frac{\theta_I(m_0 \pm a) \prod_{l=1}^{16} \theta_I(m_l)}{2\eta^{12} \theta_1(\epsilon_1) \theta_1(\epsilon_2) \theta_1(\pm m_0 - \epsilon_+)\theta_I(\epsilon_+ \pm a)} \, ,
\end{align}
where $ 1/2 $ factor comes from the Weyl group.

\paragraph{2-string}
There one continuous sector and 6 discrete sectors of the $ O(2) $ flat connections. The continuous sector contribution is
\begin{align}
Z_2^{(0)} &= \oint \frac{2\pi \eta^2 du}{i} \frac{i\theta_1(2\epsilon_+)}{\eta} \frac{(i\eta)^6}{\theta_1(\epsilon_{1,2}) \theta_1(\epsilon_{1,2}\pm 2u)} \nonumber \\
&\quad \cdot \frac{(i\eta)^6}{\theta_1(\pm m_0 - \epsilon_+) \theta_1(\pm m_0 - \epsilon_+ + 2u) \theta_1(\pm m_0 - \epsilon_+ - 2u)} \frac{i^2 \theta_1(\pm m_0 + \epsilon_-)}{\eta^2} \nonumber \\
&\quad \cdot \frac{\theta_1(m_0 \pm a + u) \theta_1(m_0 \pm a - u)}{\theta_1(\epsilon_+ \pm a + u) \theta_1(\epsilon_+ \pm a - u)} \prod_{l=1}^{16} \frac{i^2\theta_1(m_l \pm u)}{\eta^2} \, .
\end{align}
The integral can be evaluated by summing over the following JK-residues:
\begin{itemize}
	\item $ \epsilon_{1,2} + 2u = 0, 1, \tau, \tau+1 $
	\begin{align}
	Z_2^{(1)} &= \sum_{J=1}^4 \frac{\prod_{l=1}^{16} \theta_J(m_l \pm \frac{\epsilon_1}{2})}{2\eta^{24} \theta_1(\epsilon_{1,2}) \theta_1(2\epsilon_1) \theta_1(\epsilon_2-\epsilon_1) \theta_1(\pm m_0 - \epsilon_+) \theta_1(\pm m_0 - \epsilon_+ - \epsilon_1)} \nonumber \\
	& \qquad\quad \cdot \frac{\theta_J(m_0 + \frac{\epsilon_1}{2} \pm a) \theta_J(m_0 - \frac{\epsilon_1}{2} \pm a)}{\theta_J(\epsilon_+ + \frac{\epsilon_1}{2} \pm a) \theta_J(\epsilon_+ - \frac{\epsilon_1}{2} \pm a)} + (\epsilon_1 \leftrightarrow \epsilon_2)
	\end{align}
	
	\item $ \pm m_0 - \epsilon_+ + 2u = 0, 1, \tau, \tau+1 $
	\begin{align}
	Z_2^{(2)} &= -\sum_{J=1}^4 \frac{\prod_{l=1}^{16} \theta_J(m_l \pm \frac{m_0 + \epsilon_+}{2})}{2\eta^{24} \theta_1(\epsilon_{1,2}) \theta_1(2m_0) \theta_1(2m_0+2\epsilon_+) \theta_1(\epsilon_+ \pm m_0) \theta_1(m_0 + \epsilon_+ + \epsilon_{1,2})} \nonumber \\
	&\qquad \quad \cdot \frac{\theta_J(\frac{3m_0}{2} + \frac{\epsilon_+}{2} \pm a)}{\theta_J(\frac{m_0}{2} + \frac{3}{2}\epsilon_+ \pm a)} + (m_0 \to -m_0)
	\end{align}
	
	\item $ \epsilon_+ \pm a + u = 0 $
	\begin{align}
	Z_2^{(3)} &= \frac{\prod_{l=1}^{16} \theta_1(m_l \pm (\epsilon_+ + a))}{\eta^{24} \theta_1(\epsilon_{1,2}) \theta_1(2a) \theta_1(\epsilon_{1,2}+2a) \theta_1(2\epsilon_+ + 2a) \theta_1(2\epsilon_+ + \epsilon_{1,2} + 2a)} \nonumber \\
	&\quad \cdot \frac{\theta_1(\pm m_0 + \epsilon_-)}{\theta_1(\pm m_0 -3\epsilon_+ - 2a)} + (a \to -a) \, .
	\end{align}
\end{itemize}
The contributions coming from the discrete sectors are
\begin{align}
Z_2^{(I,J)} &= \frac{i \eta(u^I+u^J)}{\eta} \frac{i\theta_1(2\epsilon_++u^I+u^J)}{\eta} \frac{(i\eta)^6}{\theta_1(\epsilon_{1,2}+2u^{I,J}) \theta_1(\epsilon_{1,2}+u^I+u^J)} \nonumber \\
& \quad \cdot \frac{(i\eta)^6}{\theta_1(\pm m_0 - \epsilon_+ + 2u^{I,J}) \theta_1(\pm m_0 - \epsilon_+ + u^I+u^J)} \frac{i^2\theta_1(\pm m_0 + \epsilon_- + u^I + u^J)}{\eta^2} \nonumber \\
&\quad \cdot \frac{\theta_1(m_0 \pm a + u^{I,J})}{\theta_1(\epsilon_+ \pm a + u^{I,J})} \prod_{l=1}^{16} \frac{i^2 \theta_1(m_l + u^{I,J})}{\eta^2} \, ,
\end{align}
where $ (I,J)=(1,2),(1,3),(1,4),(2,3),(2,4),(3,4) $ corresponds to six flat connections $ (u^I, u^J) \allowbreak = (0, \frac{1}{2}), (0, \frac{\tau+1}{2}), (0, \frac{\tau}{2}), (\frac{1}{2}, \frac{\tau+1}{2}), (\frac{1}{2}, \frac{\tau}{2}), (\frac{\tau+1}{2}, \frac{\tau}{2}) $. This can be written as
\begin{align}
\begin{aligned}
Z_2^{(I,J)} &= \frac{\theta_{\sigma(I,J)}(0) \theta_{\sigma(I,J)}(2\epsilon_+) \theta_{\sigma(I,J)}(\pm m_0 + \epsilon_-) \theta_I(m_0\pm a)\theta_J(m_0 \pm a)}{\eta^{24} \theta_1(\epsilon_{1,2})^2 \theta_{\sigma(I,J)}(\epsilon_{1,2}) \theta_1(\pm m_0 - \epsilon_+)^2 \theta_{\sigma(I,J)}(\pm m_0 - \epsilon_+)} \\
&\quad \cdot \frac{\prod_{l=1}^{16} \theta_I(m_l) \theta_J(m_l)}{\theta_I(\epsilon_+ \pm a) \theta_J(\epsilon_+ \pm a)} \, ,
\end{aligned}
\end{align}
where $ \sigma(I,J) $ is defined in \eqref{sigma}. In total, the 2-string elliptic genus is
\begin{align}\label{SO32-2str}
Z_2
= \frac{1}{2} \sum_{I=1}^3 Z_2^{(I)} + \frac{1}{4} \sum_{I < J}^4 Z_2^{(I,J)} \, ,
\end{align}
where $ 1/2 $ and $ 1/4 $ are Weyl group factors.

\subsection{Elliptic genus of \texorpdfstring{$ SU(3) + 1\mathbf{sym} + 1\mathbf{\Lambda}^2 $}{SU3+1Sym+1Anti}}\label{app:AOA}

From the 2d gauge theory description given in Figure~\ref{fig:SU3-brane}, the elliptic genus of $ k $-string can be written as
\begin{align}
Z_k &= \frac{1}{k!} \frac{1}{(2\pi i)^k} \oint \qty(\prod_{I=1}^k \frac{2\pi \eta^2 du_I}{i} \frac{i \theta_1(2\epsilon_+)}{\eta}) \qty(\prod_{I\neq J} \frac{i\theta_1(u_{IJ})}{\eta} \frac{i\theta_1(2\epsilon_++u_{IJ})}{\eta}) \nonumber \\
&\quad \qty(\prod_{I,J} \frac{(i\eta)^2}{\theta_1(\epsilon_{1,2}+u_{IJ})}) \qty(\prod_I \prod_{j=1}^N \frac{(i\eta)^2}{\theta_1(\epsilon_+ \pm (u_I-a_j))}) \nonumber \\
&\quad \qty(\prod_{I \leq J} \frac{(i \eta)^2}{\theta_1(-\epsilon_+ \pm (u_I + u_J + m_2))}) \qty(\prod_{I < J} \frac{i^2 \theta_1(-\epsilon_- \pm (u_I + u_J + m_2))}{\eta^2}) \nonumber \\
&\quad  \qty(\prod_I \prod_{j=1}^N \frac{i\theta_1(u_I+a_j+m_2)}{\eta}) \qty(\prod_{I < J} \frac{(i\eta)^2}{\theta_1(-\epsilon_+ \pm (u_I + u_J + m_1))}) \nonumber \\
&\quad \qty(\prod_{I \leq J} \frac{i^2 \theta_1(-\epsilon_- \pm (u_I + u_J + m_1))}{\eta^2}) \qty(\prod_I \prod_{j=1}^N \frac{i\theta_1(u_I+a_j+m_1)}{\eta}) \, ,
\end{align}
where $ u_{IJ} = u_I - u_J $, $ a_{1,2,3} $ are $ U(3) $ chemical potentials, $ m_1 $ and $ m_2 $ are $ U(1)_S $ and $ U(1)_A $ chemical potentials. Here we focus on $ N=3 $ case, which gives the elliptic genera of strings in the $ SU(3) + 1\mathbf{sym} + 1\mathbf{\Lambda}^2 $ LST.

\paragraph{1-string}
The relevant poles are $ \epsilon_+ + u_1 - a_j = 0 $ and $ -\epsilon_+ + 2u_1 + m_2 = 0 $, and the contributions from these poles are
\begin{itemize}
	\item $ u_1 = -\epsilon_+ + a_j $
	\begin{align}
	-\sum_{j=1}^3 \frac{\theta_1(2a_j+m_1-\epsilon_+) \theta_1(2a_j + m_1 - \epsilon_+ - \epsilon_{1,2})}{\theta_1(\epsilon_{1,2}) \theta_1(2a_j+m_2-3\epsilon_+)} \prod_{k \neq j}^3  \frac{\theta_1(a_j+a_k + m_{1,2} - \epsilon_+)}{\theta_1(a_{jk}) \theta_1(2\epsilon_+ - a_{jk})}
	\end{align}
	\item $ u_1 = \frac{\epsilon_+ - m_2}{2} + x $, where $ x = 0,\frac{1}{2},\frac{\tau+1}{2},\frac{\tau}{2} $
	\begin{align}
	\sum_{I=1}^4 \frac{\theta_1(m_1-m_2+\epsilon_{1,2})}{2\theta_1(\epsilon_{1,2})} \prod_{j=1}^3 \frac{\theta_I(a_j + m_1 - \frac{m_2}{2} + \frac{\epsilon_+}{2})}{\theta_I(a_j-\frac{3\epsilon_+-m_2}{2})}
	\end{align}
\end{itemize}
The 1-string elliptic genus $ Z_1 $ is given by the summation of these two contributions.

\paragraph{2-string}
Let $ x_1, x_2 $ be $ 0, \frac{1}{2}, \frac{\tau}{2}, \frac{\tau+1}{2} $. The 2-string elliptic genus is given by summation of the following contributions from the JK-residues:
\begin{itemize}
	\item $ \epsilon_+ + u_1 - a_j = 0 $, $ -\epsilon_+ + u_1 + u_2 + m_1 = 0 $
	\begin{align}
	& \sum_{j=1}^3 \frac{\theta_1(2\epsilon_+) \theta_1(m_1-m_2-\epsilon_{1,2}) \theta_1(2a_j + m_1 - \epsilon_+)}{2\theta_1(\epsilon_{1,2}) \theta_1(m_1-m_2) \theta_1(2a_j + m_2 - 3\epsilon_+) } \nonumber \\
	& \quad \cdot \frac{\theta_1(2a_j+m_1-3\epsilon_+) \theta_1(2a_j + m_1 - 5\epsilon_+)}{\theta_1(2a_j + 2m_1 - m_2 - 3\epsilon_+) \theta_1(2a_j + 2m_1 - m_2 - 5\epsilon_+)} \\
	& \quad \cdot \prod_{k \neq j}^3 \frac{\theta_1(a_{jk} + m_1 - m_2 - 2\epsilon_+) \theta_1(a_j+a_k+m_2-\epsilon_+)}{\theta_1(a_{jk}) \theta_1(a_j+a_k+m_1-3\epsilon_+)} \nonumber
	\end{align}
	
	\item $ \epsilon_+ + u_1 - a_j = 0 $, $ -\epsilon_+ + u_1 + u_2 + m_2 = 0 $
	\begin{align}
	-&\sum_{j=1}^3 \frac{\theta_1(2\epsilon_+) \theta_1(m_1-m_2+\epsilon_{1,2}) \theta_1(2a_j+m_1-\epsilon_+)}{2\theta_1(\epsilon_{1,2}) \theta_1(m_1-m_2) \theta_1(2a_j + m_2 - 3\epsilon_+)} \nonumber \\
	&\quad \cdot \frac{\theta_1(2a_j + m_1 - \epsilon_+ - \epsilon_{1,2}) \theta_1(2a_j - m_1 + 2m_2 - 3\epsilon_+ - \epsilon_{1,2})}{\theta_1(2a_j + m_2 - \epsilon_+ - \epsilon_{1,2}) \theta_1(2a_j + m_2 - 3\epsilon_+ - \epsilon_{1,2})} \\
	&\quad \cdot \prod_{k\neq j}^3 \frac{\theta_1(a_{jk} - m_1 + m_2 - 2\epsilon_+) \theta_1(a_j+a_k+m_1-\epsilon_+)}{\theta_1(a_{jk}) \theta_1(a_j+a_k+m_2-3\epsilon_+)} \nonumber
	\end{align}
	
	\item $ -\epsilon_+ + 2u_1 + m_2 = x_1 $, $ -\epsilon_+ + u_1 + u_2 + m_1 = 0 $
	\begin{align}
	\begin{aligned}
	&\sum_{I=1}^4 \frac{\theta_1(m_1-m_2) \theta_1(m_1-m_2-\epsilon_{1,2}) \theta_1(m_1-m_2+2\epsilon_+)}{4\theta_1(\epsilon_{1,2}) \theta_1(2m_1-2m_2) \theta_1(2m_1-2m_2-2\epsilon_+)} \\
	&\quad \cdot \prod_{j=1}^3 \frac{\theta_I(a_j + \frac{m_2}{2} + \frac{\epsilon_+}{2}) \theta_I(a_j - m_1 + \frac{3m_2}{2} + \frac{\epsilon_+}{2})}{\theta_I(a_j + \frac{m_2}{2} - \frac{3\epsilon_+}{2}) \theta_I(a_j + m_1 - \frac{m_2}{2} - \frac{3\epsilon_+}{2})}
	\end{aligned}
	\end{align}
	
	\item $ \epsilon_+ + u_1 - a_j = 0 $, $ \epsilon_+ + u_2 - a_k = 0 $ $ (j \neq k) $
	\begin{align}
	\begin{aligned}
	&\sum_{j \neq k}^3 \frac{\theta_1(2a_{j,k}+m_1-\epsilon_+) \prod_{i=1}^2 \theta_1(2a_{j,k}+m_1-\epsilon_+-\epsilon_i)}{2\theta_1(\epsilon_{1,2})^2 \theta_1(a_{jk} + \epsilon_{1,2}) \theta_1(a_{jk} - \epsilon_{1,2}) \theta_1(2a_{j,k} + m_2 - 3\epsilon_+)} \\
	&\quad \cdot \frac{\prod_{i=1}^2 \theta_1(a_j+a_k+m_i-\epsilon_+) \theta_1(a_j+a_k+m_i-\epsilon_+-\epsilon_{1,2})}{\theta_1(a_j+a_k+m_{1,2}-3\epsilon_+)} \\
	&\quad \cdot \prod_{l \neq j,k}^3 \frac{\theta_1(a_j+a_l+m_{1,2}-\epsilon_+) \theta_1(a_k+a_l+m_{1,2}-\epsilon_+)}{\theta_1(a_{jl}) \theta_1(a_{kl}) \theta_1(a_{jl}-2\epsilon_+) \theta_1(a_{kl}-2\epsilon_+)}
	\end{aligned}
	\end{align}
	
	\item $ \epsilon_+ + u_1 - a_j = 0 $, $ -\epsilon_+ + 2u_2 + m_2 = x_2 $ and $ -\epsilon_+ + 2u_1 + m_2 = x_1 $, $ \epsilon_+ + u_2 - a_j = 0 $
	\begin{align}
	-2&\sum_{I=1}^4 \sum_{j=1}^3 \frac{\theta_1(m_1-m_2+\epsilon_{1,2}) \theta_1(2a_1+m_1-\epsilon_+) \theta_1(2a_j + m_1 - \epsilon_+ - \epsilon_{1,2}) }{4\theta_1(\epsilon_{1,2})^2 \theta_1(2a_j+m_2-3\epsilon_+) \theta_I(a_j+m_1-\frac{m_2}{2}-\frac{3\epsilon_+}{2})} \nonumber \\
	&\qquad \cdot \frac{\theta_I(a_j+m_1-\frac{m_2}{2}+\frac{\epsilon_+}{2}-\epsilon_{1,2}) \theta_I(a_j + \frac{m_2}{2} - \frac{7\epsilon_+}{2})}{\theta_I(a_i + \frac{m_2}{2}-\frac{3\epsilon_+}{2}-\epsilon_{1,2})} \\
	&\qquad \cdot \prod_{k \neq j}^3 \frac{\theta_1(a_j+a_k+m_{1,2}-\epsilon_+) \theta_I(a_k + m_1 - \frac{m_2}{2} + \frac{\epsilon_+}{2})}{\theta_1(a_{jk}) \theta_1(a_{jk}-2\epsilon_+) \theta_I(a_k+\frac{m_2}{2}-\frac{3\epsilon_+}{2})} \nonumber
	\end{align}
	
	\item $ -\epsilon_+ + 2u_1 + m_2 = x_1 $, $ -\epsilon_+ + 2u_2 + m_2 = x_2 $
	\begin{align}
	\begin{aligned}
	&\sum_{I,J=1}^4 \frac{\theta_{\sigma(I,J)}(0) \theta_{\sigma(I,J)}(2\epsilon_+) \theta_1(m_1-m_2+\epsilon_{1,2})^2 \theta_{\sigma(I,J)}(m_1-m_2+\epsilon_{1,2})}{8\theta_1(\epsilon_{1,2})^2 \theta_{\sigma(I,J)}(\epsilon_{1,2}) \theta_{\sigma(I,J)}(m_1-m_2) \theta_{\sigma(I,J)}(m_1-m_2+2\epsilon_+)} \\
	&\quad \cdot \prod_{j=1}^3 \frac{\theta_I(a_j+m_1-\frac{m_2}{2}+\frac{\epsilon_+}{2}) \theta_J(a_j + m_1 - \frac{m_2}{2} + \frac{\epsilon_+}{2})}{\theta_I(a_k + \frac{m_2}{2} - \frac{3\epsilon_+}{2}) \theta_J(a_k + \frac{m_2}{2} - \frac{3\epsilon_+}{2})}
	\end{aligned}
	\end{align}
	
	\item $ \epsilon_+ + u_1 - a_j = 0 $, $ \epsilon_{1,2} - u_1 + u_2 = 0 $ and $ \epsilon_{1,2} + u_1 - u_2 = 0 $, $ \epsilon_+ + u_2 - a_j = 0 $
	\begin{align}
	2& \sum_{j=1}^3 \frac{\theta_1(2a_j + m_1 - \epsilon_+) \theta_1(2a_j + m_1 - 3\epsilon_+) \theta_1(2a_j + m_1 - 3\epsilon_+ + \epsilon_{1,2})}{2\theta_1(\epsilon_{1,2}) \theta_1(2\epsilon_1) \theta_1(\epsilon_2-\epsilon_1) \theta_1(2a_j + m_2 - 3\epsilon_+ - \epsilon_1)} \nonumber \\
	& \quad \cdot \frac{\theta_1(2a_j+m_1-\epsilon_+-2\epsilon_1) \theta_1(2a_j+m_1-\epsilon_+-3\epsilon_1)}{\theta_1(2a_j + m_2 - 3\epsilon_+ - 2\epsilon_1)}  \\
	&\quad \cdot \prod_{k \neq j}^3 \frac{\theta_1(a_j+a_k+m_{1,2}-\epsilon_+) \theta_1(a_j+a_k+m_{1,2}-\epsilon_+-\epsilon_1)}{\theta_1(a_{jk}) \theta_1(a_{jk} - \epsilon_1) \theta_1(a_{jk}-2\epsilon_+) \theta_1(a_{jk}-2\epsilon_+-\epsilon_1)} + (\epsilon_1 \leftrightarrow \epsilon_2) \nonumber
	\end{align}
	
	\item $ \epsilon_{1,2} + u_1 - u_2 = x_1 $, $ -\epsilon_+ + u_1 + u_2 + m_2 = 0 $
	\begin{align}
	\begin{aligned}
	&\sum_{I=1}^4 \frac{\theta_1(m_1-m_2+\epsilon_{1,2}) \theta_1(m_1-m_2+2\epsilon_1) \theta_1(m_1-m_2 - \epsilon_1 + \epsilon_2)}{4\theta_1(\epsilon_{1,2}) \theta_1(2\epsilon_1) \theta_1(\epsilon_2-\epsilon_1)} \\
	&\quad \cdot \prod_{j=1}^3 \frac{\theta_I(a_j+m_1-\frac{m_2}{2}+\frac{\epsilon_+}{2}\pm\frac{\epsilon_1}{2})}{\theta_I(a_j+\frac{m_2}{2}-\frac{3\epsilon_+}{2}\pm\frac{\epsilon_1}{2})} + (\epsilon_1 \leftrightarrow 
	\epsilon_2)
	\end{aligned}
	\end{align}
	
	\item $ -\epsilon_+ - 2u_2 - m_2 = x_1 $, $ -\epsilon_+ + u_1 + u_2 + m_1 = 0 $
	\begin{align}
	\begin{aligned}
	&\sum_{I=1}^4 \frac{\theta_1(m_1-m_2-\epsilon_{1,2}) \theta_1(m_1-m_2-2\epsilon_+) \theta_1(m_1-m_2-4\epsilon_+)}{4\theta_1(\epsilon_{1,2}) \theta_1(2m_1-2m_2-2\epsilon_+) \theta_1(2m_1-2m_2-4\epsilon_+)} \\
	&\quad \cdot \prod_{j=1}^3 \frac{\theta_I(a_j - m_1 + \frac{3m_2}{2} + \frac{3\epsilon_+}{2})}{\theta_I(a_j + m_1 - \frac{m_2}{2} - \frac{5\epsilon_+}{2})}
	\end{aligned}
	\end{align}
\end{itemize}

\bibliographystyle{JHEP}
\bibliography{ref}
\end{document}